\documentclass[preprint]{aastex}

\usepackage{isotope}
\usepackage{capt-of}
\usepackage{multirow}
\usepackage{longtable}
\usepackage{lscape}
\usepackage{morefloats}
\usepackage{color}
\usepackage[section] {placeins}
\newcommand{\fig}[1]{Fig.\,\ref{#1}}

\newcommand{\tab}[1]{Tab.\,\ref{#1}}

\newcommand{\spr}{\mbox{$s$-process}}

\newcommand{\cdr}{\isotope[13]{C}}
\newcommand{\czw}{\isotope[12]{C}}

\newcommand{\nvi}{\ensuremath{^{14}\mem{N}}}

\newcommand{\ofu}{\ensuremath{^{15}\mem{O}}}
\newcommand{\ose}{\isotope[16]{O}}

\newcommand{\oac}{\ensuremath{^{18}\mem{O}}}

\newcommand{\fac}{\ensuremath{^{18}\mem{F}}}
\newcommand{\fne}{\ensuremath{^{19}\mem{F}}}

\newcommand{\nezw}{\ensuremath{^{22}\mem{Ne}}}

\newcommand{\mgfu}{\ensuremath{^{25}\mem{Mg}}}

\newcommand{\msun}{\ensuremath{\, M_\odot}}

\newcommand{\mppnp}{\textsf{mppnp}}

\newcommand{\MESA}{\texttt{MESA}}

\newcommand{\beq}{\begin{equation}}
\newcommand{\beqa}{\begin{eqnarray}}
\newcommand{\eeq}{\end{equation}}
\newcommand{\eeqa}{\end{eqnarray}}
\newcommand{\bedis}{\begin{displaymath}}
\newcommand{\edis}{\end{displaymath}}

\newcommand{\mem}[1]{\ensuremath{\mathrm{ #1}}}

\newcommand{\ov}{{\it overshoot}}

\graphicspath{{figures/}}

\slugcomment{DRAFT: \today}

\shorttitle{Single-degenerate scenario role in $p$-process}
\shortauthors{}

\begin{document}

\title{Heavy Elements Nucleosynthesis On Accreting White Dwarfs: building seeds for the $p$-process}
\author{U. Battino\altaffilmark{1,6},
  M. Pignatari\altaffilmark{2,5,6,8},
  C. Travaglio\altaffilmark{3,6},
  C. Lederer-Woods\altaffilmark{1,6},
  P. Denissenkov\altaffilmark{4,5,6},
F. Herwig\altaffilmark{4,5,6},
F. Thielemann\altaffilmark{7},
T. Rauscher\altaffilmark{7,9}
}

\altaffiltext{1}{School of Physics and Astronomy, University of Edinburgh, UK}
\altaffiltext{2}{E.A. Milne Centre for Astrophysics, Dept of Physics and Mathematics, University of Hull, HU6 7RX, United Kingdom}
\altaffiltext{3}{INFN - Istituto Nazionale Fisica Nucleare, Turin, Italy}
\altaffiltext{4}{Department of Physics \& Astronomy, University of Victoria, Victoria, BC, V8W 2Y2 Canada.}
\altaffiltext{5}{Joint Institute for Nuclear Astrophysics, Center for the Evolution of the Elements, Michigan State University, 640 South Shaw Lane, East Lansing, MI 48824, USA}
\altaffiltext{6}{The NuGrid collaboration, \url{http://www.nugridstars.org}}
\altaffiltext{7}{Department of Physics, University of Basel, Klingelbergstrasse 82, CH-4056 Basel, Switzerland}
\altaffiltext{8}{Konkoly Observatory, Research Centre for Astronomy and Earth Sciences, Hungarian Academy of Sciences, Konkoly Thege M. út 15-17, 1121, Budapest, Hungary}
\altaffiltext{9}{Centre for Astrophysics Research, University of Hertfordshire, College Lane, Hatfield AL10 9AB, United Kingdom}

\begin{abstract}
The origin of the proton-rich trans-iron isotopes in the solar system
is still uncertain.  Single-degenerate thermonuclear supernovae (SNIa)
with n-capture nucleosynthesis seeds assembled in the external layers
of the progenitor's rapidly accreting white dwarf phase may produce
these isotopes.  We calculate the stellar structure of the accretion
phase of five white dwarf models with initial masses $\geq$ 0.85\msun\ using the
stellar code \MESA.  The near-surface layers of the 1, 1.26, 1.32 and
1.38\msun\ models are most representative of the regions in which the
bulk of the $p$ nuclei are produced during SNIa explosions, and for
these models we also calculate the neutron-capture nucleosynthesis in
the external layers. Contrary to previous rapidly-accreting white dwarf models at lower mass, we find that the H-shell flashes are the main site of n-capture nucleosynthesis.
We find high neutron densities up to several 10$^{15}$cm$^{-3}$ in the
most massive WDs. Through the recurrence of the H-shell flashes these
intermediate neutron densities can be sustained effectively for a long
time leading to high neutron exposures with a strong production up to
Pb.  Both the neutron density and the neutron exposure increase with
increasing the mass of the accreting WD.  Finally, the SNIa
nucleosynthesis is calculated using the obtained abundances as seeds.
We obtain solar to super-solar abundances for $p$-nuclei with
A$>$96. Our models show that SNIa are a viable $p$-process production
site.
\end{abstract}


\section{Introduction}
\label{intro}



At the end of the Asymptotic Giant Branch (AGB) evolution,
low mass and massive AGB stars (M $<$ 8 \msun) loose all their envelope,
and cool down as carbon-oxygen (CO) white dwarfs (WD) \citep[e.g.,][]{herwig:05,karakas:14}.
However, a fraction of those CO WDs that are part of a binary system can evolve differently,
and explode as thermonuclear supernovae \citep[SNIa, e.g.,][]{hillebrandt:13},
if the CO mass can reach the Chandrasekhar limit (1.39 M$_{\odot}$).
Since the mid-twentieth century, accreting WDs have been the subject of a large number of studies \citep[e.g.,][]{mestel:52}. In this section, we focus on those studies most relevant to this work.
Neglecting any possible effect of rotation of the stellar progenitor,
a new born CO WD cannot be more massive than 1.1 \msun\ \citep{herwig:05}.
Therefore a successful accretion of at least 0.3 \msun\ from a companion is required in order to reach the Chandrasekhar limit.
This can mainly happen in two ways:
1) The WD accretes material  from a  main-sequence  or  evolved  companion \citep[single-degenerate scenario, hereafter SD, e.g.][]{whelan:73,nomoto:84b};
2) The WD merges with another CO WD being its companion in the binary system \citep[double-degenerate scenario, hereafter DD, e.g.][]{iben:84c,webbink:84}.
Observational evidence has been found for both channels.
\citet{mazzali:07} reported a spectral analysis of a large sample of SNIa,
indicating that all the supernovae considered here burned similar masses and suggesting that their progenitors had the same mass,
thus favoring a SD channel. On the other hand,
\citet{palkmor:12} showed how synthetic spectra from violent mergers of massive white dwarfs can closely resemble
spectra of the bulk of SNIa \citep[see also][]{maoz:14,seitenzahl:19}.


\citet{sion:79} and \citet{paczynski:83} modelled the long-term evolution in close binary systems of accreting WDs,
focusing on H-shell flash occurrence on the WD surface during the accretion and on their dependence on the accretion rate. 
\citet{paczynski:83} introduced a one-zone formalism that has been
adopted later by e.g., \citet{shen:07}.
In these and similar studies \citep[e.g.][]{nomoto:07, ma:13, starrfield:12}, the evolution of the deeper He-burning layers,
which are ignited when enough Helium is accumulated from the above H-burning layers,
is not followed in detail,
and only the feedback to the He luminosity is considered (see also \citet{nomoto:07}, \citet{ma:13} and \citet{starrfield:12}).
In this work, stable H-burning conditions on accreting WDs are considered,
and the dependence of various H-burning regimes (unstable, nova-recurrent, stable burning and super-Eddington limit) on the accretion rate,
on the accreted material metallicity and on the WD mass is defined.
In particular, at the same time we follow the evolution of the He-burning layers, simultaneously simulating H and He flashes.
One of the first works to model both H and He flashes was that presented by \citet{jose:93},
performing a numerical two-zone study. Later, \citet{tornambe:00} highlighted that, during the accretion phase,
He-flash thermal pulses (hereafter TPs) strongly reduce the capability of the star to retain the accreted mass,
making it more difficult for the SNIa SD scenario to reach the Chandrasekhar mass.

\citet{langer:00} studied the evolution of close binary systems consisting of a main sequence star and a white dwarf,
resolving both components of the binary system, but treating the WD as a material point, i.e. without resolving its structure.
With this setup, they investigated the properties of the systems as a function of the initial donor star mass, initial WD mass,
initial revolution period, and chemical composition. 
They obtained that, in order to reach the Chandrasekhar limit,
the initial mass of the WD when the accretion starts should be at least 0.7 M$_\odot$.
However, the authors did not consider the mass-retention limiting effect given by He-flashes.


\citet{denissenkov:17} modelled rapidly accreting white dwarfs (RAWDs) that follow He-shell flashes.
This study indicated that the He retention efficiency after a flash is less than 10\% for solar metallicity,
thereby suggesting that the SD channel for SNIa is unlikely. 
This result is consistent with observations by \citet{woods:13},
who showed how the measured emission by the He II recombination is much lower than expected
if the SD scenario was the dominant SNIa channel.
In particular, for photospheric temperatures between 10$^{5}$ and 10$^{6}$ K,
characteristic of 1.1 \msun\ CO WDs, accreting  above  the  steady  nuclear-burning  limit,
they showed how the SD channel would imply a substantial contribution to the HeII-ionizing continuum.
Consistent with \citet{woods:13}, \citet{johansson:16} found that
the SD channel should contribute to about 3\%-6\% of the total SNIa rate,
based on spectroscopic observations of a large sample of early-type galaxies.
On the other hand, \citet{kobayashi:20} looked at the galactic chemical evolution (GCE) of Mn with respect to Fe.
Indeed, a single-degenerate channel is needed to reach the observed [Mn/Fe] in the Sun,
since the high densities ($\sim$2$\times$10$^{8}$ g cm$^{-3}$ ), essential for Mn production \citep[see][]{thielemann:86},
are only found in near-Chandrasekhar-mass WD explosions, which are naturally formed in the SD scenario.
They concluded that up to 75\% of all SNIa must be near-Chandrasekhar-mass in order to explain the present Mn content in the Sun.
This is in disagreement with a similar study recently presented by \citet{eitner:20},
who suggested that only 25\% of SNIa comes from near-Chandrasekhar-mass exploding WDs.
Despite the disagreement between these GCE studies,
even the lower SD contribution obtained by \citet{eitner:20} is about a factor of five larger than what obtained by \citet{johansson:16}.

In conclusion,
studies using different observables predict highly discrepant estimations on the fraction of SNIa originating from the SD or the DD channels. 
While this issue must be addressed by future studies,
the results presented in this work have been obtained assuming a successful SD channel scenario in reaching the Chandrasekhar limit.


SNIa are fundamental sources for galactical chemical evolution.
They produce iron group elements in the ejecta exposed to the most extreme SN conditions \citep[e.g.,][]{iwamoto:99,brachwitz:00}.
In the ejecta exposed to less extreme conditions, intermediate mass elements like Si and Ca are made,
as also confirmed from optical spectra of recent SNIa remnants \citep[e.g.,][and references therein]{filippenko:97,hillebrandt:13}.
Moreover, SNIa are considered potential sources for the so called $p$-process,
the nucleosynthesis process responsible for the production of a small fraction of the total abundances beyond iron in the Solar System,
made up of proton-rich isotopes called the $p$-nuclei.
\citet{howard:91}, for the first time,
considered Carbon-Oxygen WD explosions as able to provide the right conditions for $p$-nuclei production,
provided a fundamental assumption is made: heavy element (56$<$A$<$209) rich material is present,
at the onset of the explosion, in the outer accreted layers of the SNIa progenitor.
In this way,
a chain of photo-induced reactions \citep[the $\gamma$-process][]{audouze:75,arnould:76} synthesises the $p$-nuclei during the explosion,
starting from the initial heavy element seed distribution. 
This assumption was tested also by \citet{kusakabe:11},
who analyzed the effects of different seed distributions on $p$-production using the one-dimensional W7 model \citep{nomoto:84}.
They found that, in all cases of seed distribution considered,
about 50\% of $p$-nuclides had an abundance ratio $p$/$^{56}$Fe consistent with the observed solar value (where the $p$ in the ratio is the abundance of a given $p$-isotope, produced during the SNIa explosion).
On the other hand, they were not able to produce, at the same level as $^{56}$Fe, the Mo and Ru $p$-nuclei.
Previous studies \citep[e.g.][]{woosley:78,rauscher:02} found the same underproduction of $^{92}$Mo, $^{94}$Mo, $^{96}$Ru and $^{98}$Ru,
even when considering different explosive scenarios, like Type II Supernovae.

\citet{travaglio:11} (hereafter TRV11) and more recently \citet{travaglio:15} calculated high-resolution 2D hydrodynamic models of SNIa.
Differently from \citet{kusakabe:11}, they adopted a multi-dimensional approach and highly resolved the most external layers,
as a good resolution of the outer zones of the SNIa, where $p$-process takes place, is fundamental for $p$-nuclei production.
TRV11 obtained that almost all the $p$-process nuclei could be produced with similar enhancement factors relative to $^{56}$Fe,
including the puzzling $^{96}$Ru and $^{98}$Ru and a much higher production of $^{92}$Mo, $^{94}$Mo compared to previous studies.
On the other hand, they confirmed the crucial role of \citet{howard:91} assumption about the initial \spr\ seed distribution,
which is meant to be produced as a consequence of the recurrent He-flashes during the accretion phase considering a SD scenario.
Indeed, the $p$-process in SNIa can be efficient only if there is a previous heavy-elements enrichment during the accretion stage,
which will act as seed for the $p$-process during the SNIa explosion. 
Furthermore, this enrichment should be located in the most external layers of the SNIa progenitor,
which will experience the right temperature range (1.5 $<$ T $<$ 3.8 GK)to effectively produce $p$-nuclei via photo-disintegration reactions.
If they are located too deep,
the temperature conditions are too extreme and the SN shock will completely photo-disintegrate all abundances heavier than the Fe group,
including any $p$-process isotope.

Possible sources of heavy element seed abundances in the SNIa progenitor are the \spr\ abundances accumulated during the previous AGB phase.
However, these abundances will be buried inside the progenitor during the accretion phase,
at mass coordinate $\sim$ 0.6-0.7 M$_{\odot}$.
In the SD case, this region will be first convectively mixed during the simmering phase before exploding as SNIa,
where a convective region grows at the WD center over a timescale of $\sim$ 1000 yr \citep{piro:08}, including up to about 1\msun\ .
The \spr\ enrichment built in the AGB He intershell is mixed over the WD structure, with a strong dilution of heavy element enrichment.
Therefore, in this scenario there will be no $p$-process material ejected from this stellar region.

TRV11 assumed that, during the accretion phase,
the \spr\ material was formed and accumulated in the external 0.2-0.3 M$_{\odot}$ of the SNIa progenitor,
with a distribution typical of the main \spr\ component.
\citet{iben:81} proposed that heavy elements could actually be made in these conditions,
where helium accumulates below the accreting H-burning layers until a He-flash occurs. 
They suggested the \nezw\ ($\alpha$,n)\mgfu\ as the main neutron-source during these flashes.
That work was a first attempt to access and discuss this possibility, without simulating the accretion phase.

\citet{denissenkov:17} demonstrated for the first time in multi-cycle
He-shell flash simulations that RAWDs can be prodigious producers of
neutron-capture elements. The mechanism is a continued H-ingestion
during the He-shell flash that does not lead to the dramatice
H-ingestion flash observed in single-star post-AGB He-shell flash
models \citet[e.g.][]{Herwig:1999uf,MillerBertolami:2006dr}. In the
post-AGB situation the H-ingestion into He-shell flashes leads to an
immediate and violent convective feedback of the H ingested into the
He-convection zone. In 1D models this leads to a split of the
convection zone preventing $^{13}$N from the H + $^{12}$C reaction to
reach the bottom of the He-convection zone where it is hot enough to
release neutrons. However, observations of post-AGB star Sakurai's
object \citep{Asplund:1999ta} demonstrated first-peak heavy-element
enhancements of $\approx 3\mathrm{dex}$. These observed n-capture
elements can be reproduced in models of the H-ingestion into the
He-shell flash convection zone if it is assumed that the split of the
convection zone due to energetic feedback of the H + $^{12}$C reaction
is delayed by approximately 60 convective turn-over times or $\approx
900\mathrm{min}$ \citep{Herwig:2011dj}. Initial attempts to simulate
the H-ingestion flash have shown a complex dynamic interaction between
nuclear energy release and a convective response via the Global
Oscillation of Shell H-ingestion \citep{Herwig:2014cx}. In the RAWD models on the other hand, the degeneracy of the older WD
is much higher and the He-shell flashes are much stronger than in the
single post-AGB case. Therefore the H ingestion in the relatively
low-mass WDs simulated by Denissenkov et al (2017) do not show the
split of the convection zone and can maintain the H-ingestion
conditions and the associated high neutron density for many month,
thereby creating the conditions for heavy-element production in
RAWDs. \citet{2019MNRAS.488.4258D} presented i-process yields for
seven metallicities from solar to [Fe/H] = -2.6 and showed that
low-metallicity RAWD models can naturally explain the observed
abundance of most CEMP-r/s stars. \citet{cote:18} showed that
RAWDs can have significant contributions to several first-peak
neutron-capture species. These models show that RAWDs are efficient producers of trans-iron
elements. However, the processes in high-mass accreting WDs have not
yet been investigated.





Other $p$-process sites were also considered in addition to SNIa from H-accreting WDs.
\citet{goriely:02} found that He-detonation in He-accreting CO WD is accompanied with an efficient $p$-process.
Most of the $p$-nuclei, including the puzzling cases of Mo and Ru isotopes, 
were found to be co-produced in these conditions in relative quantities close to solar.
Unfortunately they were underproduced (except $^{78}$Kr) compared to the Ca to Fe species.
He-accretor progenitors were considered by \citet{piersanti:14}.
They explored the impact of different He accretion-rates on the WD structure,
without investigating the nucleosynthesis coming from a SNIa event with this kind of progenitor.
On the other hand, they extensively explored the parameter space given by different combinations of
WD mass and Helium accretion rates, determining the mass retention efficiency as a function of the accretor total mass and accretion rate.
Additionally, they derived interpolation formulae, which could be directly used in population synthesis codes,
in order to describe the evolution of WDs accreting He-rich matter.
Finally, we must mention the potential role of Type II Supernovae in explaining Solar System abundances of $p$-nuclei:
\citet{travaglio:18} considered the $\gamma$-process in Type II Supernovae,
which is very efficient for a wide range of progenitor masses at solar metallicity.
Since it is a secondary process, its contribution is strongly reduced below solar
metallicity. In particular, \citet{travaglio:18} found that the contribution from Type II Supernovae to the Solar System content of
the $p$-nuclei is less than 10 \% , with the only exception of light $p$-nuclei up to $^{92}$Mo.
Another very interesting feature has been found by \citet{ritter:18} in O-C shell mergers during the pre-SN of massive stars,
with a $p$-process production increased by more than an order of magnitude compared to the explosive $\gamma$-process component.
So far this effect has been found at metallicities equal or larger than Z=0.01,
while a contribution from lower metallicities, essential to boost the Galactic $\gamma$-process contribution of massive stars,
is currently under investigation.
For this reason, is essential to consider a complementary $p$-process source that could be represented by SNIa.

In this work we will present our simulations of WDs accreting solar-composition matter.
We will investigate the neutron-capture nucleosynthesis of heavy elements during the accretion phase,
testing the assumption of TRV11.
This work is organized as follow. In Section \ref{wd:mod} we describe stellar code and post-processing nucleosynthesis tools.
In Section \ref{sec:mesa} the stellar models are presented,
and in Section \ref{sec:postpro} we describe the nucleosynthesis results of the seeds for the $p$-process.
In the same section, we use the seeds we obtained as initial abundances to calculate the explosive nucleosynthesis
using our multi-D SNIa model \citep[described in][]{travaglio:11}, focusing on p-nuclei formation.
Our conclusions are given in Section \ref{sec:conclusions}.

\section{Accreting WD models: main stellar model properties and computational tools}
\label{wd:mod}


The stellar models presented in this section are computed using the 1D
stellar code \MESA\ \citep[\MESA\ revision 4219,][]{mesa}.
The accretion itself as well as the mass loss process and the
  shell burning and mixing process are all subject to important
  non-spherically symmetric effects. The accretion process is thought
  to proceed through a disk. In order to justify the spherical
  approximation, the assumption is that the disk accretion quickly
  spreads across the white dwarf.  \citet{macdonald:83} showed that
  shear effects, acting on the surface of WDs accreting material from
  a viscous disk, can drive dynamical instabilities. Accreted material
  rapidly spread over the whole stellar surface. This is aided by the
  fact that the thermal time scale of the outermost layers, where the
  accretion happens, is shorter than the nuclear time scale on which
  enough material is accumulated to launch a He-shell flash.
  In that regard spherically-symmetric simulations are likely well
  justifies. Another 3D aspect of the problem is the reverse
  common-envelope evolution that drives the mass loss when the
  accreting white dwarf expands into the orbit of the companion. This
  process is usually modeled with a simple Roche-lobe
  prescription. The uncertainty introduced by this assumption is
  larger in lower-mass accreting WDs because of the high-mass WDs
  considered here the super-Eddington wind is dominating the mass
  loss. Finally, 3D effects are important for the treatment of
  convective boundaries and the energetic feedback of the H-ingestion
  process itself as discussed in the introduction. The example of
  Sakurai's object shows \citep{Herwig:2011dj} that observations can
  be explained if the split of the convection zone is delayed. Such a
  delay would always enhance the heavy-element production. In our
  models we do not apply any delay to the convection split induced by
H ingestion. In this way the contribution of H ingestion into He-shell flash convection to the overall heavy-element abundance enrichment in our models is underestimated when H ingestion takes place. However, in our models the main contribution to heavy element enrichment in high-mass RAWDs comes from H-shell flashes.

The solar distribution used as a reference is given by \cite{grevesse:93}.
The CO-enhanced opacities are used throughout the calculations, using OPAL tables \citep{Iglesias:96}.
For lower temperatures, the corresponding opacities from \cite{ferguson:05} are used.
Mass loss is only considered when Super-Eddington wind conditions are met,
which is triggered when the luminosity of the accreting WD exceeds the Eddington luminosity, defined as

\begin{equation}
L_{Edd} = \frac{4\pi\ GMc}{\kappa}
\end{equation}

where $M$ is the WD mass and $\kappa$ is the opacity \citep{nomoto:82,shen:07,ma:13}.
Considering a pure ionized H plasma,
a simple derivation of the Eddington limit is obtained by setting the force of the outward radiation pressure equal to the inward gravitational force, thus giving:

\begin{equation}
L_{Edd} = 3.2 \times 10^{4} \frac{M}{\msun\ } L_{\odot}
\end{equation}

Once L$_{Edd}$ is defined, if the stellar luminosity L exceeds  L$_{Edd}$ mass-loss is calculated according to \citet{paczynski:86}, 
who determined an analytical relation between the stellar luminosity and the mass outflow rate:

\begin{equation}
\frac{dM}{dt } = 1.67 \times 10^{18} gs^{-1} (\frac{L}{ L_{Edd}} -1)
\end{equation}

Convective mixing follows the standard mixing length theory \citep{cox:68}. 
For \MESA\ simulations the convective boundary mixing (CBM) is computed using the 
exponential overshooting of \citep{herwig:00}:

\beq\
D(dr) = D_{0} \times \exp(-2dr/f_1Hp_0)
\label{cbm:falk}
\eeq\

where dr is the geometric distance to the convective boundary. 
The term f$_1$$\times$Hp$_0$ identifies the scale height of the \ov\ regime.
The values D$_0$ and Hp$_0$ are respectively the diffusion coefficient and the 
pressure scale height at the convective boundary.







For the simulations the following nuclear networks are adopted:

1) \emph{wd-accr.net}, including 33 isotopes from protons to $^{26}$Mg linked by nuclear-reactions including the pp chains, 
the CNO tri-cycle, the triple-$\alpha$ and the $\alpha$-induced reactions
\czw\ ($\alpha$, $\gamma$)\ose\ , \nvi\ ($\alpha$, $\gamma$)\fac\ (e$^+$$\nu$)\oac\ , \oac\ ($\alpha$,
$\gamma$)\nezw\ , \cdr\ ($\alpha$, n)\ose\ and \fne\ ($\alpha$, p)\nezw\ . We use the NACRE \citep{angulo:99}
reaction rate compilation for most reactions. For the \czw\ ($\alpha$,$\gamma$)\ose\ we adopt the rate by
\citep{kunz:02}, whose values at He-burning temperatures are consistent with \citet{deboer:17} (see their figure 29).
\nvi\ (p, $\gamma$)\ofu\ by \citep{imbriani:04} and the triple-$\alpha$ by \citep{fynbo:05}.
This network is used for the main calculations in this work, presented in section \ref{sec:wd_accr};

2) $nova.net$, which includes 48 isotopes from H to \isotope[30]{Si} coupled by 120 reactions.
The same network was used by e.g., \citet{denissenkov:13}, and, compared to \emph{wd-accr.net}, it 
differs by including He-burning reactions. This network is used for the tests presented in section \ref{sec:burning_regime};

3) $cno-extras.net$, including 13 isotopes from H to \isotope[24]{Mg} coupled by 56 reactions.
This network is used for the tests presented in section \ref{sec:burning_regime}.

We tested an extended version of \emph{wd-accr.net}, including and linking  nuclear species up to \isotope[56]{Fe}.
No significant impact on stellar structure and nucleosynthesis was observed,
confirming \emph{wd-accr.net} to be a suitable network to be adopted for this study.

The initial WD models used to start the simulations are included in the \MESA\ revision 4219, in the $data$ folder.
For the models presented in section \ref{sec:burning_regime}, we use the \MESA\ WDs models 0.639from3.0z2m2.mod, 0.856from5.0z2m2.mod, 1.025from7.0z2m2.mod and 1.316from8.5z2m2.mod.
For the models in section \ref{sec:wd_accr},
we use 0.639from3.0z2m2.mod (initial WD mass M$_{ini}$ = 0.639 \msun\ ) and 0.856from5.0z2m2.mod (M$_{ini}$ = 0.856 \msun\ ).
In the center they are made of about 30$\%$, 68$\%$ and 2$\%$ on C, O, Ne respectively.
They are characterized by an He-rich cap over the CO core, which is a relic of the He-burning layers on top of the CO core,
with typical mass fractions 55$\%$ $\le$ He $\le$ 65$\%$, 
35$\%$ $\le$ C $\le$ 45$\%$ and 2$\%$ $\le$ O $\le$ 10$\%$.

We also use the following ONeMg WDs: $M = 1.259 \msun$, with 60\% O, 30\% Ne, 7\% C and 3$\%$ Mg, and $M = 1.376 \msun$, with 60\% O, 37\% Ne, and 3$\%$ Mg.


Notice that we use ONeMg WDs, that are not SNIa progenitors.
This does not impact our results since the physics of interest is in the accreted envelope and independent of interior \citep[][]{wolf:13}.
On the other hand, it is plausible to use these WD models as progenitors assuming that they reached that mass by previous accretion, or by forming recently as hybrid WDs \citep[][]{denissenkov:15}.

\section{Description of the stellar models}
\label{sec:mesa}
\subsection{Identification of different burning regimes}
\label{sec:burning_regime}

In accreting WDs, different burning regimes are possible, mainly depending on the accretion rate, the metallicity and composition of the accreted material. 
Five main accretion regimes are identified: strong H-shell flashes, mild H-shell flashes, 
steady H-burning, red-giant and super-Eddington wind.

We modelled accretion adopting an accretion-mass rate within the steady H-burning regime,
as this ensures the most favorable conditions to efficiently accrete material, and hence reach the Chandrasekhar-mass limit, are met.
Given a specific composition of the accreted material and assuming constant accretion rate,
stable burning of H requires accretion rates within a narrow range. 
Below this range unstable H burning occurs, and above this range the accreted material piles-up building a red giant size envelope, 
or eventually a strong overflow is set if the nuclear Eddington limit is reached for higher rates.
Super-soft X-ray sources have been proposed to be astrophysical sites where steady burning of H accreted onto WDs is taking place \citep{vandenheuvel:92}. For such systems,
theoretical stellar calculations found H-rich matter transfer rates ranging around 10$^{-7}$ \msun\ yr$^{-1}$ \citep[e.g.,][]{shen:07,nomoto:07}.

In this section, we present a preliminary study done to identify steady H-burning conditions with the MESA code.
The first set of stellar models,
focused specifically on the H-burning phase (i.e. they experience no He-flash instabilities),
and their basic properties are presented in Table \ref{tab:WDmodel_name_burn_regime}.
All the models are computed starting from WDs with different masses, described in section \ref{wd:mod}.

The main goal is to verify the value of the critical mass-accretion rate,
which is the transition point from unstable to stable H-burning, as a function of the accreting WD mass.
This information will then be implemented in our second set of stellar models,
described in the next section and in table \ref{tab:WDmodel_name},
where the stellar structure is computed over a longer timescale, including during the onset of He-flash instabilities.
In both tables \ref{tab:WDmodel_name_burn_regime} and \ref{tab:WDmodel_name},
models' name are given consistently to the fundamental initial models conditions:
the initial mass of the accreting WD in \msun\ units is given after the initial 'M',
while the metallicity of the accreted material is given after the `Z`. For example,
M0p856.Z1m2 in table \ref{tab:WDmodel_name} simulates the accretion of Z=0.01 material onto a WD with initial mass M=0.856 M\msun\ .
In table \ref{tab:WDmodel_name_burn_regime},
models denoted by $bare$ were calculated using the \MESA\ CNO-cycle network  $cno-extras.net$ and no CBM applied.
Models denoted by $cbm1$ and $cbm2$ were calculated using $nova.net$,
with the $cbm1$ models using the CBM prescription of \citet{denissenkov:13}, 
i.e. a single-exponential decay overshooting scheme with f=0.004 applied to every convective boundary, 
and the $cbm2$ ones using f=0.014.

\begin{figure}[htbp]
\begin{center}
\includegraphics[scale=0.3]{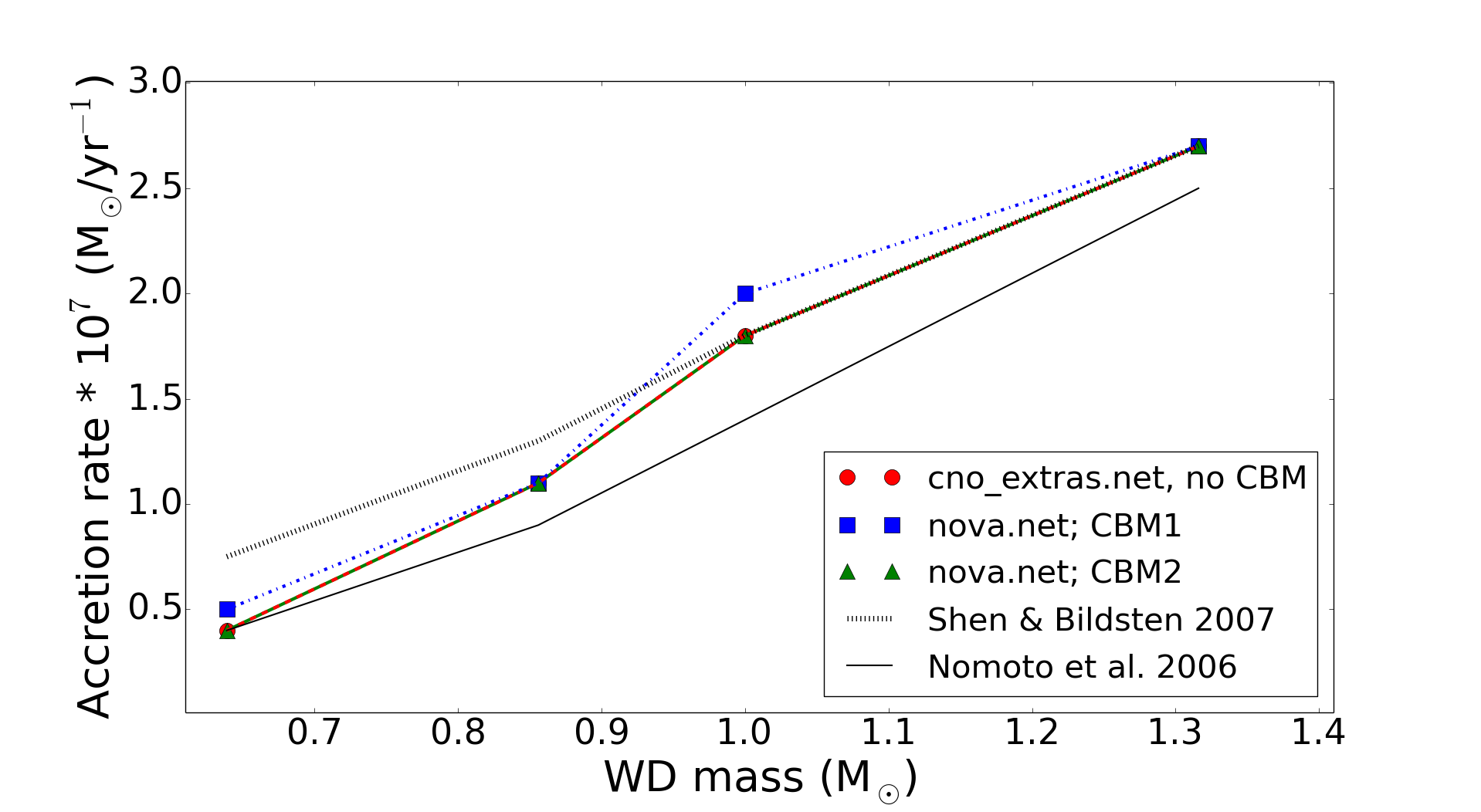}
\end{center}
\caption{Critical mass accretion rate marking in a transition from unstable to stable hydrogen burning as a function of the accreting WD mass. 
All the accretion rates lower than the critical values result in unstable hydrogen burning, making the reaching of the Chandrasekhar limit more difficult. 
A comparison between our accretion models and literature is also provided.}
\label{accrate:mwd}
\end{figure}

These calculations allow to explore the impact of different nuclear networks and CBM schemes.
A summary of the results is given in figure \ref{accrate:mwd}.
For comparison, the results from \citet{nomoto:07} and \citet{shen:07} are also included. 
The results show that: 1) the details of the stable-burning accretion are only marginally affected by the CBM,
or 2) by using different networks; 
3) a good agreement is obtained between these calculations and \citet{nomoto:07} and \citet{shen:07},
despite using different stellar codes and simulation setup.

As indicated in the previous section, the accreted material has Z=0.01 metallicity.
The results obtained above depend on the metallicity of the accreted material \citep{shen:07}.
In the next section, we will use the fixed accretion rates indicated in table \ref{tab:WDmodel_name} consistently with the WD initial mass. 
From the table and from figure \ref{accrate:mwd},
the accretion rate needed to burn H steady increases with increasing WD mass as expected \citep{shen:07,ma:13}. 
This will also imply a decrease of the interpulse period, between two He-flash episodes,
with the WD mass. \citet{paczynski:74} found the same result in AGB stars,
where in that case the relation is between the interpulse period and the CO core mass.

\subsection{Accretion models}
\label{sec:wd_accr}

During the accretion phase, H is efficiently burned via the CNO cycle, accumulating He on stellar layers just below the surface.
Eventually, in this way the He shell is compressed and heated in thin-shell conditions,
until a thermonuclear run-away occurs, developing a pulse-driven convection zone \citep[hereafter PDCZ, e.g.,][]{tornambe:00}. 
The accretion phase appears to have similarities with the convective TPs occurrence during the evolution of AGB stars.
But the main qualitative difference is that during the AGB phase there is a large H-rich envelope on top of the He-burning region,
while in this case there is no envelope. 
As also indicated by previous works \citep[e.g.,][]{kato:99}, during the accretion regular convective TPs occur in the He-burning layers.
The duration of the accretion phase to reach the Chandrasekhar-mass implies thousands of TPs. However,
here we simulate only a limited number of TPs for each initial WD mass, to study the different conditions obtained along the accretion phase.

In table \ref{tab:WDmodel_name} the list of models analyzed in this work together with their main parameters setup, is presented.
All the models were calculated using the reaction network $wd-accr.net$ and the CBM prescription described by equation \ref{cbm:falk}.
Each model is characterized by a fixed accretion rate value, which ensures a stable H-burning.
Initial mass, metallicity, CBM parameters and number of TPs simulated are also given in the table.
In table \ref{tab:WDm3z2m2_TPprop}, for each model the main stellar properties after each TP event are given.
In addition to the mass coordinate at the top and bottom of the He-flash convective zone, 
for every TP the largest temperature (in logarithm) at the bottom of the flash-convective zone is presented. 
This is one of the most crucial stellar properties from a nucleosynthetic point of view, 
since it directly determines the efficiency of the \nezw($\alpha$,n)\mgfu\ and the resulting neutron density.
Finally, the stellar mass at and after the TP is given, allowing to estimate the mass retention efficiency for each model.


As we mentioned, 
only a limited number of TPs is simulated for each model. 
The only exception is model M1p025.Z1m2, where 137 TPs are calculated to explore relevant changes of the TPs properties in a longer sequence, and to study if the accreting WD is actually increasing in mass.   

\begin{figure}[htbp]
\centering
\resizebox{10.8cm}{!}{\rotatebox{0}{\includegraphics{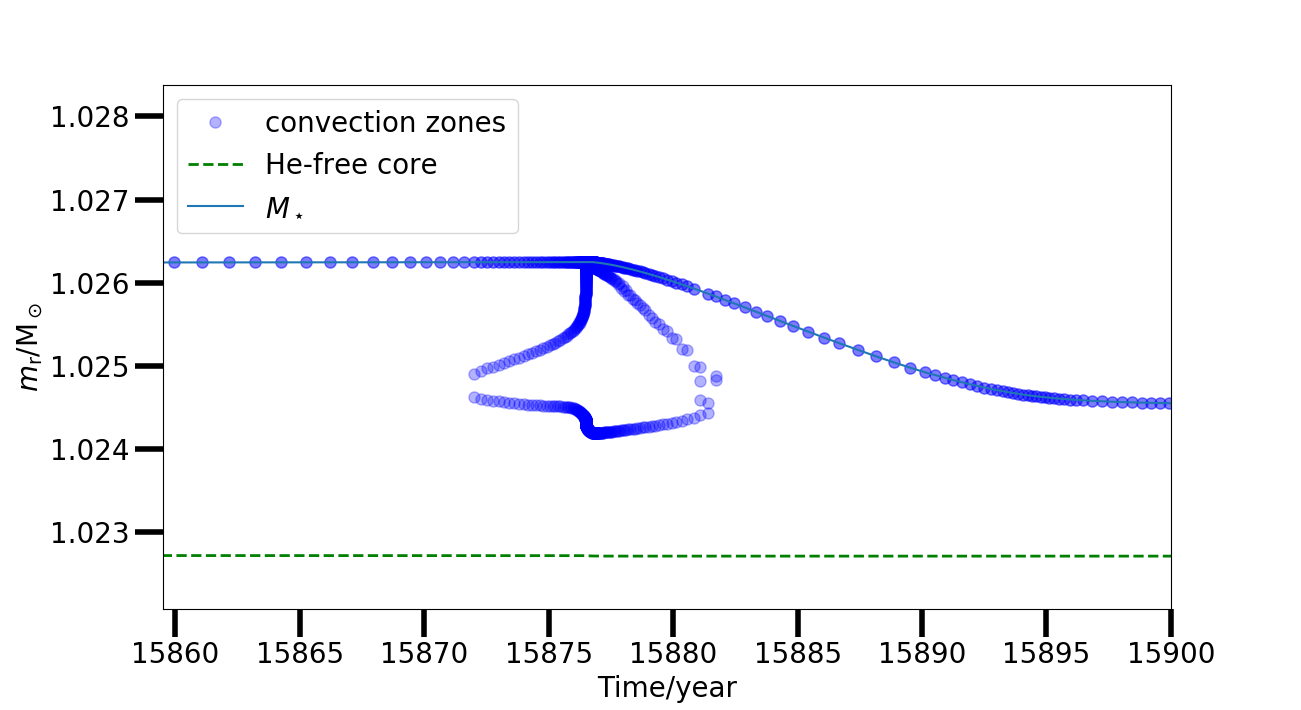}}}
\resizebox{10.8cm}{!}{\rotatebox{0}{\includegraphics{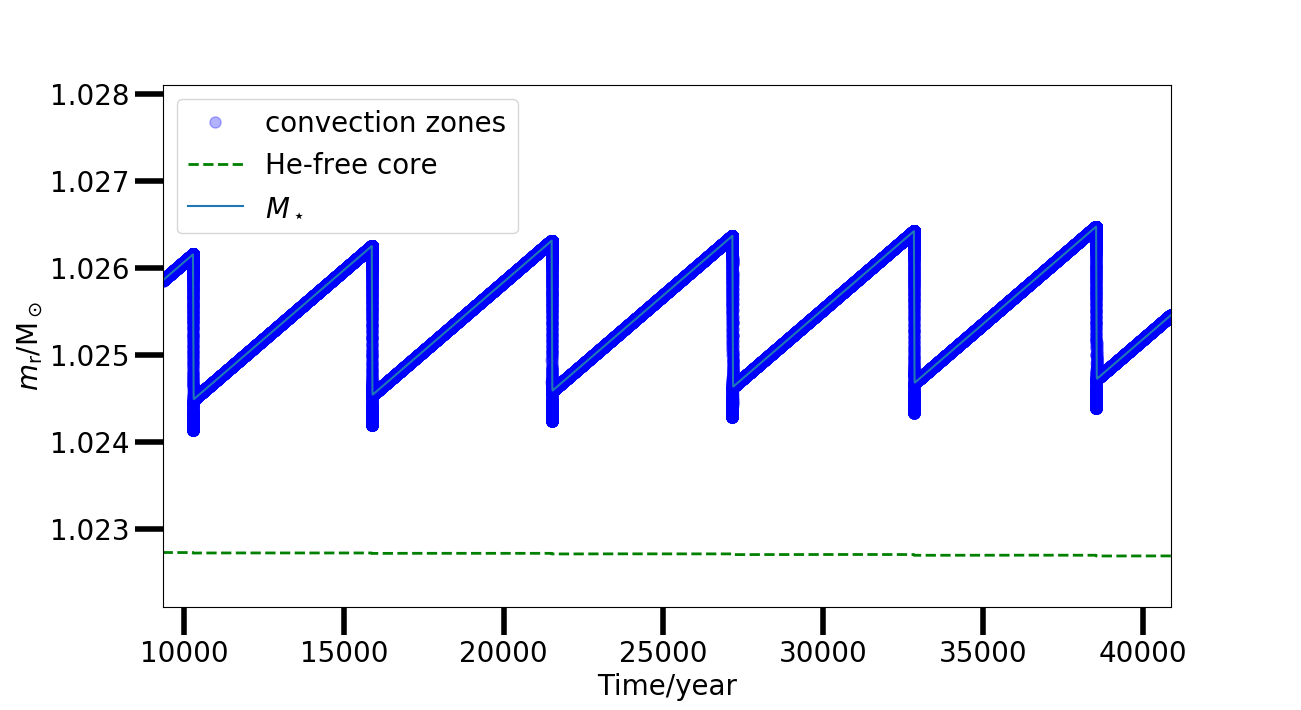}}}
\resizebox{10.8cm}{!}{\rotatebox{0}{\includegraphics{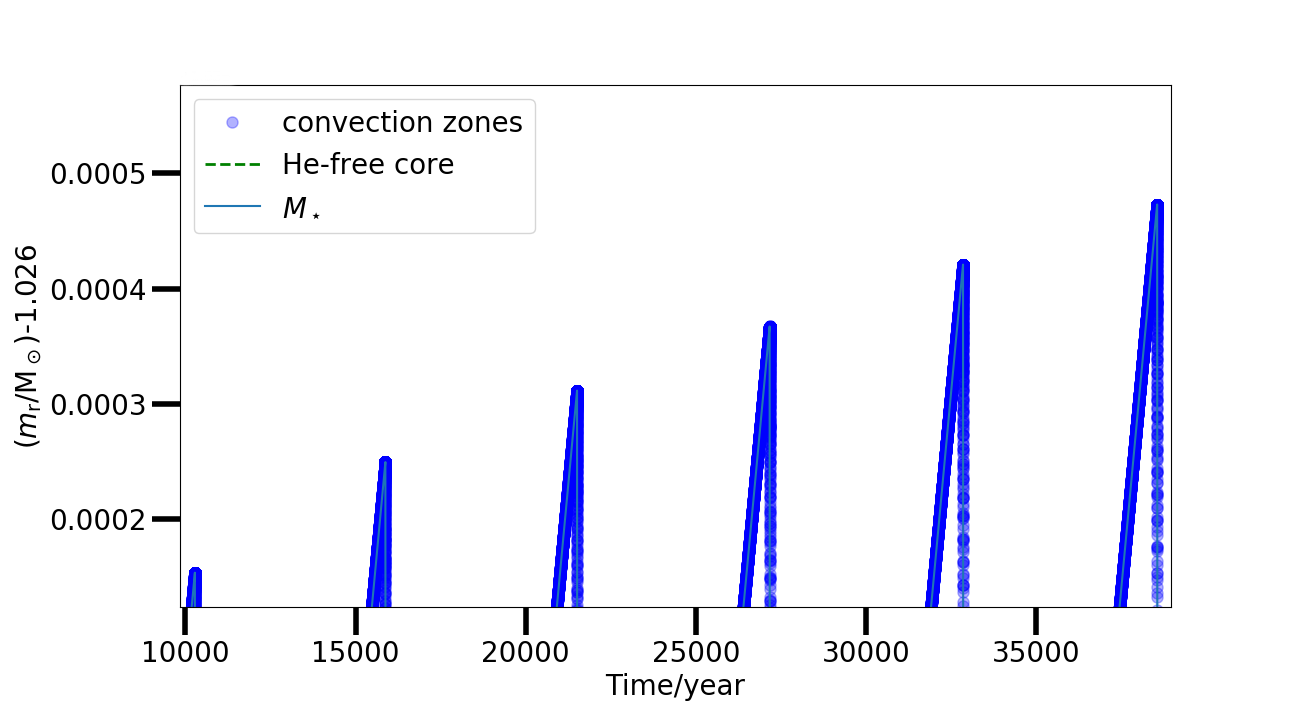}}}
\caption{Upper panel: Kippenhahn diagram of a Typical Helium-flash event during the accretion phase of our 1 M$_\odot$ accreting white dwarf model. M$_{r}$ is the lagrangian mass-coordinate of the star (being zero in the star center and equal to the total star mass on the surface).
 In particular, note the fast mass-loss following the flash due to super-Eddington wind.
 Middle panel: Kippenhahn diagram of a sequence of Helium-flashes.
Lower panel: A zoom of the plot in the middle panel, showing the WD mass increase during the accretion. The fast mass loss after each TP is visible.}
\label{TPs:acc}
\end{figure}

Figure \ref{TPs:acc} shows the Kippenhahn diagram for a TP and a sequence of TPs of the model M1p025.Z1m2.
The results of this model (and of all the models in table \ref{tab:WDmodel_name}) confirm the calculations of \citet{tornambe:00} and \citet{denissenkov:17} :
the Eddington limit \citep{rybicki:79} is easily exceeded immediately after each He flash, ejecting almost all the matter accumulated during 
the previous interpulse phase. This reduces the efficiency of the accretion process toward Chandrasekhar mass.
Despite these powerful winds, in figure \ref{TPs:acc} it is highlighted that the star is still growing in mass.
Higher WD mass will finally result in higher surface gravity,
thus reducing the net amount of mass loss after the super-Eddington wind phase which follows every TP.
It is important to notice how one of the main differences, between the WD models presented in \citet{denissenkov:17} and in this work,
is the different WD mass range considered. Specifically, the WD initial mass range is extended in this study,
including up to near-Chandrasekhar WDs.
According to the models presented here, the average net amount of mass loss after every TP in M0p856.Z1m2 is around 0.006 \msun,
while it is around 0.001 \msun\ in M1p025.Z1m2 model (figure \ref{TPs:acc}), 2$\times$10$^{-4}$ \msun\ in M1p256.Z1m2,
1.5$\times$10$^{-5}$ \msun\ for M1p316.Z1m2 and 10$^{-5}$ \msun\ for M1p376.Z1m2.

\begin{figure}[htbp]
\begin{center}
\includegraphics[scale=0.3]{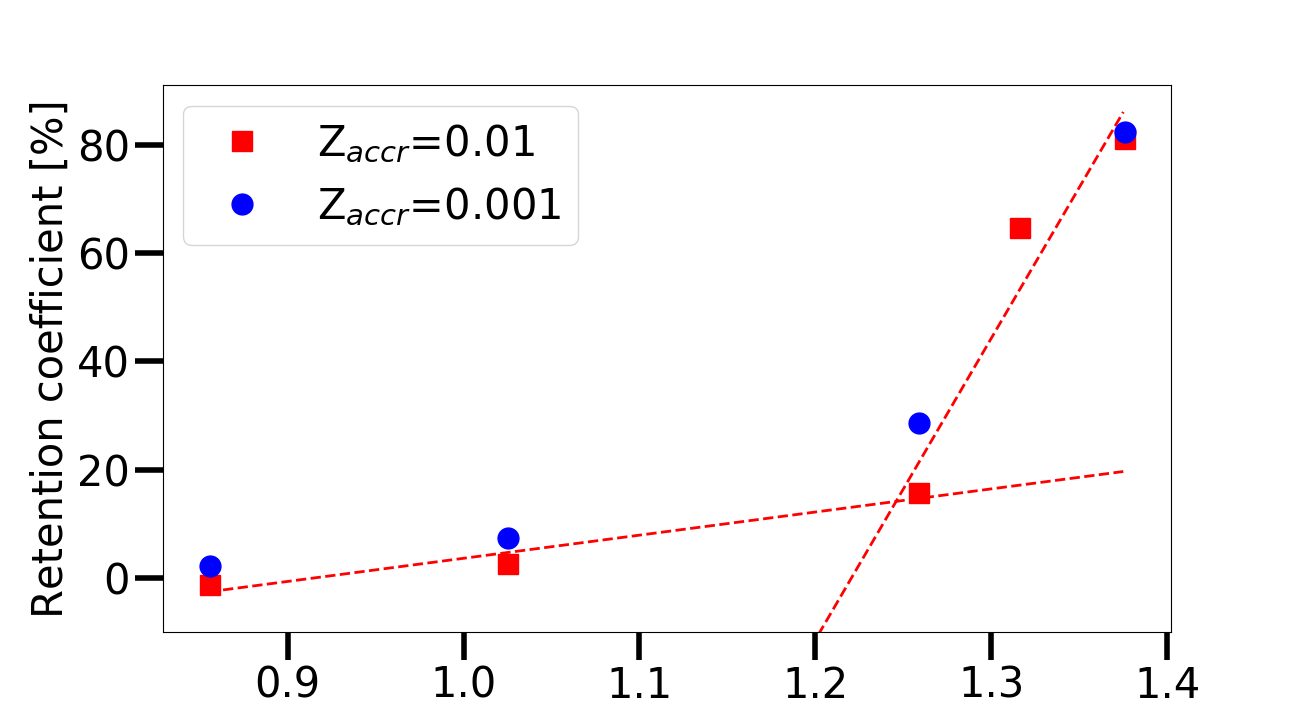}
\end{center}
\caption{Mass retention efficiency (ratio of the stellar masses  before and after the mass loss by super-Eddington wind, following each He-flash) as a function of the WD mass for two different metallicities of the accreted material. Mass-loss is purely accounted by super-Eddington wind. Notice how a higher amount of retained material is a consequence of both an increase of the star mass, as surface gravity is increased, and and a decrease in metal content in the accreted material, as the wind efficiency drops with the opacity. Two linear fits are shown to interpolate the Z=0.01 points for WD masses M$_{WD}$ $<$ 1.25 M$_{\odot}$ and M$_{WD}$ $>$ 1.25 M$_{\odot}$, showing a strong variation of the mass retention increase as a function of M$_{WD}$ as soon as M$_{WD}$ becomes larger than 1.25 M$_{\odot}$ }
\label{ret:WDmass}
\end{figure}

Looking at figure \ref{ret:WDmass},
we can estimate the minimum initial mass a WD must have in order to be able to reach the Chandrasekhar limit (hereafter M$_{inimin}$).
The plot shows the retention efficiency,
defined as the ratio between the mass gained before and the mass lost after each TP, as a function of initial WD mass. 
Current stellar evolution theory predicts that a newly formed CO core cannot exceed about 1.1 \msun\ without igniting carbon,
that would convert the star into Oxygen and Neon, which are not SNIa progenitors \citep{becker:79,dominguez:93,dominguez:96}.
Furthermore, since stellar lifetimes exponentially decrease with increasing initial mass,
the stellar donor must be less massive than the WD progenitor star.
This ensures to have a less-evolved companion (i.e. not yet a WD), still with a H-rich envelope able to provide the accreted matter.
The direct consequence of this constrain, is that the amount of mass that can be accreted is not infinite,
as it must be lower than the companion's envelope mass.
Adopting these conditions, we estimate  M$_{inimin}$ integrating the double-linear fit shown in figure \ref{ret:WDmass}.
For 0.85 \msun\ $<$ M$_{WD}$ $<$ 1.25 \msun\ , mass retention efficiency ($\eta_{He}$) is fitted by

\begin{equation}
  \eta_{He} = 0.43 \times M_{WD} - 0.39
\end{equation}

While for 1.25 \msun\ $<$ M$_{WD}$ $<$ 1.40 \msun\ the linear fit becomes much steeper:

\begin{equation}
  \eta_{He} = 5.57 \times M_{WD} - 6.80
\end{equation}

The total accreted material (M$_{acc}$) will be given by the following mass integration:

\begin{equation}
M_{acc} = (M_{com} - M_{c}) \times \int_{M_{ini}}^{M_{Ch}} \eta_{He} dm
\label{int:toch}
\end{equation}

where M$_{ch}$ is the Chandrasekhar mass,  M$_{ini}$ is the WD mass at the beginning of the mass transfer,
M$_{com}$ is the companion mass and M$_{c}$ the companion's core mass once it evolves past the main-sequence,
when it fills its Roche lobe and mass transfer starts.
Hence, (M$_{com}$ - M$_{c}$) gives the mass from the H-rich envelope that can be accreted onto the WD.
From equation \ref{int:toch}, M$_{acc}$ is just enough to reach the Chandrasekhar mass when M$_{ini}$ = M$_{inimin}$ = 0.91 \msun\ .
This kind of WD would need to accrete $\sim$ 0.34\msun\ to become at least as massive as 1.25\msun\ .
At that point, the retention efficiency will grow much faster as the WD increases, making the evolutionary path to SNIa much easier.
On the other hand,
\citet{woods:13} showed how WD with photospheric temperature T  $\sim$ 2 $\times$ 10$^{5}$ K
(typical of accreting WD with masses  around $\sim$ 0.9 \msun\ )
could only accrete up to $\sim$ 0.1 \msun\ to be compatible with the lack of detection of He II recombination lines.
We anyway notice how our M1p256.Z1m2 model has a photospheric temperature T  $\sim$ 9 $\times$ 10$^{5}$ K,
which would allow it to reach the Chandrasekhar limit without contradicting the null detection of the He II line
(see figure 8 in \citet{woods:13}).
This means that accreting $\sim$ 0.1 \msun\ may be enough for an accreting WD with M$_{WD}$ $\sim$ 1.1 \msun\ to reach a mass comparable to M1p256.Z1m2, hence being consistent with the observational constraints of \citet{woods:13}.
We therefore set M$_{inimin}$ $\sim$ 1.1 \msun\ , compatible with the highest mass a newly formed CO WD can have.

Moreover, two more aspects with opposite potential impact must be mentioned:
1) lowering the metallicity of the accreted material leads to a global increase of the retention efficiency\citep[see][]{denissenkov:19}:
in this case,
by setting Z=0.001 the retention efficiency increases by around a factor of two in the initial mass range between 1 and 1.25 \msun\ ;
2) Roche-Lobe overflow may lower the retention efficiency, but it's not considered here,
both for computational time requirements and for the small impact on  heavy-elements nucleosynthesis on top layers of the WD,
which is the main target of the present study.
It is important to consider that in our simulations we resolve the accreting WD structure but we do not simulate the evolution of the donor, 
using instead a constant accretion rate on the WD surface.

\begin{figure}[htbp]
\centering
\resizebox{12.3cm}{!}{\rotatebox{0}{\includegraphics{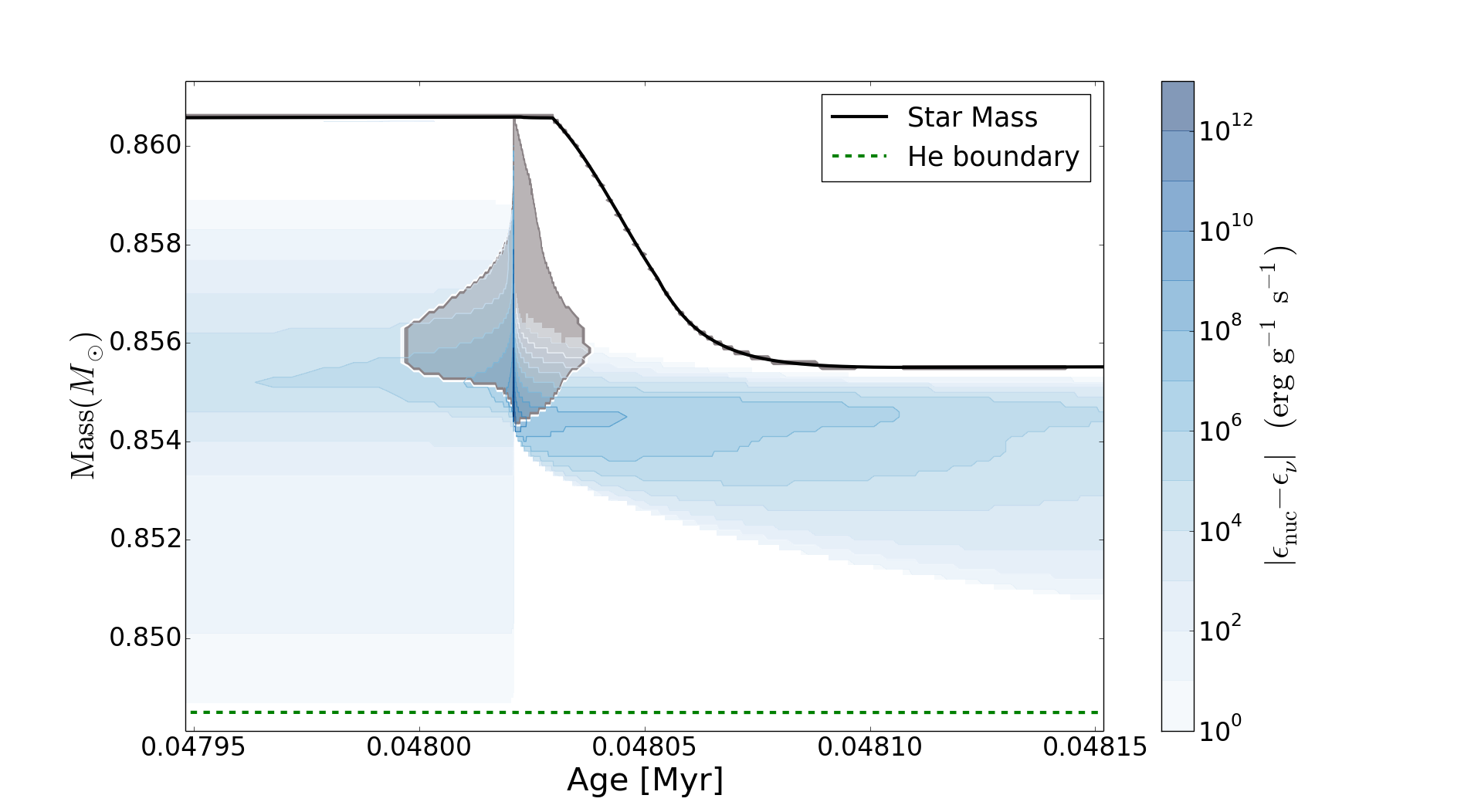}}}
\resizebox{12.3cm}{!}{\rotatebox{0}{\includegraphics{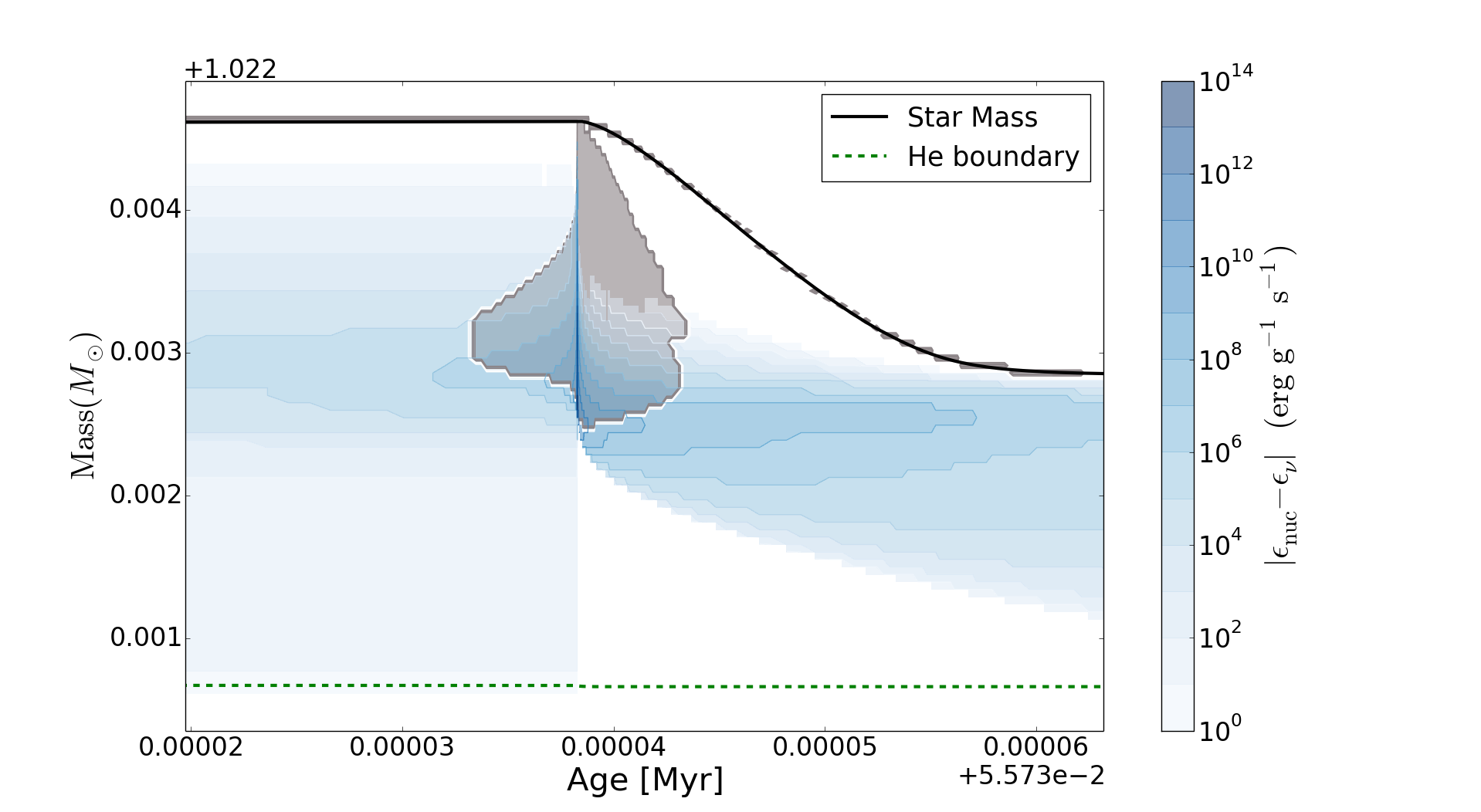}}}
\resizebox{12.3cm}{!}{\rotatebox{0}{\includegraphics{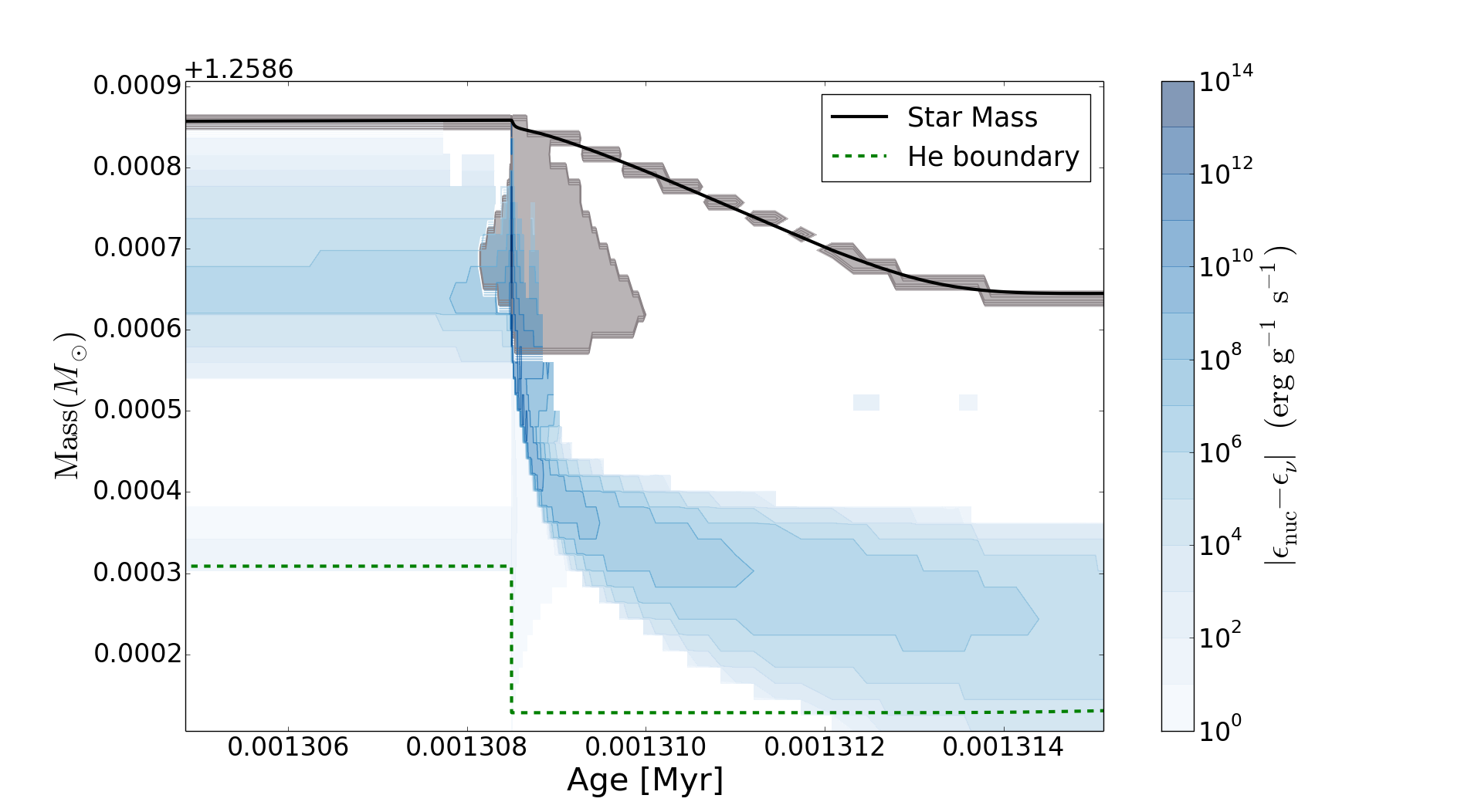}}}
\caption{Kippenhahn diagram of a convective TP for the models M0p856.Z1m2 (top panel), M1p025.Z1m2 (middle panel) and M1p259.Z1m2 (bottom panel). The energy generation (blue shaded areas) and convective zones (grey-shaded areas) are indicated.}
\label{fig:summarykipcont}
\end{figure}

In figure \ref{fig:summarykipcont} the Kippenhahn diagram for three convective TPs of different models are shown.
As in figure \ref{TPs:acc}, the strong mass loss is clearly visible, forcing the structure to loose most of the mass accreted in the last interpulse phase.
The highest energy generation coincides with the highest downward extension of the PDCZ. It is also interesting to observe that the energy generation from He burning continues to be significant also when the H-burning has already started.

\begin{figure}[htbp]
\centering
\includegraphics[width=1.0\textwidth]{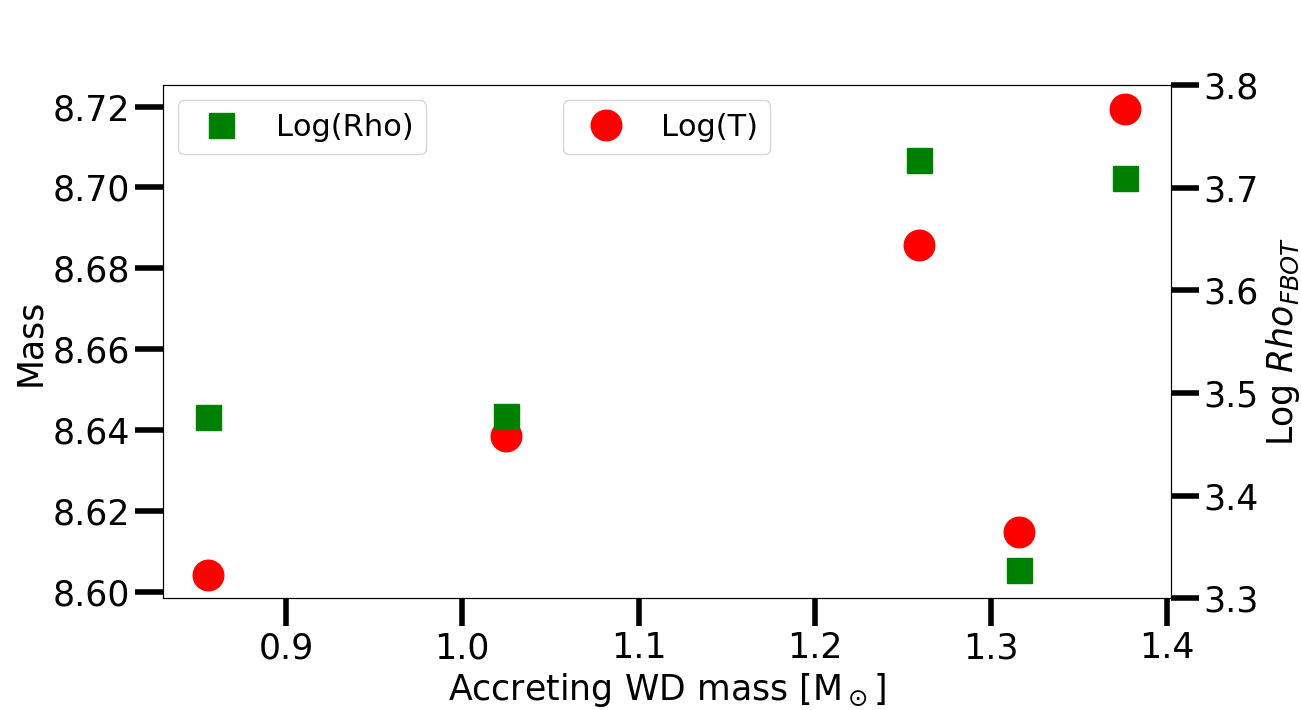}
\caption{Temperature and density at the bottom of the PDCZ with respect to the WD mass for the models M0p856.Z1m2, M1p025.Z1m2, M1p259.Z1m2, M1p316.Z1m2 and M1p376.Z1m2. As expected, the more massive the WD is, the higher the temperature and density are the TP onset. Details about  the important exception of model M1p316.Z1m2 are given in the text.}
\label{Temp:mwd}
\end{figure}
\begin{figure}[htbp]
\centering
\resizebox{8.3cm}{!}{\rotatebox{0}{\includegraphics{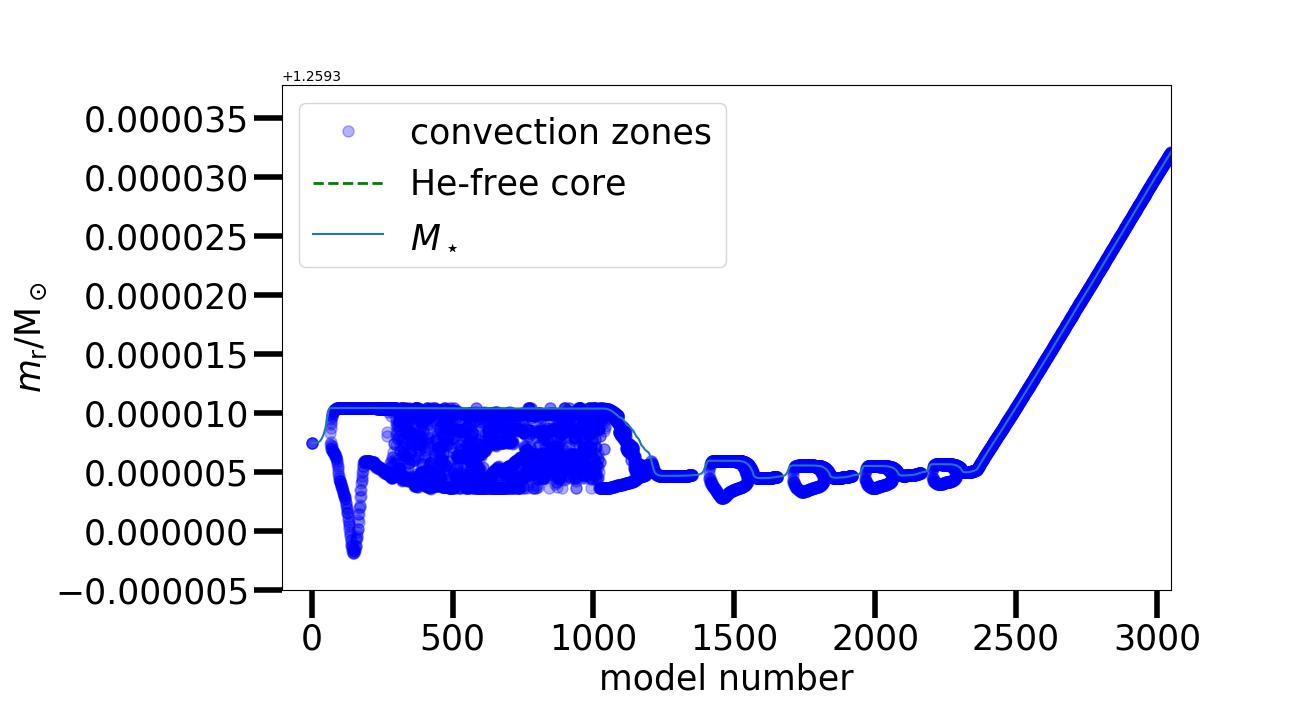}}}
\resizebox{8.3cm}{!}{\rotatebox{0}{\includegraphics{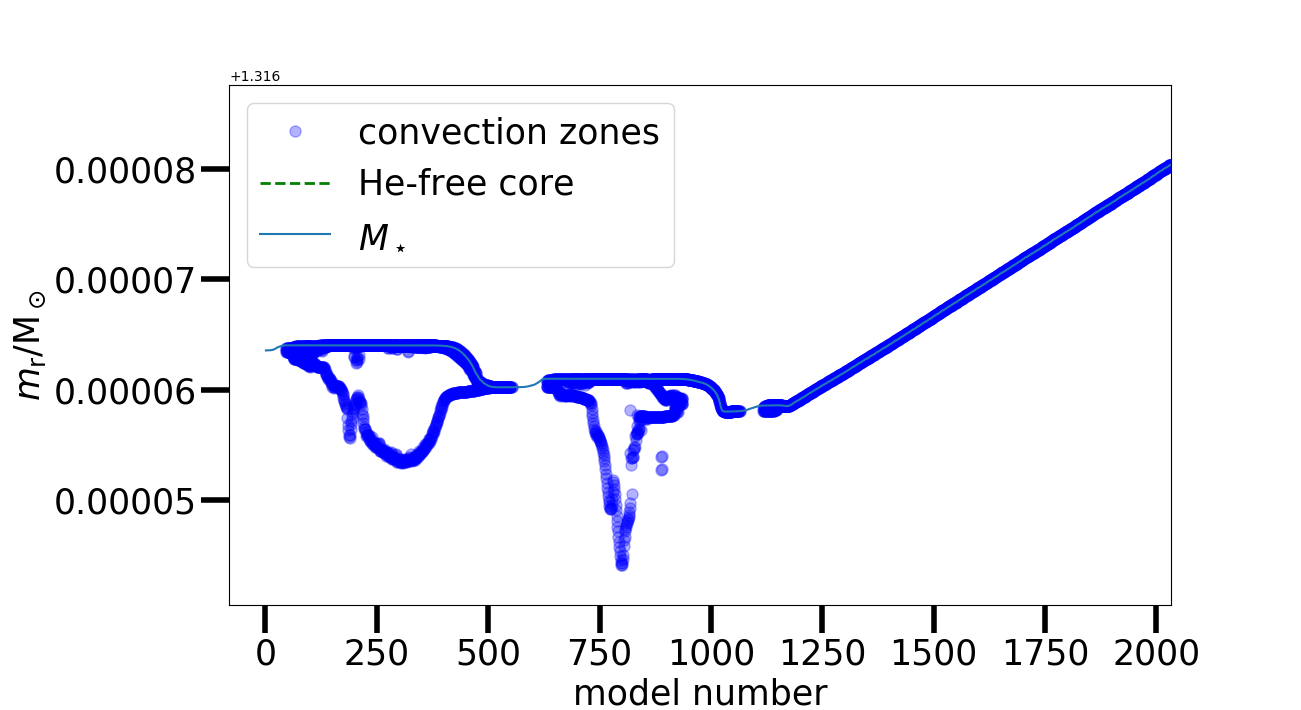}}}
\resizebox{8.3cm}{!}{\rotatebox{0}{\includegraphics{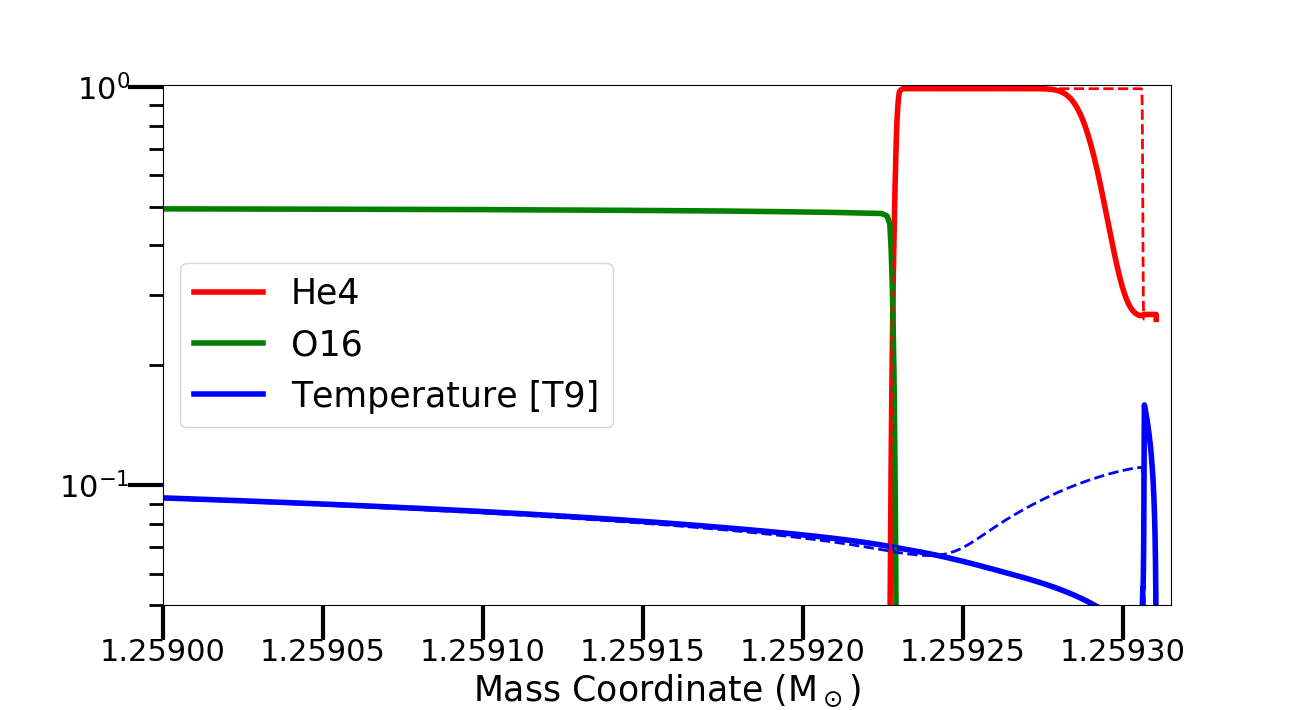}}}
\resizebox{8.3cm}{!}{\rotatebox{0}{\includegraphics{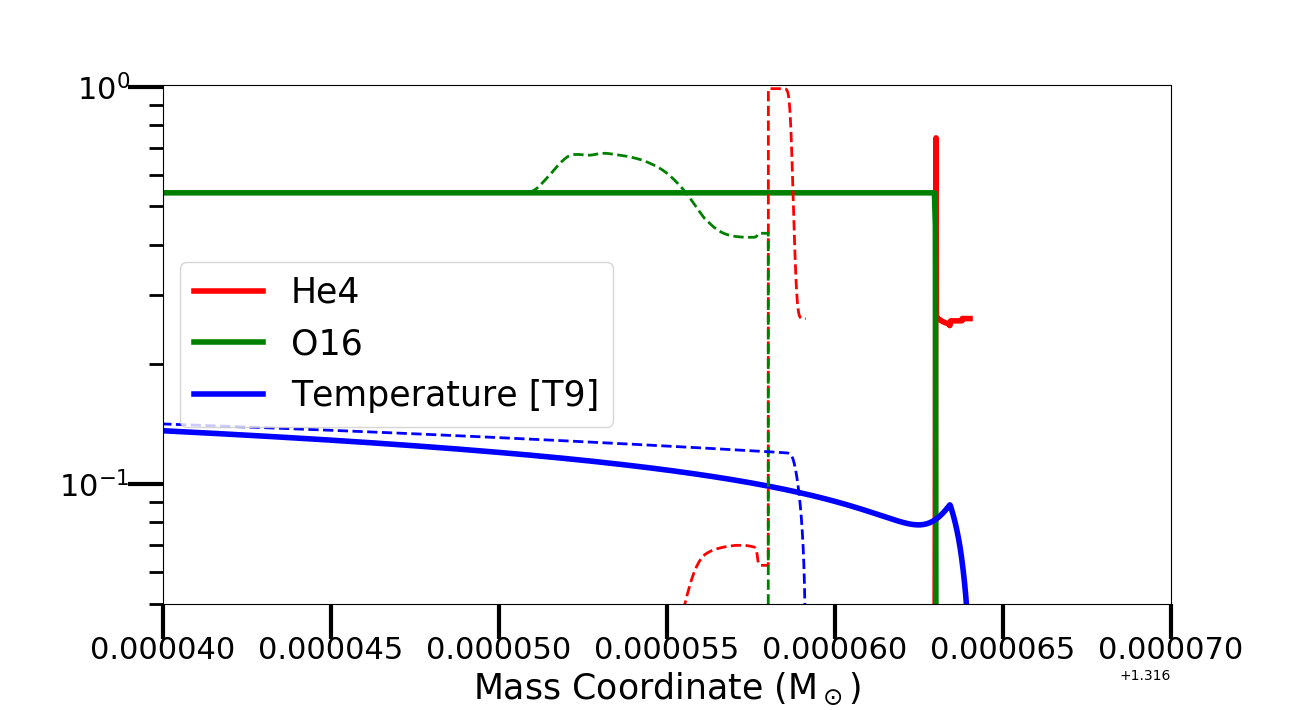}}}
\caption{Top panels: Kippenhahn diagrams of models M1p259.Z1m2 and M1p316.Z1m2  showing the H-flashes before the onset of the steady accretion and the following TP. Bottom panels: temperature and reference-elements profiles of the same models showing the impact on the temperature around the top boundary of the He-free core, showing the temperature before (continue line) and after (dashed line) the H-flashes. No impact is seen in the M1p259.Z1m2 model.}
\label{fig:flfeedback}
\end{figure}

In figure \ref{Temp:mwd}, the evolution of temperature and density at the bottom of the PDCZ
(T$_{FBOT}$, see table \ref{tab:WDm3z2m2_TPprop}) with respect to the WD mass is shown.
The temperatures T$_{FBOT}$ are taken from one TP, but they are representative of others TPs calculated with the same WD mass.
Indeed, according to table \ref{tab:WDm3z2m2_TPprop}, the T$_{FBOT}$ variation between following TPs is marginal.
Even for the more extended M1p025.Z1m2 model with more than a hundred simulated TPs, T$_{FBOT}$ variation is less than 3\%.
Generally, the logarithm of the temperature increases linearly with the WD mass (see figure  \ref{Temp:mwd}).
The main reason of the temperature increase with the progenitor mass is the thin shell instability  and partial degeneracy \citep{kippenhahn:90,herwig:05}: 
the higher the accreting WD mass, the lower the shell thickness will be, making thin shell instability more efficient,
as the expanding shell pushed by triple $\alpha$ will need more time to reach a thickness large enough to restore hydrostatic equilibrium,
causing a longer temperature rise.
A similar variation with the WD mass is also visible for the density.
This trend is consistent for all the mass range explored, with the important exception of model M1p316.Z1m2,
where both temperature and  density at the bottom of the PDCZ are significantly lower than what expected from the general trend.
The reason behind this apparent strange behavior, is shown in Figure \ref{fig:flfeedback}: in the top two panels,
the Kippenhahn diagram of models M1p259.Z1m2 and M1p316.Z1m2 are given, showing some initial Hydrogen-flashes,
anticipating the steady-accretion before the onset of the TP takes place:
the much higher temperature of the flashes experienced by M1p316.Z1m2 (0.27 GK), compared to those taking place in  M1p259.Z1m2 (0.17 GK),
has an impact on the temperature right at the top boundary of the He-free core (see bottom panels),
where the He-flash will ignite, increasing it by $\sim$ 20\%.
On the other hand, this fingerprint is not left by the weaker Hydrogen-flashes of  M1p259.Z1m2,
with the net result of a lower degeneracy state of the He-intershell in the  M1p316.Z1m2 model,
which causes both weaker He-flashes and the faster increase of the mass retention efficiency visible in figure \ref{ret:WDmass}.

We have mentioned that the thickness of the He intershell decreases with the increase of the WD mass.
This is shown in figure \ref{mh:mint}. Between models M0p639.Z1m2 and M1p376.Z1m2, the mass of the He intershell decreases by almost three orders of magnitude, from few 10$^{-2}$\msun\ down to a few 10$^{-5}$\msun.
Figure \ref{mh:mint} shows the amount of H-rich mass which is not processed by H-burning, before the next convective TP occurs.
Its mass is reduced from M $\sim$ 10$^{-4}$\msun\
for model M0p639.Z1m2 to M $\sim$ 10$^{-7}$\msun\ for model M1p376.Z1m2.
These last numbers need to be compared to the mass accreted during the same interpulse phase, that is comparable to the He-intershell mass. This means that at the onset of the convective TP, only a few per cent of the H-rich material accreted still has to burn.
In general, for the mass range explored in our simulations and for all the convective TPs, the size of the He intershell mass is about two orders of magnitude larger than the unburned H-rich material. Similar relative proportions between He-rich layers and H-rich material at the surface are found in the very late thermal pulse (VLTP) or late thermal pulse (LTP) events in post-AGB stars. In these stars, the H is ingested in the He-burning regions causing important effects on the nucleosynthesis of these objects \citep[e.g.,][]{herwig:99,miller-bertolami:07}. 
We will discuss this effect in more detail in the next sections. 

\subsection{M1p025.Z1m2: extended stellar calculations}
\label{subsec:comp_1msun_WD}

As reported in table \ref{tab:WDmodel_name}, model M1p025.Z1m2 includes 137 convective TPs. 
This is the longest accreting WD model in the set presented in this work.
For higher WD masses, convergence criteria for stellar calculations become more and more difficult to achieve,
and require more computing time. For example, with the simulations setup adopted in this work,
a typical TP of the M1p025.Z1m2 model is calculated in about 7000 timesteps.
More than 25000 timesteps are needed to complete a TP of the M1p376.Z1m2 model.
The larger number of timesteps needed can be explained by the higher temperatures at the bottom of the PDCZ.
For all the simulations presented in this work, the accretion interpulse phase is the most difficult to resolve,
with up to 10$^4$ mass zones needed.
The largest resolution, down to 10$^{-12}$\msun\ ,
is required to have convergence in the top layers where accretion and H burning is taking place.
These limitations are more severe for larger WD masses and larger accretion rates.

The simulations details for different TPs of M1p025.Z1m2 are given in table \ref{tab:WDm3z2m2_TPprop}.
Here only the numbers for the first seven TPs and the last three are reported.
Their variation during the evolution is extremely slow, and local variations between TPs are not relevant.
In general,
the thermodynamics conditions at the bottom of the TP, as well as the amount of mass lost after each TP event,
are not changing significantly during the evolution of the M1p025.Z1m2 model. 

\subsection{H-ingestion events in WD-accretion models}
\label{subsec:comp_1msun_WD_HIF}

\begin{figure}[htbp]
\centering
\resizebox{14.8cm}{!}{\rotatebox{0}{\includegraphics{kip_cont_M1p02}}}
\resizebox{14.8cm}{!}{\rotatebox{0}{\includegraphics{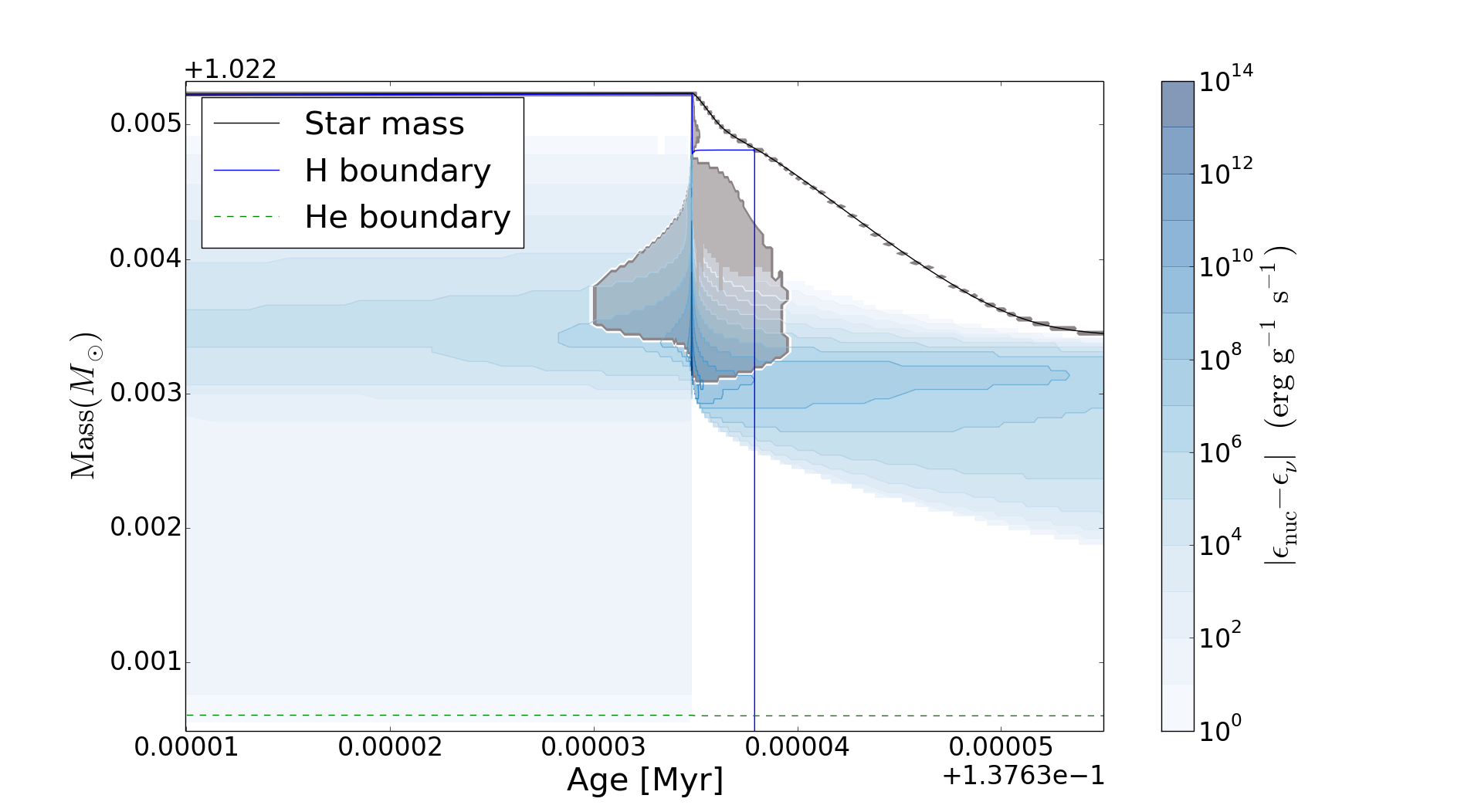}}}
\caption{The Kippenhahn diagrams at the 6th TP (upper panel) and at the 34th TP (lower panel) for the model M1p025.Z1m2 are shown for comparison. 
The 6th TP is also included in \fig{fig:summarykipcont}. The upper part of the 34th TP experiences H ingestion. 
The energy generation (blue shaded areas) and convective zones (grey-shaded areas) are indicated.}
\label{fig:comp_1WD_HIF}
\end{figure}

In \fig{fig:comp_1WD_HIF}, we compare the Kippenhahn diagrams obtained for the 6th and the 34th convective TPs 
of the model M1p025.Z1m2, where the last TP was affected by ingestion of H. 
The behavior of the convective region is affected in the upper part of the TP,
where the H burning activated by the ingestion of H-rich material, splits the He intershell in two parts:
about 90\% of the He intershell region is affected by the convection driven by the He burning at the bottom of the TP,
while the remaining upper part undergoes a short convective episode,
triggered by H burning (see horizontal blue segment in \fig{fig:comp_1WD_HIF}). 
This result is qualitatively consistent with theoretical predictions of H-ingestion episodes in post-AGB stars \citep[e.g.,][]{herwig:00a}.

In this specific case, both the two convective regions below and above the split end before the H ingested is exhausted and the super-Eddington winds start to eject material. After about three-four months from when the split is formed, the star becomes H-free by losing a large fraction of the He intershell (the vertical blue segment in \fig{fig:comp_1WD_HIF} indicates when the stellar surface becomes H-free).
This behavior may change for different TPs, but at least for model M1p025.Z1m2 it seems to become quite regular.
For a given WD mass, similar thermodynamic conditions at the bottom of the 
PDCZ, He-intershell size and the amount of H-rich material accreted are the main causes of such a regular behavior.
However, H-ingestion events for different WD masses may look quite different because of the differences seen in the previous sections.

It is important to keep in mind that
multi-dimensional hydrodynamic calculations are needed to better simulate H ingestion episodes \citep[][]{herwig:11,herwig:14}. 
Nevertheless,
one-dimensional stellar models can be adapted and constrained by gathering information from hydrodynamics simulations and observations.
For instance, this was the approach followed to study the H-ingestion event in the post-AGB star Sakurai's object by \citet{herwig:11}.

Unfortunately, there are no observations or specific hydrodynamics simulations available yet to do the same with massive accreting-WD models.
It would be plausible to assume that H-ingestion events for WD masses of $M = 0.6-0.8 \msun$ are similar to the VLTP observed in post-AGB stars,
since the conditions of the VLTP and of the TP are in the two cases quite similar,
including the amount of H available at the surface of the star \citep[see \fig{mh:mint} and e.g.,][]{miller-bertolami:07}.
However, for this work we are interested in WD masses of $M \gtrsim 1\msun$, where the main physics properties of the TPs are different. 

In the next sections we will discuss the neutron capture nucleosynthesis in these models, also
discussing the relevance of H-ingestion flashes for the conclusions of this work:
in general,
the occurrence of H-ingestion events is a crucial source of uncertainty.
This needs to be considered to provide a comprehensive picture of neutron-capture nucleosynthesis in accreting WDs.

\section{Post-processing nucleosynthesis calculations}
\label{sec:postpro}

\subsection{Building the seeds for the p-process}
\label{sec:seeds_postpro}

The nucleosynthesis during the accretion is calculated for up to 7 TPs and their interpulse phase for each model
(see table \ref{tab:WDmodel_name}).
The post-processing code \mppnp\ was used, which is described in detail in \cite{pignatari:16}.
The stellar structure evolution data were computed and saved with \MESA\ for all zones at all time steps,
as described in section \ref{sec:mesa}, and then used as input and processed with \mppnp.
Therefore the full nucleosynthesis and the stellar structure are computed separately,
hence requiring less computing time and resources.
The network we adopted for our full nucleosynthesis simulations includes up to about 5000 isotopes between H and Bi,
and more than 50000 nuclear reactions. A self-controlled dynamical network defines the number of species and reactions
considered in calculations, based on the strength of nucleosynthesis flows producing and destroying each isotope.
Rates are collected from different data sources:
NACRE compilation \citep[][]{angulo:99} and \cite{iliadis:01}, and more recent experimental information,
if available \citep[e.g.,][]{fynbo:05,kunz:02,imbriani:05}.
In particular, for the \isotope[13]{C}($\alpha$,n)\isotope[16]{O} and \isotope[22]{Ne}($\alpha$,n)\isotope[25]{Mg} rates we use \cite{heil:08} and \cite{jaeger:01}, respectively. For experimental neutron capture rates of stable isotopes and available rates for unstable isotopes we
use mostly the Kadonis compilation version 0.3 \citep[see][]{dillmann:14}. Exceptions relevant for this work are the neutron-capture cross sections of \isotope[90,91,92,93,94,95,96]{Zr}: we used instead the rates recommended by \cite{lugaro:14},
which are based on recent experimental measurements \citep{tagliente:08a,tagliente:08b,tagliente:10,tagliente:11a,tagliente:11b,tagliente:13}. For stellar $\beta$-decay and electron-capture weak rates we use \cite{fuller:85}, \cite{oda:94}, \cite{langanke:00}
and \cite{goriely:99}, depending on the mass region.
If not available from one the resources mentioned above, rates recommended in the JINA reaclib library were adopted \citep[][]{cyburt:10}.

\begin{figure}[htbp]
\centering
\resizebox{13.8cm}{!}{\rotatebox{0}{\includegraphics{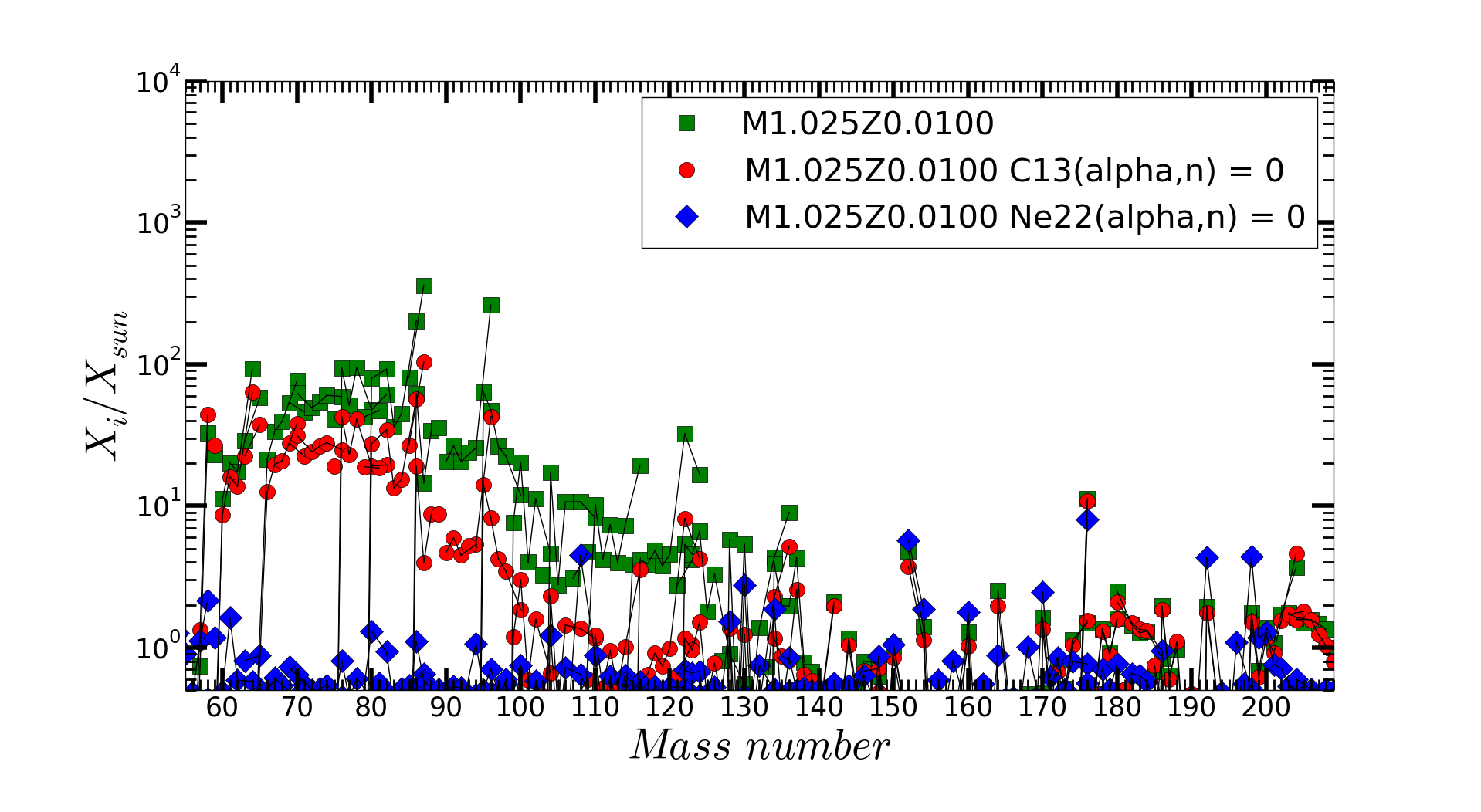}}}
\resizebox{13.8cm}{!}{\rotatebox{0}{\includegraphics{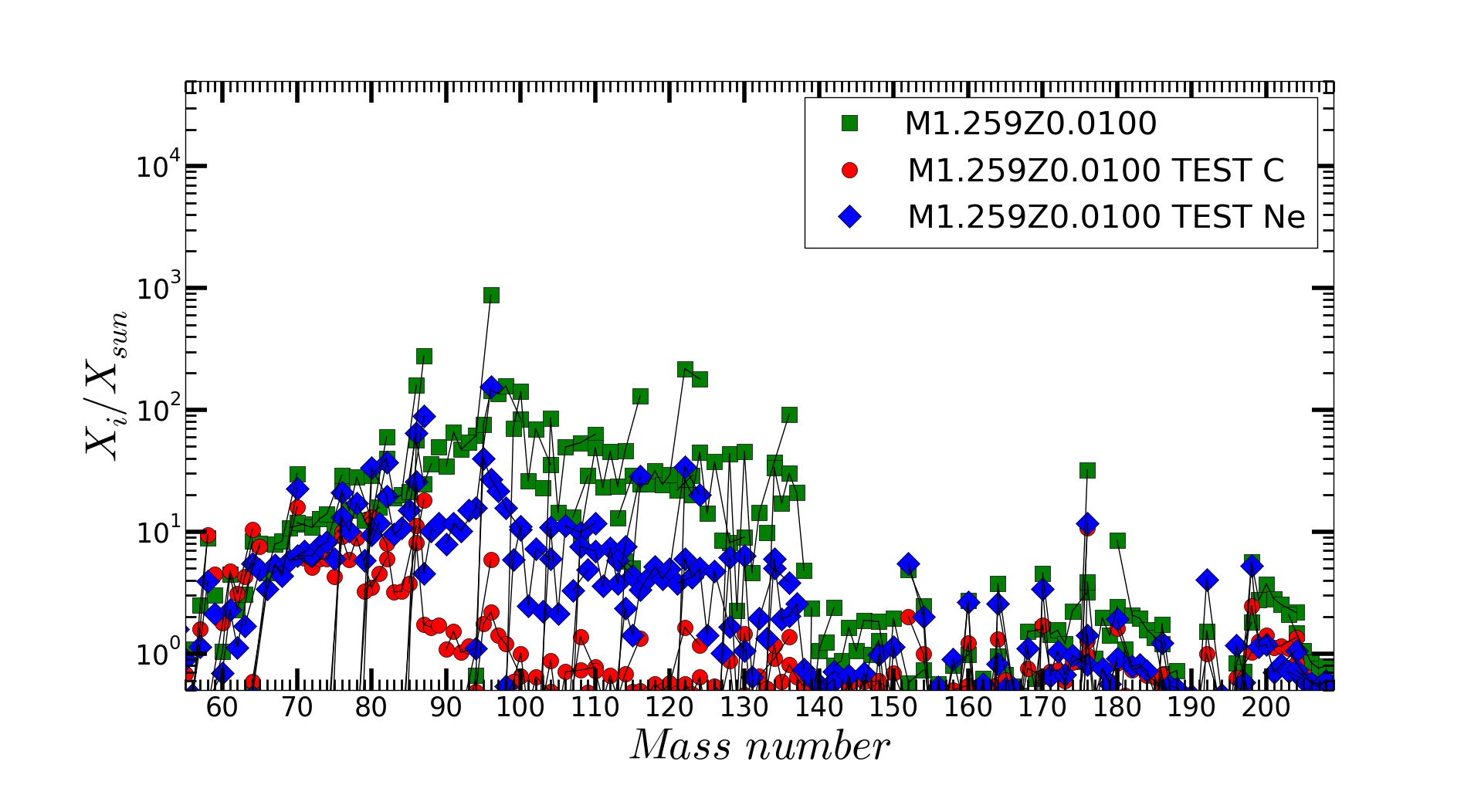}}}
\caption{Upper panel: Final isotopic distributions in M1p025.Z1m2 obtained setting separately the \isotope[13]{C}($\alpha$,n)\isotope[16]{O} and \isotope[22]{Ne}($\alpha$,n)\isotope[25]{Mg} to zero. The standard case (both neutron sources fully operative) is also shown in comparison).
Lower panel: same as in the upper panel, but referred to model M1p259.Z1m2}
\label{nuc:test}
\end{figure}

One of the main assumptions made by TRV11 was a starting seed distribution for the $p$-process similar to the $s$-process main component.
As predicted by \citet{iben:81},
in our simulations with WD masses up to 1.26 \msun\ the \nezw($\alpha$,n)\mgfu\ reaction is the main neutron source,
activated at the bottom of the intershell during the convective TP.
For higher masses, \cdr($\alpha$,n)$^{16}$O plays the main role due to H-ingestion events that lead to \cdr\ formation and burning at high temperature during the TP, causing a considerably high neutron-density peak. This is demonstrated in \fig{nuc:test}, where we show the final isotopic distribution obtained in models  M1p025.Z1m2 and M1p259.Z1m2 switching separately off the \cdr($\alpha$,n)$^{16}$O and the \nezw($\alpha$,n)\mgfu\ neutron sources: the heavy element distribution in M1p025.Z1m2 drops almost completely when the \nezw($\alpha$,n)\mgfu\ rate is set to zero, indicating that this reaction is the main responsible of the final neutron-capture elements production,  while the same happens in M1p259.Z1m2 when the \cdr($\alpha$,n)$^{16}$O is switched off.

\begin{figure}[htbp]
\centering
\includegraphics[width=1.0\textwidth]{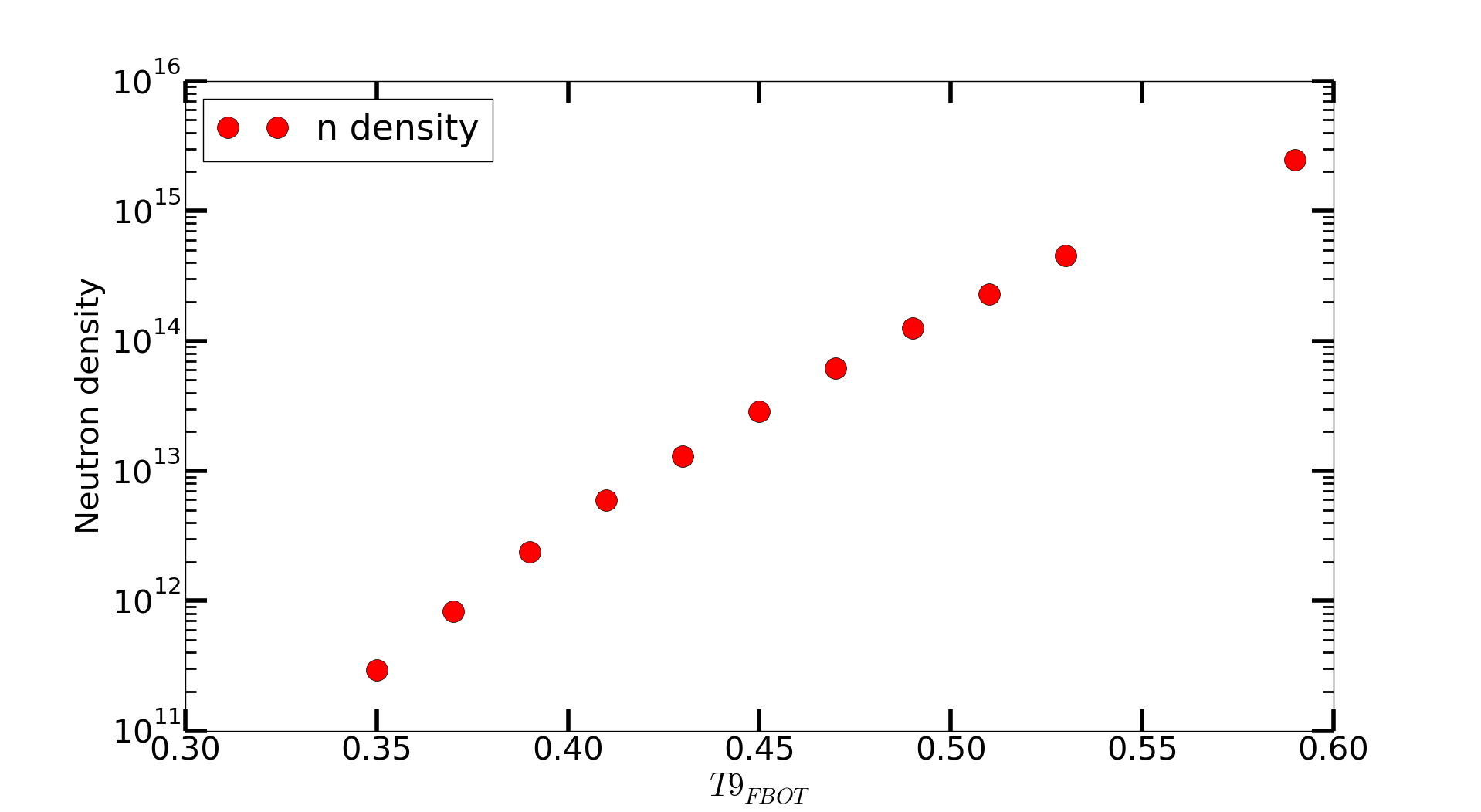}
\caption{Neutron density from \nezw($\alpha$,n)\mgfu\ as a function of temperature at the bottom of the TP during the accretion phase.}
\label{neut:dens}
\end{figure}

The temperature at the bottom of the PDCZ increases with the increase of the WD mass. This is why larger neutron densities are obtained.
According to table \ref{tab:WDm3z2m2_TPprop} and \fig{Temp:mwd}, typical T$_{FBOT}$ ranges from 3.2$\times$10$^{8}$ K (model M0p639.Z1m2) up to 5.9$\times$10$^{8}$ K (model M1p376.Z1m2).
The higher temperatures obtained cause an efficient \nezw\ and \cdr\ depletion by $\alpha$ capture, producing neutron densities up to four orders of magnitude higher than typical TPs in low mass AGB stars. In \fig{neut:dens},
we show the neutron density calculated in the temperature range of interest,
by using realistic abundances from the He intershell abundances of model M1p025.Z1m2 at the onset of the TP,
which is similar to the other models presented here.
The initial $^{22}$Ne abundance in mass fraction is X($^{22}$Ne) = 0.00393. 
The neutron density peak obtained ranges from few 10$^{11}$ cm$^{-3}$ up to few 10$^{15}$ cm$^{-3}$,
which is well beyond the typical $s$-process conditions.
Combining \fig{Temp:mwd} with the results shown in \fig{neut:dens}, TPs for WD masses of $M \sim 1.2\msun$ or larger are characterized by temperatures of about 5$\times$10$^{8}$ K or higher, which means the neutron density peaks will be few 10$^{14}$ cm$^{-3}$ or higher.

\begin{figure}[htbp]
\centering
\resizebox{14.8cm}{!}{\rotatebox{0}{\includegraphics{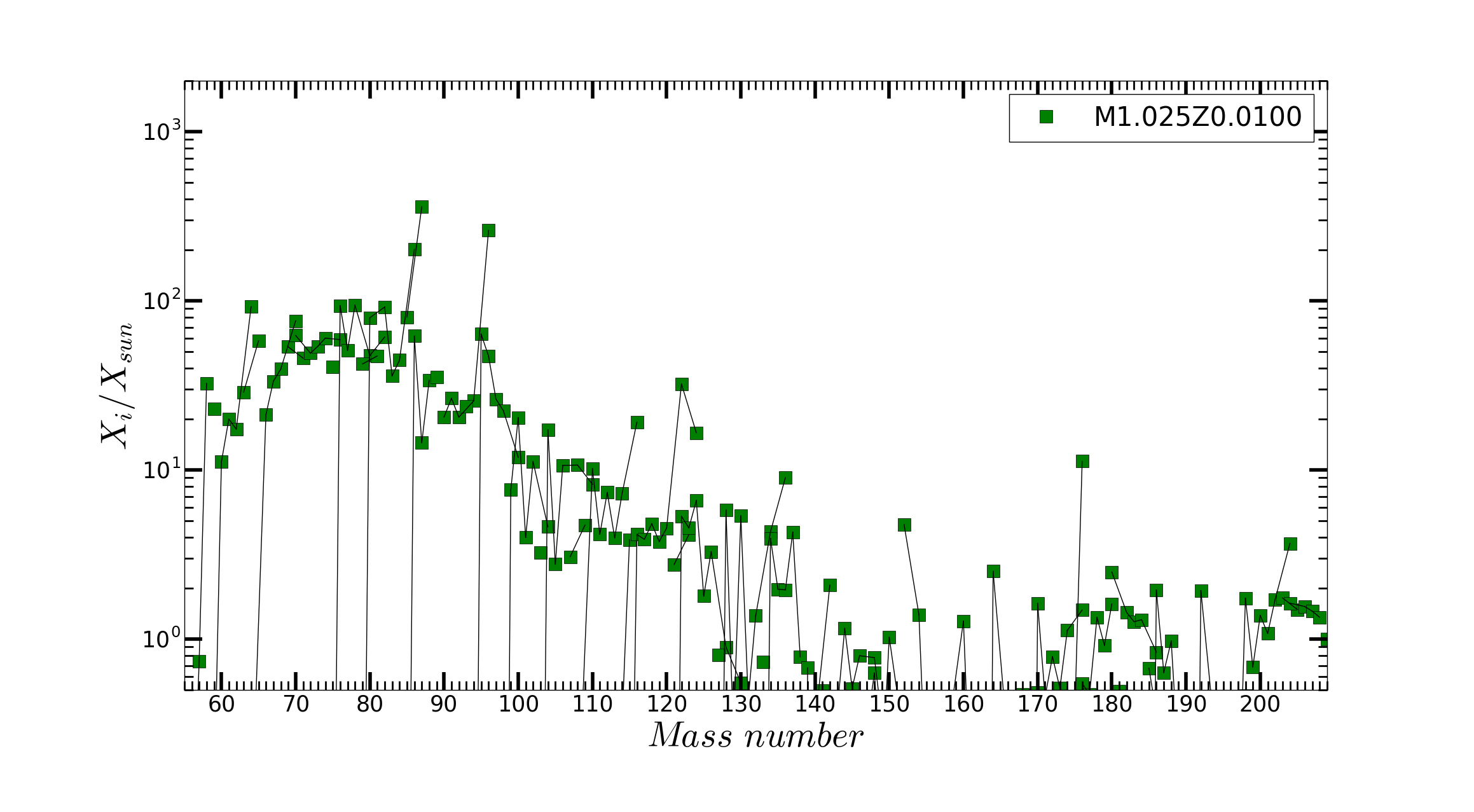}}}
\caption{Isotopic distribution beyond Fe at the 7th TP of model M1p025.Z1m2. The abundances are plotted assuming complete decay of unstable isotopes.}
\label{TPs:abu1p03}
\end{figure}

In \fig{TPs:abu1p03}, the abundance distribution beyond Fe in the He intershell  is shown at the end of the 4th TP of M1p025.Z1m2.
The largest production is obtained for elements between Fe and Zr,
in the mass region typical of the weak $s$-process \citep[e.g.,][]{the:07,pignatari:10,kaeppeler:11}.
On the other hand, neutrons are mostly released at neutron densities larger than in massive stars.
The isotopes produced the most are $^{86}$Kr, $^{87}$Rb and $^{96}$Zr. Also $^{70}$Zn, $^{76}$Ge and $^{82}$Se,
that are classically considered as $r$-process isotopes,
are efficiently synthesized at similar abundances with other isotopes nearby that are less neutron-rich.
The heaviest isotope showing a production factor in the order of 100 is $^{123}$Sb, but overall the production efficiency decreases beyond Zr.

\begin{figure}[htbp]
\centering
\resizebox{13.8cm}{!}{\rotatebox{0}{\includegraphics{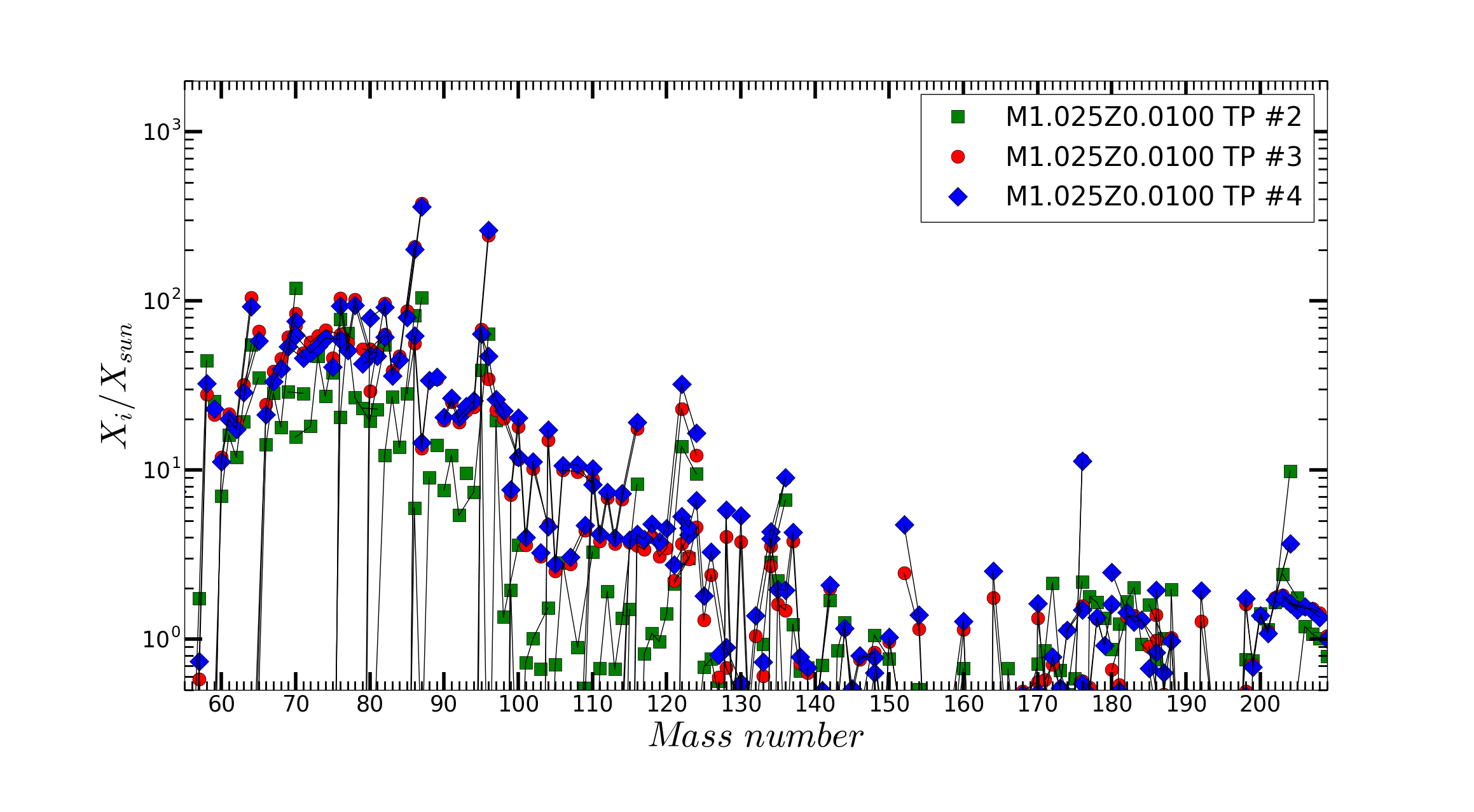}}}
\resizebox{13.8cm}{!}{\rotatebox{0}{\includegraphics{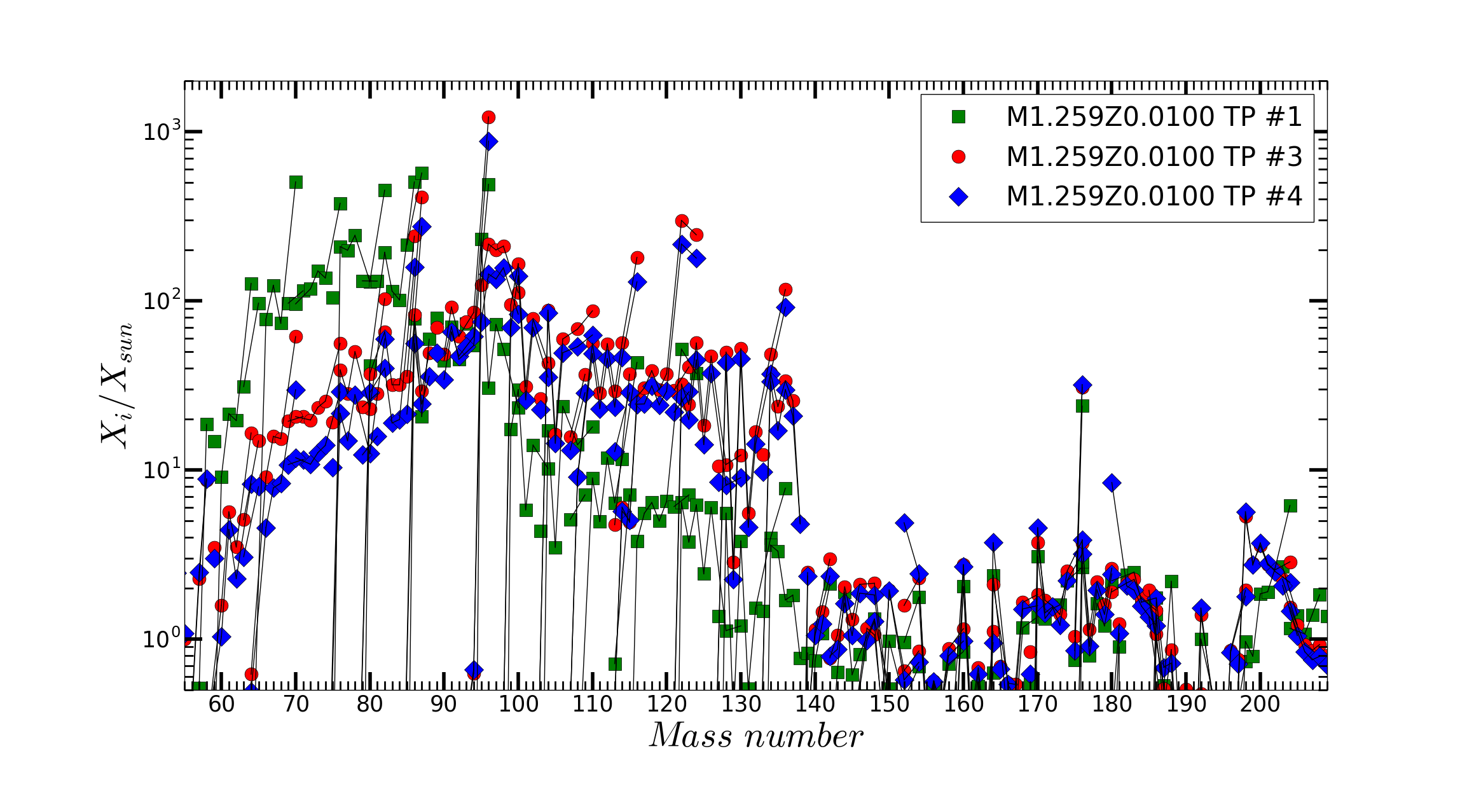}}}
\caption{Upper panel: abundance distribution of M1p025.Z1m2 after the 2nd, 3rd and 4th TP.
Interestingly, the abundances increase up to about the 3rd TP, and then they saturate to some production factor in the following TPs.).
Lower panel: the same as in the upper panel, but for the M1p259.Z1m2 model.}
\label{TPs:abu}
\end{figure}

\begin{figure}[htbp]
\centering
\resizebox{14.8cm}{!}{\rotatebox{0}{\includegraphics{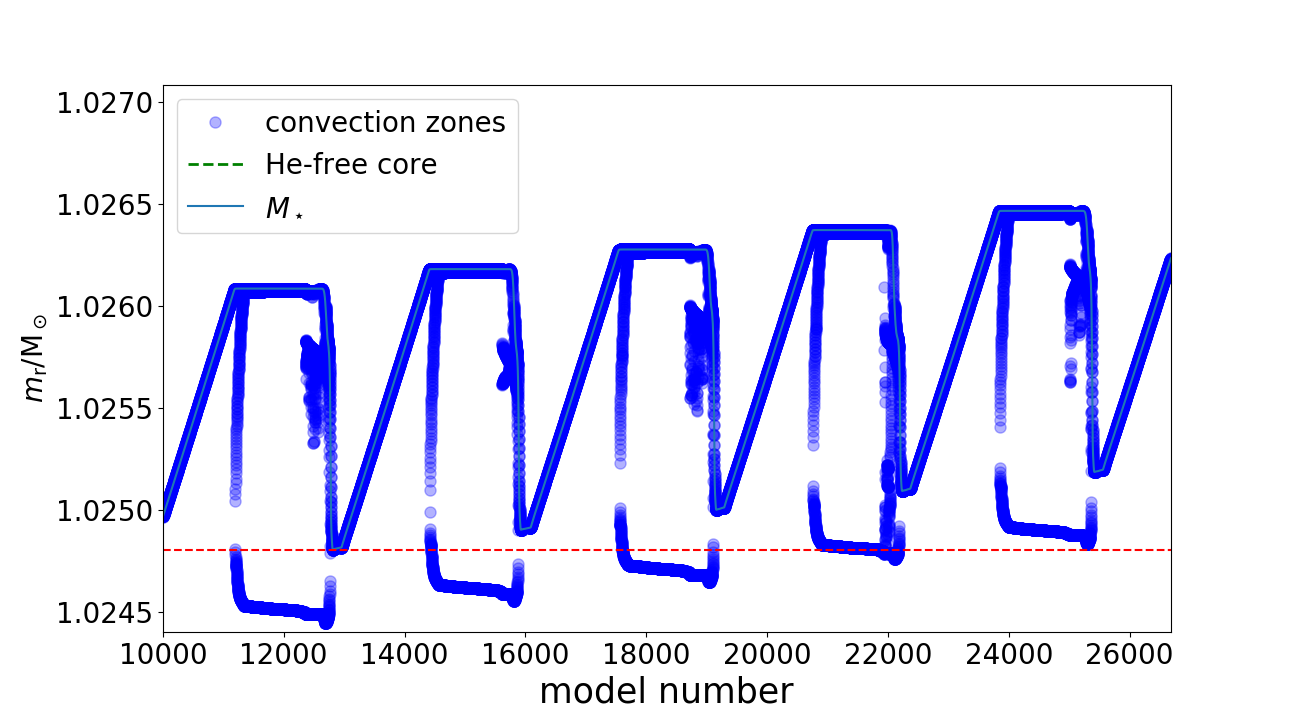}}}
\caption{Kippenhahn diagram of M1p025.Z1m2 showing five consecutive TPs are shown in a Kippenhahn diagram of M1p025.Z1m2.
 The X-axis shows the timestep number, in order to make the convective zones more visible for the particular purpose of this plot. 
The red-dashed line shows the mass coordinate location of the most external zone which has been enriched in heavy elements by the first TP,
without being ejected by the super-Eddington wind.}
\label{ash:int}
\end{figure}

\begin{figure}[htbp!]  
\centering    
\includegraphics[width=1.0\textwidth]{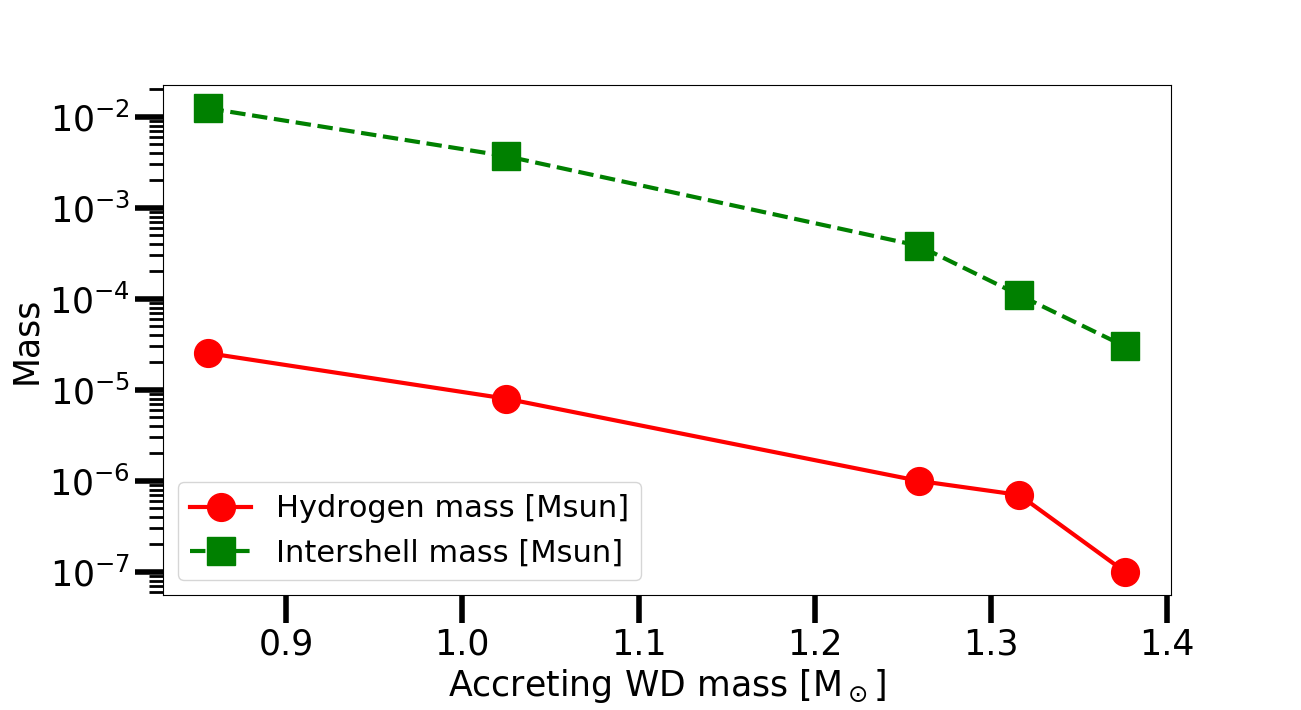}
\caption{Mass of the H-rich material accumulated at the surface of the star and He Intershell mass at the onset of the convective TP are shown, as a function of the accreting-WD mass.}
\label{mh:mint}
\end{figure}

The larger production of the elements at the neutron magic peak N=50 (Sr, Y and Zr),
compared to heavier elements at the neutron magic peak N=82 (e.g., Ba and La), is not surprising.
Even the complete depletion of $^{22}$Ne by $\alpha$-captures in massive stars,
during core He-burning and C-shell burning, does not efficiently produce Ba-peak elements \citep[e.g.,][]{kaeppeler:11}.
The main reason is that $^{22}$Ne is also efficiently capturing neutrons via (n,$\gamma$),
and as it releases neutrons via ($\alpha$,n) reactions, it also produces \mgfu,
which is a strong neutron poison \citep[e.g.,][]{pignatari:10},

In \fig{TPs:abu}, upper panel, the abundance distribution of M1p025.Z1m2 is shown again.
but this time after the 2nd, 3rd and 4th TP.
Interestingly, the abundances increase up to about the 3rd TP, and then they saturate to a constant overabundance in the following TPs.
This feature is explained in \fig{ash:int}, where five consecutive TPs are shown in a Kippenhahn diagram of M1p025.Z1m2.
The red-dashed line shows the mass coordinate of the most external zone which has been enriched in heavy elements by the first TP,
without being ejected by the super-Eddington wind. It is visible how this mass coordinate is enriched by the the second and third TP as well,
while is barely touched by the fourth and does not see any contribution by the fifth TP,
whose flash-driven convective zone is located above the dashed line.
This is why the heavy element abundance distribution of M1p025.Z1m2 grows up to the 4th TP and then stops.
Therefore, there is no need to simulate all the 137 TPs of model M1p025.Z1m2 with the post-processing code,
to compute the typical abundance distribution at mass coordinate $\sim$ 1 \msun\ .

In the lower panel,
we show the same kind of comparison for the M1p259.Z1m2 model.
Also in this case the saturation of the production is obtained after reaching a maximum after the 3rd TP,
when the distribution stabilizes.
Therefore, for a given WD mass it is possible to obtain a good estimate of the heavy isotope abundances in the He intershell,
by only simulating few convective TPs.
As we have shown in the previous sections,
the He intershell conditions mainly depend on the WD mass, evolving at the same rate of the net mass increase.
For example, the WD mass in M1p025.Z1m2 increases by just 0.5$\%$ over 137 TPs,
so the thermodynamic conditions and nucleosynthesis stay the same in the short-term evolution of the star
(i.e. as long as the WD mass will not increase significantly),
and the heavy elements production after few TPs can be considered representative of WD mass $\sim$ 1 \msun\ .
The same applies to the other models discussed in this work.
Since the $p$-process seeds are synthesized only close to the WD surface,
once the WD mass changes significantly over long enough timescales,
they will be buried under the newly accreted mass and will not change anymore.

\begin{figure}[htbp!] 
\centering    
\includegraphics[width=1.0\textwidth]{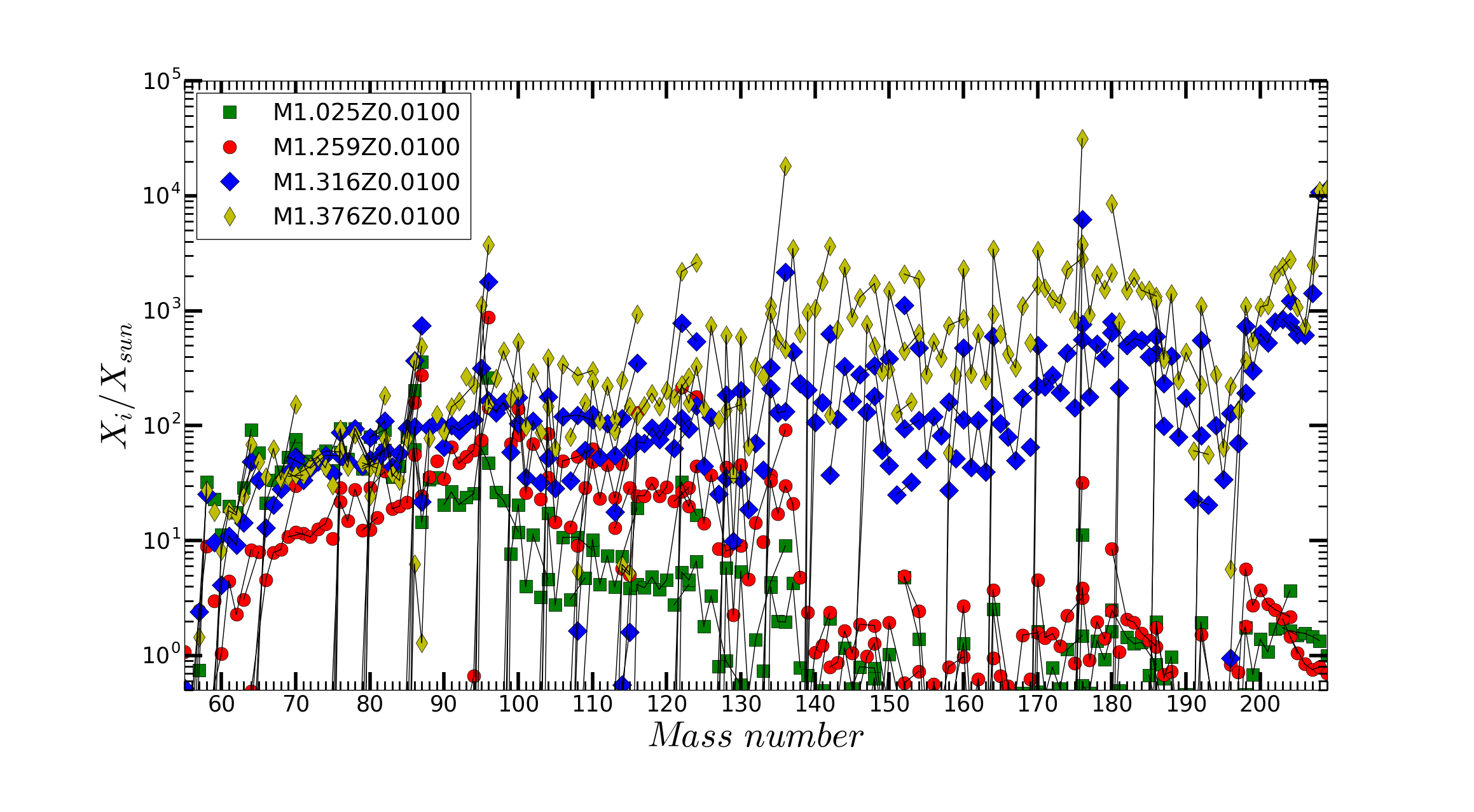}
\caption{Final abundance distribution calculated for models M1p025.Z1m2, M1p259.Z1m2, M1p316.Z1m2 and M1p376.Z1m2.}
\label{final:dis}
\end{figure}

\begin{figure}[htbp]
\centering
\resizebox{13.8cm}{!}{\rotatebox{0}{\includegraphics{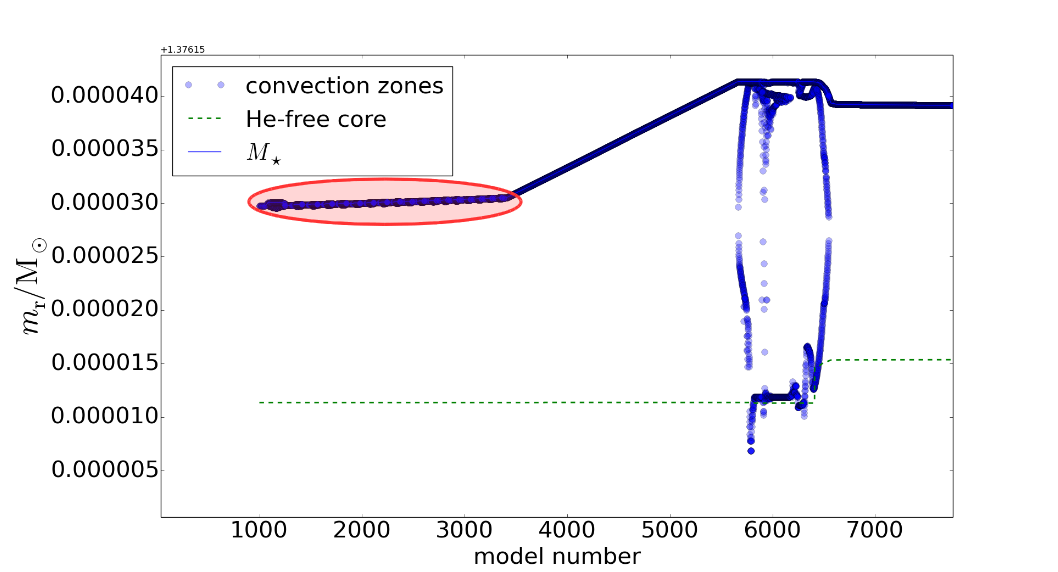}}}
\resizebox{13.8cm}{!}{\rotatebox{0}{\includegraphics{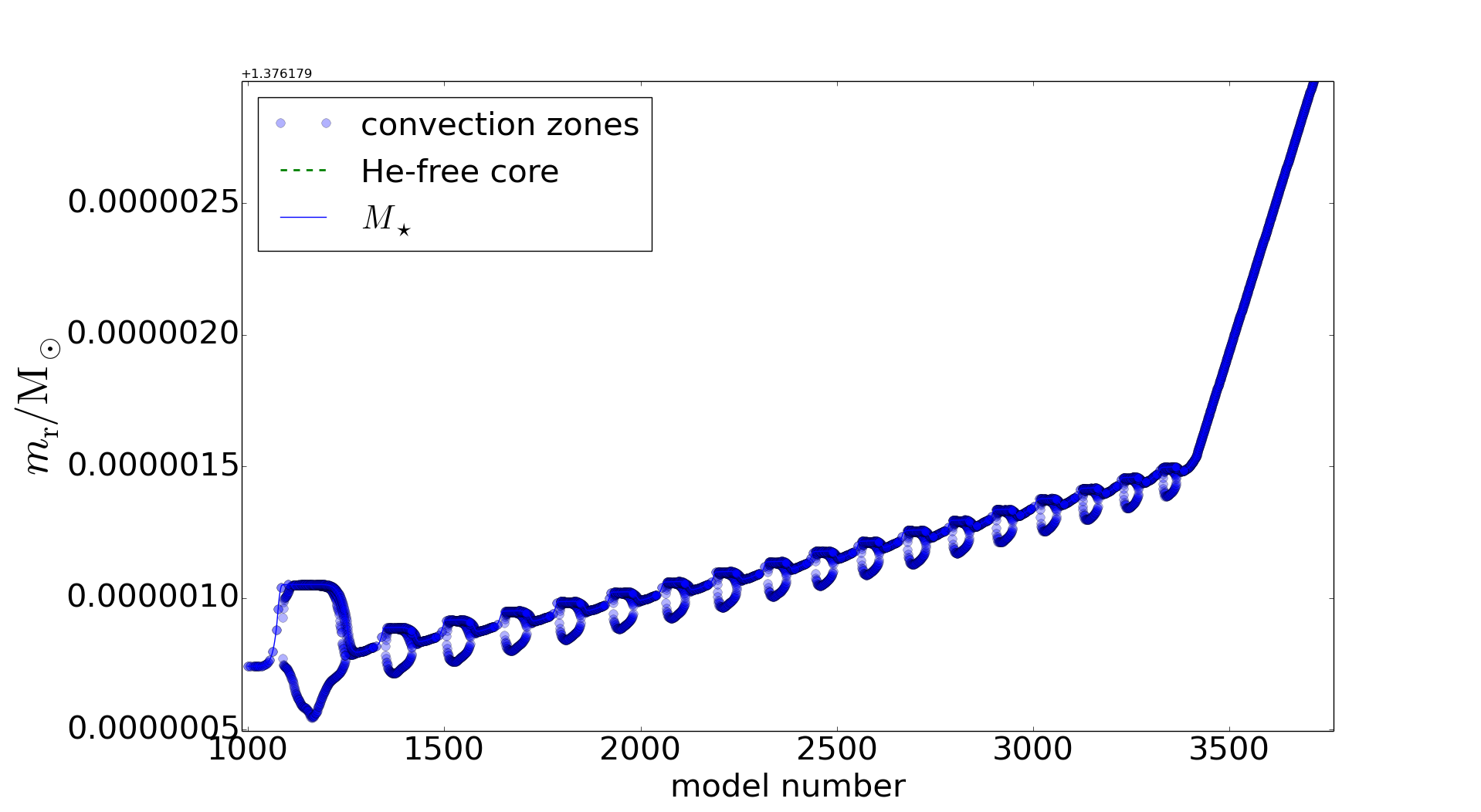}}}
\caption{Upper panel: Kippenhahn diagram of the first He-flash of model M1p376.Z1m2. The red ellipse selection identifies the recurrent H-flashes that trigger the synthesis of $^{13}$C and its convective consumption via \isotope[13]{C}($\alpha$,n)\isotope[16].
Lower panel: zoom into the surface zones where H-flashes occur}
\label{h:flash}
\end{figure}

During the entire accretion phase up to the Chandrasekhar mass, thermodynamic conditions close to the surface are evolving significantly.
The neutron density peak changes by four orders of magnitude,
also indicating a more and more efficient depletion of the available \nezw\ and,
for more massive WDs, \cdr.
The production factors shown in \fig{TPs:abu} calculated for the M1p259.Z1m2 model are larger than the ones calculated for the M1p025.Z1m2 model.
This can be better seen in \fig{final:dis}, where the abundance distribution calculated for different WD masses is shown. The complete abundance distributions are also given in \tab{tab:pf}.
The production factors tend to increase with the increase of the WD mass, due to the more efficient production of neutrons.
The abundance distribution for the M0p856.Z1m2 and M1p025.Z1m2 models is quite similar,
with the largest efficiency in the mass region between Fe and Zr.
On the other hand, the distributions obtained for the M1p259.Z1m2 model show a significant production up to $^{136}$Xe,
while the one obtained in M1.376.Z1m2 and  M1.316.Z1m2 continues with large efficiency up to the Pb region.
This is due to  larger neutron exposures obtained for those models, caused by numerous
H-flashes, as shown in \fig{h:flash}. These flashes trigger proton captures onto the abundant \isotope[12]{C} producing \cdr,
which is then completely burned via \cdr($\alpha$,n)$^{16}$O in a region of the order of 10$^{-7}$ \msun\ at very high temperature
(T$\sim$0.3 GK), with a resulting neutron density around 2$\times$10$^{15}$ n cm$^{-3}$ and a very high neutron exposure.
We find that these H-shell flashes are the main site of n-capture nucleosynthesis in our most massive accreting WD models.
Notice that this happens during the time interval before the onset of the TP.
The most produced isotopes in the Pb region are $^{204}$Hg and $^{209}$Bi.

Therefore, these calculations show that, in the SNIa progenitor,
the heavy seeds of the $p$-process are changing between $M \sim 1\msun$ in mass-coordinate and the surface:
the abundance distribution will be enriched between Fe and Zr in the deepest (and hottest) layers;
instead, the abundance distribution will be enriched up to and over the Ba mass region approaching the surface
(where the bulk of the $p$-process nuclei are made, see TRV11).  
Moving from $M \sim 1\msun$ outward, the production factors are also increasing up to more than one order of magnitude between Fe and Zr,
and even two orders of magnitude between Zr and Xe.
Elements heavier than Ba can increase up to three order of magnitude,
while over the mass-coordinate range between 0.85 \msun\ and 1.26 \msun\ ,
they are always less abundant compared to the lighter neutron capture products.

\subsection{SNIa nucleosynthesis calculation}
\label{sec:snia_postpro}

In this Section we discuss the nucleosynthesis results obtained from post
processing the SNIa hydrodynamic model described in details in TRV11.
The  scenario  is  based  on  the explosion
triggered once  the CO-WD approaches  the Chandrasekhar  mass.
Thermonuclear  burning  starts  out  as a  subsonic deflagration  and  later turns into a supersonic detonation. We
refer to the SD scenario in which the white dwarf  accretes
material  from a  main-sequence  or  evolved  companion  star. The  explosion
is modelled using a two-dimensional hydrodynamic simulations following the
delayed detonation model (hereafter DDT-a).
This model was tested and compared with early spectra, near-maximum spectra and light-curve of SN 2011fe by \citet{ropke:12}.

It is important to stress that the multi-dimensional simulations reach a qualitatively different
level of predictive power than 1D models. In particular, the amount of material burned at a
given density cannot longer be fine-tuned but is determined by the fluid motions on the
resolved scales. Therefore, once the flame model has been fixed, numerical simulations of the
thermonuclear explosion of a given white dwarf can be done by just choosing the ignition
conditions, including the chemical composition of the WD, the only remaining (physical)
parameter. In addition the 1D thermonuclear supernovae models are commonly not well
resolved in the outermost layers of the WD, where $p$-process nucleosynthesis occurs.

\begin{figure}[htbp!] 
\centering    
\resizebox{12.2cm}{!}{\rotatebox{0}{\includegraphics{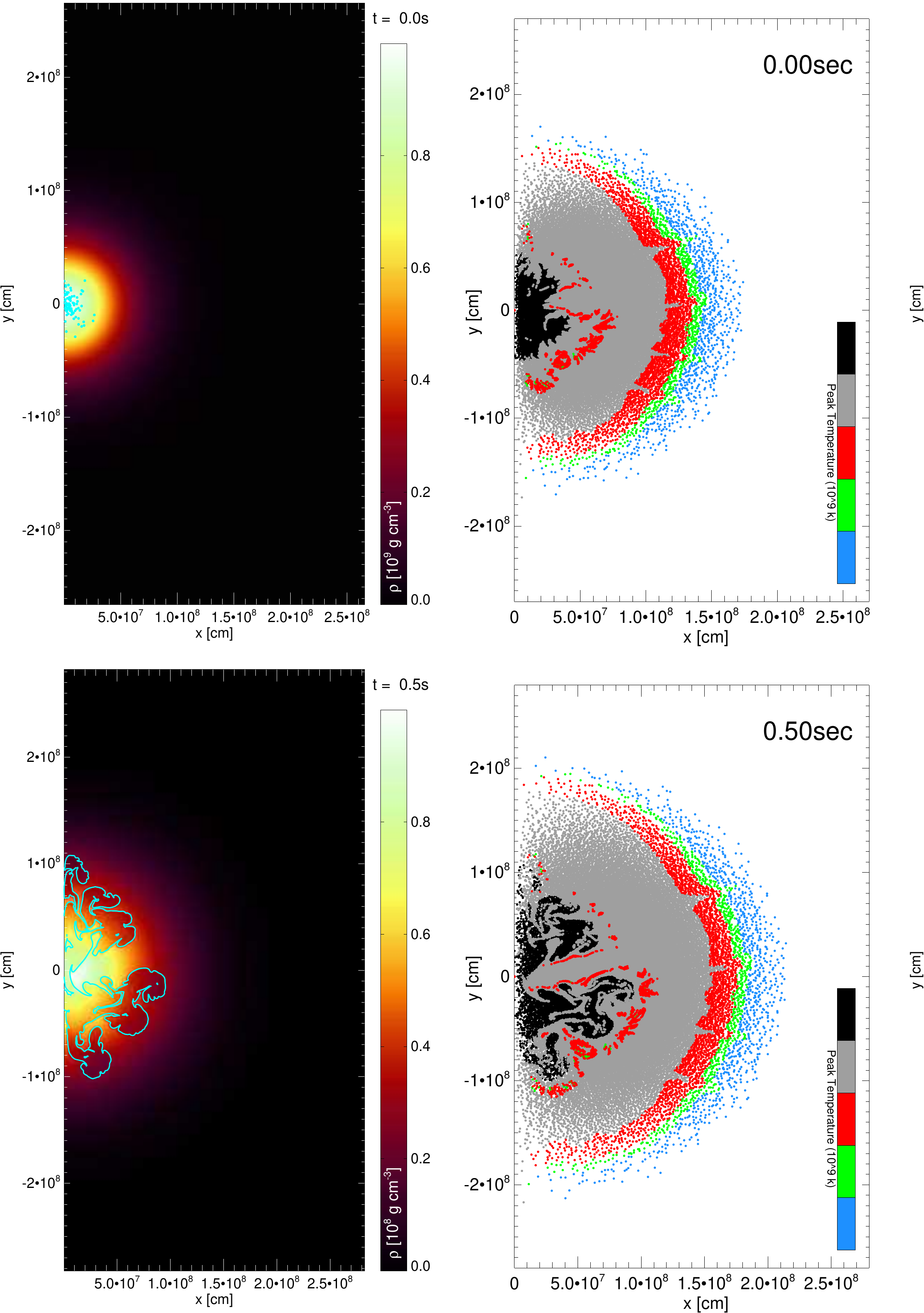}}}
\caption{Snapshots from the SNIa hydrodynamic model described in details in TRV11 at 0.0 s and 0.5 s after ignition. On the left, the hydrodynamic evolution is illustrated by color-coded density and the locations of the deflagration flame and the detonation front (cyan and blue contours respectively). The tracer distribution is given on the right-hand side. While the locations correspond to the time given, the color coding is according to the maximum temperature reached during the entire explosion: black tracers peak with T $>$ 7.0 GK; gray tracers with 3.7 GK $<$ T $<$ 7.0 GK; tracers marked in blue (1.5 GK $<$ T $<$ 2.4 GK), green (2.4 GK $<$ T $<$ 3.0 GK), and red (3.0 $<$ T $<$ 3.7) are peak temperatures reached in ranges where the $p$-process nucleosynthesis is possible. See \citet{travaglio:11} for more snapshots up to 1.45 seconds after the ignition.}
\label{tracers:tpeak}
\end{figure}

Since only a crude description of the thermonuclear burning is applied in the
hydrodynamic explosion simulation, details on the nucleosynthesis are
recovered in a post-processing step. A Lagrangian component in the form of
tracer particles is introduced over the Eulerian grid in order to follow and
store the temperature and density evolution of the fluid. Nevertheless it is
impossible (due to limited computational resources) to consider mixing processes between tracers,
which anyway are not expected to significantly impact our results given the small mass of each tracer of about 3.0 $\times$ 10$^{-5}$M$_\odot$.
In particular for the slower tracers in the
outermost part of the star (responsible for the $p$-process nucleosynthesis)
this is certainly a very good approximation. For the DDT-a model we used
51,200 tracer particles, uniformly distributed over the star.
The distribution of the tracer particles for our model is shown in Figure \ref{tracers:tpeak} in two snapshots illustrating the evolution. Different colors are used for different ranges of peak temperature of the tracers. The tracers marked in black have a maximum temperature above 7 GK. Hence, they attain nuclear statistic equilibrium (NSE) conditions and most of the nucleosynthesis goes to $^{56}$Ni or Fe-group nuclei. The gray tracers are instead the main producers of the lighter $\alpha$-isotopes. Tracers marked in blue (1.5 GK $<$ T $<$ 2.4 GK), green (2.4 GK $<$ T $<$ 3.0 GK), and red (3.0 $<$ T $<$ 3.7) have peak temperatures in ranges where the $p$-process nucleosynthesis is possible. These three ranges are connected to three different behaviors identified by TRV11 for the production of $p$-nuclei (see their Section 5).

During the hydrodynamic simulation for each tracer particle,
T and ${\rho}$ histories are recorded along their paths.
The nuclear post-processing calculations are then performed separately for each particle.
The $p$-process nucleosynthesis is calculated using a nuclear network with
1024 species from neutrons and protons up to $^{209}$Bi combined with $\beta$-decays, neutron,
proton, and $\alpha$-induced reactions and their inverse.
The set of thermonuclear reaction rates we adopted is based
on the theoretical values and the Hauser–Feshbach statistical model
NON-SMOKER \citep{rauscher:97,rauscher:02}.
The $p$-process nucleosynthesis occurs in SNIa starting on a
pre-explosive heavy-element enrichment; therefore, it is essential to determine
this enrichment in the exploding WD. As described in Section \ref{sec:wd_accr},
we found recurrent He-flashes occurring in the He-shell during the accretion
phase, providing the right conditions to synthesize an heavy-element seeds distribution.

In order to include properly the seeds as a function of the WD mass we use the calculated mass
fraction applying them to specific mass ranges, i.e.: M $<$ 1.1 M$_{\odot}$ uses mass fractions from our 1.03
Msun model; 1.1 M$_{\odot}$ $<$ M $<$ 1.29 mass fractions from our 1.26 M$_{\odot}$ model; 1.29 $<$ M $<$ 1.35 M$_{\odot}$
uses mass fraction from our 1.32 M$_{\odot}$ model and anything at higher masses uses 1.38 M$_{\odot}$ WD.
Only the outermost 0.08 M$_{\odot}$ contribute consistently to the $p$-nuclei,
while masses between 1.1 and 1.3 M$_{\odot}$ do give a contribution but definitely much less important.

\begin{figure}[htbp!] 
\centering    
\includegraphics[width=1.0\textwidth]{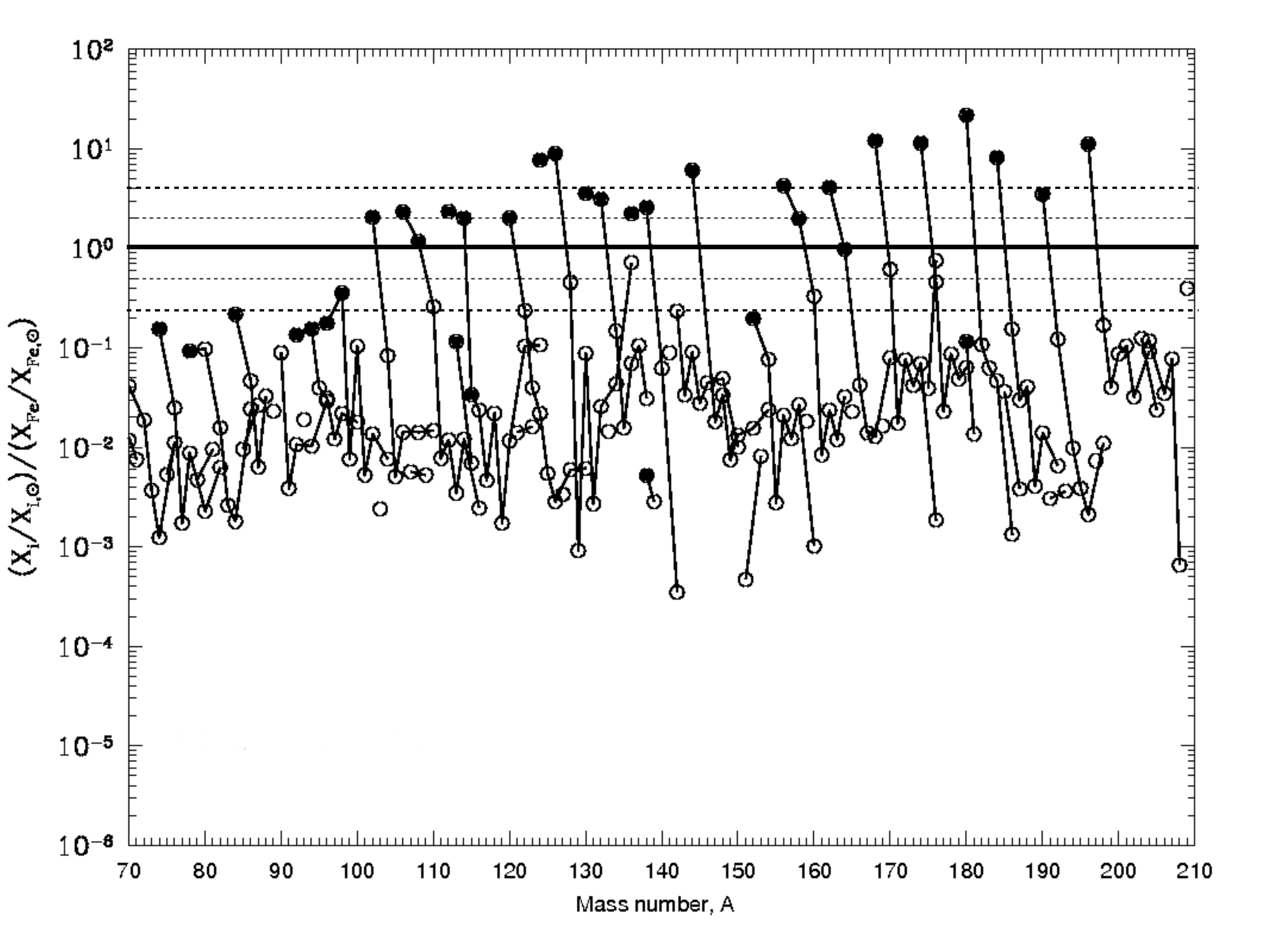}
\caption{Final abundance distribution results obtained post processing the SNIa hydrodynamic model described in details in TRV11. The heavy-element seeds shown in figure \ref{final:dis} are used as initial abundances at the appropriate mass coordinate. Filled circles represent $p$-only isotopes. The thick solid horizontal line indicates the Solar-System production level relative to iron, while the dashed lines  are located a factor of two and four up and down.}
\label{final:dis_exp}
\end{figure}

Differently to TRV11 where the recurrent He-flashes occurring in the He-shell
during the accretion phase were assumed as an hypothesis, in this work they
were naturally obtained along the WD evolution.
TRV11 tested different metallicities Z, but they inferred that the
($p$/$^{56}$Fe)/($p$/$^{56}$Fe)$_{\odot}$ ratio is not so much dependent on metallicity.
This suggests a primary nature of $p$-process.
For this work we therefore consider one metallicity Z=0.01 as representative of a general trend.
In Figure \ref{final:dis_exp} we show our results using the heavy-element seeds described in Section \ref{sec:seeds_postpro},
abundances are normalized to their solar values and to $^{56}$Fe and plotted
versus their mass number.
The solid horizontal line indicates the Solar-System production level relative to $^{56}$Fe.
As discussed by \citet{nishimura:18}, the impact of nuclear uncertainties on $p$-process abundances is generally within a factor of two.
We then show four dashed lines, located a factor of two and four above and below the Solar System level.
This allows to see if the $p$-only isotopes are consistent with the observed Solar System level within the typical nuclear uncertainties.
Additionally, in table \ref{tab:prepost} we provide pre- and post-explosion abundances of the $p$-nuclei normalized to solar. Globally, $p$-nuclei are heavily depleted by neutron captures during the accretion phase. There are anyway three exceptions ($^{108}$Cd, $^{114}$Sn and $^{180}$W), whose abundance ratios compared to solar slightly exceeds unity. This is the fingerprint of proton capture reactions, triggered by the H-ingestion events discussed in the previous chapters.
As previously discussed by TRV11 and
\citet{travaglio:15}, few nuclei originally ascribed to the $p$-only group
($^{113}$In, $^{115}$Sn, $^{138}$La, $^{152}$Gd, and $^{180}$Ta) are far below
the average of the other $p$-nuclei production also in this work. This is a
further indication for a different nucleosynthetic origin for them.
Concerning $^{113}$In and $^{115}$Sn, as discussed by \citet{dillmann:08} and \citet{nemeth:94},
they can get important contributions from ${\beta}$-delayed $r$-process decay chains.
As for $^{138}$La, \citet{woosley:90} demonstrated that the $\nu$-nucleus
interaction can contribute appreciably to the synthesis of $^{138}$La in the
neon shell of core collapse supernovae. $^{152}$Gd instead is predominantly of
$s$-process origin, as it was demonstrated by Kaeppeler et al.~(2011).
Like $^{152}$Gd, also $^{164}$Er is of predominant $s$-process origin, driven by
the $\beta$-decay channel at $^{163}$Dy, which becomes unstable at stellar
temperature \citet{takahashi:87}. Finally, $^{180}$Ta, as discussed by
TRV11 and \citet{mohr:07}, receives an important contribution from the
$s$-process due to the branching at $^{179}$Hf.
Concerning the other 30 $p$-nuclei,
\citet{travaglio:18} showed how type II SNe and SNIa could be seen as complementary sites to explain the $p$-process Solar System content,
contributing to A $<$ 92 and A $\geq$ 92 atomic-mass ranges respectively.
Looking at our $p$-process distribution, we get an efficient production for A $\geq$ 92.
We obtain solar to super-solar abundances, with about two thirds of $p$-nuclei produced consistently with the observed Solar System level,
confirming SNIa as a $p$-process production site. 
It should be noted that $^{92}$Mo, $^{94}$Mo and  $^{96}$Ru are not efficiently produced,
but multi-D models suggest ejection of high abundances of all these three isotopes in faint type-II supernovae 
\citep[see][]{eichler:18}.
In particular, according to \citet{eichler:18} $^{92,94}$Mo may be
produced by proton and neutron captures under neutron-rich conditions,
whereas another source is required for the Ru isotopes, such as a
$\nu$p-process as introduced by \citet{frohlich:06}. Its relevance
for the production of Mo and Ru isotopes has recently been studied by,
e.g., by \citep{nishimura:19} and \citep{rauscher:20}.
Finally, the $p$-nuclides below Mo may also have been produced by other stellar
core-collapse supernova components, such as $\alpha$-rich freeze-out of
hot matter (e.g., Pignatari et al. 2016; Lugaro et al. 2016; Travaglio
et al. 2018). Therefore, there seems to remain a possibility to explain
the full range of $p$-process content in the Solar System by combining
production in SNIa and in core-collapse SNe.

\section{Summary and conclusions}
\label{sec:conclusions}

In this work we have analyzed the neutron capture nucleosynthesis in accreting WD models.
We presented for the first time a heavy-element distribution calculated from
realistic simulations of WD-accretion phase in the single degenerate scenario channel to SNIa.
Such distribution arises from recurrent He- and H-flashes on the WD surface as H-rich material is accreted onto the WD from a companion.
Contrary to previous rapidly-accreting white dwarf models at lower mass, we find that the H-shell flashes are the main site of n-capture nucleosynthesis.
We mostly focused on WDs with masses of the order or larger than 1.1\msun,
since these initial masses are most likely to reach the Chandrasekhar mass,
and only the external 0.2-0.3\msun of the SNIa progenitor are relevant for the $p$-process production \citep{travaglio:11}.
The main neutron sources are the \nezw($\alpha$,n)\mgfu\ and the \cdr($\alpha$,n)\ose\ reactions for WD masses lower and larger than 1.26 \msun\ respectively.
In our simulations, those reactions are activated during convective TPs in the He intershell and during hot (T$>$0.3 GK) H-flashes WDs larger than 1.3\msun,
with neutron densities between few 10$^{12}$ cm$^{-3}$ and few 10$^{15}$ cm$^{-3}$, much higher than typical \spr\ conditions.
The final abundance distribution includes large quantities of Rb,Kr, Zr, Ba-peak isotopes,
including Pb for our M1p316.Z1m2 and M1p376.Z1m2 models .
This is therefore globally very similar to the one adopted in \cite{travaglio:11}.
When used as a starting abundance distribution, heavily affected by earlier neutron processing, p-nuclei are
significantly produced in the mass range 96$<$A$<$196.

It should be noted that Roche-Lobe over flow may lower the retention efficiency of accreting WDs.
This has not been considered here both for computational time requirements and for the small
impact on heavy-elements nucleosynthesis on top layers of the WD, which is the main target of the
present study.

A source of uncertainty for our results arises due to the occurrence of H ingestion into He-burning layers during the accretion phase,
discussed in section \ref{subsec:comp_1msun_WD_HIF}.
The capture of ingested protons on the abundant $^{12}$C (direct product of He-burning) produces the unstable isotope $^{13}$N that decays
to $^{13}$C, which acts as neutron source via  $^{13}$C($\alpha$,n)$^{16}$O reactions.
In these environments, neutron densities are a few  10$^{15}$ cm$^{-3}$, typical for so-called $i$-process environments.
In these conditions, the impact of uncertainties coming from nuclear-physics on these results could be high,
in particular concerning neutron-capture reaction rates on unstable nuclei.
Secondly, as mentioned in section \ref{subsec:comp_1msun_WD_HIF}, in order to account for the H-ingestion events and their nucleosynthesis products, one-dimensional stellar models need to be guided by full 3D-hydrodynamic simulations \citep[e.g.,][]{herwig:14}.
As pointed out by \citet{herwig:11}, one-dimensional stellar models under-estimate the production of heavy element by  at least a factor of 2 in these conditions. This impacts the production of the $p$-seeds during the accretion phase, and as a consequence also the explosive $p$-process nucleosynthesis, which directly depends on the seeds abundance at the onset of the ignition.


Similar, yet slightly different, conditions are found in the hot H-flashes before the onset of He-flashes of our M $>$ 1.3 \msun\ models,
where the $^{13}$C($\alpha$,n)$^{16}$O is strongly activated and nuclei up to the Pb region are synthetised:
these features appear to happen systematically once WD masses larger than 1.3 \msun\ are considered,
and are a natural consequence of the high temperature at which H-flash instabilities take place.
Given their crucial role in the production of heavy-elements seeds for the subsequent $p$-process nucleosynthesis during the SNIa event,
and the potential impact of modelling uncertainties very similar to the ones affecting the previously mentioned H-ingestion events,
both kind of events should be considered as important candidates to be studied in future multi-dimensional simulations.

Finally, we computed the explosive nucleosynthesis using our seed distributions as initial abundances.
We obtained that the $p$-process production in SNIa is confirmed for $p$-nuclei above A=96,
at a level between solar and super-solar relative to iron, confirming results by \citet{travaglio:11}.

\subsection{Acknowledgements}
We thank the anonymous referee, whose suggestions improved the quality of this paper.
This article is based upon work from the “ChETEC” COST Action (CA16117), supported by COST (European Cooperation in Science and Technology).
This research was enabled in part by support provided by WestGrid (www.westgrid.ca), Compute Canada Calcul Canada (www.computecanada.ca), and the University of British Columbia computer cluster “Orcinus”.
NuGrid data is served by Canfar CADC. CLW acknowledge support from the Science and Technology Facilities Council UK (ST/M006085/1).
UB and CLW acknowledge support from the European Research Council ERC-2015-STG Nr. 677497.
UB deeply thanks Sergio Cristallo for fruitful discussions which largely improved this paper.
UB acknowledges the mentoring of Roberto Gallino,
and is particularly grateful for his inspiration and support,
since the very early days of his Master’s thesis on the origin of $p$-nuclei.
This study was also supported by the Swiss National Science Foundation and the EU-FP7-ERC Advanced Grant 321263 FISH.
MP acknowledges significant support to NuGrid from STFC (through the University of Hull's Consolidated Grant ST/R000840/1),
and access to viper, the University of Hull High Performance Computing Facility.
MP acknowledges the support from the "Lendulet-2014" Program of the Hungarian Academy of Sciences (Hungary),
and the ERC Consolidator Grant funding scheme (Project RADIOSTAR, G.A. n. 724560, Hungary).
PD's research was supported by the National Science Foundation (USA) under Grant No. PHY-1430152
(JINA Center for the Evolution of the Elements)

\bibliography{astro}

\begin{thebibliography}{116}
\expandafter\ifx\csname natexlab\endcsname\relax\def\natexlab#1{#1}\fi

\bibitem[{{Angulo} {et~al.}(1999){Angulo}, {Arnould}, \& {Rayet, M. et
  al.}}]{angulo:99}
{Angulo}, C., {Arnould}, M., \& {Rayet, M. et al.} 1999, Nucl.\ Phys., A 656,
  3, {NACRE} compilation

\bibitem[{{Arnould}(1976)}]{arnould:76}
{Arnould}, M. 1976, \aap, 46, 117

\bibitem[{Asplund {et~al.}(1999)Asplund, Lambert, Kipper, Pollacco, \&
  Shetrone}]{Asplund:1999ta}
Asplund, M., Lambert, D.~L., Kipper, T., Pollacco, D., \& Shetrone, M.~D. 1999,
  A{\&}A, 343, 507

\bibitem[{{Audouze} \& {Truran}(1975)}]{audouze:75}
{Audouze}, J., \& {Truran}, J.~W. 1975, \apj, 202, 204

\bibitem[{{Becker} \& {Iben}(1979)}]{becker:79}
{Becker}, S.~A., \& {Iben}, Jr., I. 1979, \apj, 232, 831

\bibitem[{{Brachwitz} {et~al.}(2000){Brachwitz}, {Dean}, {Hix}, {Iwamoto},
  {Langanke}, {Mart{\'{\i}}nez-Pinedo}, {Nomoto}, {Strayer}, {Thielemann}, \&
  {Umeda}}]{brachwitz:00}
{Brachwitz}, F., {et~al.} 2000, \apj, 536, 934

\bibitem[{{C{\^o}t{\'e}} {et~al.}(2018){C{\^o}t{\'e}}, {Denissenkov}, {Herwig},
  {Ruiter}, {Ritter}, {Pignatari}, \& {Belczynski}}]{cote:18}
{C{\^o}t{\'e}}, B., {Denissenkov}, P., {Herwig}, F., {Ruiter}, A.~J., {Ritter},
  C., {Pignatari}, M., \& {Belczynski}, K. 2018, \apj, 854, 105

\bibitem[{{Cox} \& {Giuli}(1968)}]{cox:68}
{Cox}, J.~P., \& {Giuli}, R.~T. 1968, {Principles of stellar structure} (New
  York, Gordon and Breach [1968])

\bibitem[{{Cyburt} {et~al.}(2010){Cyburt}, {Amthor}, {Ferguson}, {Meisel},
  {Smith}, {Warren}, {Heger}, {Hoffman}, {Rauscher}, {Sakharuk}, {Schatz},
  {Thielemann}, \& {Wiescher}}]{cyburt:10}
{Cyburt}, R.~H., {et~al.} 2010, \apjs, 189, 240

\bibitem[{{deBoer} {et~al.}(2017){deBoer}, {G{\"o}rres}, {Wiescher}, {Azuma},
  {Best}, {Brune}, {Fields}, {Jones}, {Pignatari}, {Sayre}, {Smith}, {Timmes},
  \& {Uberseder}}]{deboer:17}
{deBoer}, R.~J., {et~al.} 2017, Reviews of Modern Physics, 89, 035007

\bibitem[{{Denissenkov} {et~al.}(2017){Denissenkov}, {Herwig}, {Battino},
  {Ritter}, {Pignatari}, {Jones}, \& {Paxton}}]{denissenkov:17}
{Denissenkov}, P.~A., {Herwig}, F., {Battino}, U., {Ritter}, C., {Pignatari},
  M., {Jones}, S., \& {Paxton}, B. 2017, \apjl, 834, L10

\bibitem[{Denissenkov {et~al.}(2019)Denissenkov, Herwig, Woodward, Andrassy,
  Pignatari, \& Jones}]{2019MNRAS.488.4258D}
Denissenkov, P.~A., Herwig, F., Woodward, P., Andrassy, R., Pignatari, M., \&
  Jones, S. 2019, MNRAS, 488, 4258

\bibitem[{{Denissenkov} {et~al.}(2019){Denissenkov}, {Herwig}, {Woodward},
  {Andrassy}, {Pignatari}, \& {Jones}}]{denissenkov:19}
{Denissenkov}, P.~A., {Herwig}, F., {Woodward}, P., {Andrassy}, R.,
  {Pignatari}, M., \& {Jones}, S. 2019, \mnras, 488, 4258

\bibitem[{{Denissenkov} {et~al.}(2015){Denissenkov}, {Truran}, {Herwig},
  {Jones}, {Paxton}, {Nomoto}, {Suzuki}, \& {Toki}}]{denissenkov:15}
{Denissenkov}, P.~A., {Truran}, J.~W., {Herwig}, F., {Jones}, S., {Paxton}, B.,
  {Nomoto}, K., {Suzuki}, T., \& {Toki}, H. 2015, \mnras, 447, 2696

\bibitem[{{Denissenkov} {et~al.}(2013){Denissenkov}, {Truran}, {Pignatari},
  {Trappitsch}, {Ritter}, {Herwig}, {Battino}, \&
  {Setoodehnia}}]{denissenkov:13}
{Denissenkov}, P.~A., {Truran}, J.~W., {Pignatari}, M., {Trappitsch}, R.,
  {Ritter}, C., {Herwig}, F., {Battino}, U., \& {Setoodehnia}, K. 2013, ArXiv
  e-prints

\bibitem[{{Dillmann} {et~al.}(2008){Dillmann}, {Plag}, {Heil}, {K{\"a}ppeler},
  \& {Rauscher}}]{dillmann:08}
{Dillmann}, I., {Plag}, R., {Heil}, M., {K{\"a}ppeler}, F., \& {Rauscher}, T.
  2008, ArXiv e-prints

\bibitem[{{Dillmann} {et~al.}(2014){Dillmann}, {Sz{\"u}cs}, {Plag},
  {F{\"u}l{\"o}p}, {K{\"a}ppeler}, {Mengoni}, \& {Rauscher}}]{dillmann:14}
{Dillmann}, I., {Sz{\"u}cs}, T., {Plag}, R., {F{\"u}l{\"o}p}, Z.,
  {K{\"a}ppeler}, F., {Mengoni}, A., \& {Rauscher}, T. 2014, Nuclear Data
  Sheets, 120, 171

\bibitem[{{Dominguez} {et~al.}(1996){Dominguez}, {Straniero}, {Tornambe}, \&
  {Isern}}]{dominguez:96}
{Dominguez}, I., {Straniero}, O., {Tornambe}, A., \& {Isern}, J. 1996, \apj,
  472, 783

\bibitem[{{Dominguez} {et~al.}(1993){Dominguez}, {Tornambe}, \&
  {Isern}}]{dominguez:93}
{Dominguez}, I., {Tornambe}, A., \& {Isern}, J. 1993, \apj, 419, 268

\bibitem[{{Eichler} {et~al.}(2018){Eichler}, {Nakamura}, {Takiwaki}, {Kuroda},
  {Kotake}, {Hempel}, {Cabez{\'o}n}, {Liebend{\"o}rfer}, \&
  {Thielemann}}]{eichler:18}
{Eichler}, M., {et~al.} 2018, Journal of Physics G Nuclear Physics, 45, 014001

\bibitem[{{Eitner} {et~al.}(2020){Eitner}, {Bergemann}, {Hansen}, {Cescutti},
  {Seitenzahl}, {Larsen}, \& {Plez}}]{eitner:20}
{Eitner}, P., {Bergemann}, M., {Hansen}, C.~J., {Cescutti}, G., {Seitenzahl},
  I.~R., {Larsen}, S., \& {Plez}, B. 2020, \aap, 635, A38

\bibitem[{{Ferguson} {et~al.}(2005){Ferguson}, {Alexander}, {Allard}, {Barman},
  {Bodnarik}, {Hauschildt}, {Heffner-Wong}, \& {Tamanai}}]{ferguson:05}
{Ferguson}, J.~W., {Alexander}, D.~R., {Allard}, F., {Barman}, T., {Bodnarik},
  J.~G., {Hauschildt}, P.~H., {Heffner-Wong}, A., \& {Tamanai}, A. 2005,
  Astrophysical Journal, 623, 585

\bibitem[{{Filippenko}(1997)}]{filippenko:97}
{Filippenko}, A.~V. 1997, \araa, 35, 309

\bibitem[{{Fr{\"o}hlich} {et~al.}(2006){Fr{\"o}hlich}, {Mart{\'\i}nez-Pinedo},
  {Liebend{\"o}rfer}, {Thielemann}, {Bravo}, {Hix}, {Langanke}, \&
  {Zinner}}]{frohlich:06}
{Fr{\"o}hlich}, C., {Mart{\'\i}nez-Pinedo}, G., {Liebend{\"o}rfer}, M.,
  {Thielemann}, F.~K., {Bravo}, E., {Hix}, W.~R., {Langanke}, K., \& {Zinner},
  N.~T. 2006, \prl, 96, 142502

\bibitem[{{Fuller} {et~al.}(1985){Fuller}, {Fowler}, \& {Newman}}]{fuller:85}
{Fuller}, G.~M., {Fowler}, W.~A., \& {Newman}, M.~J. 1985, ApJ, 293, 1

\bibitem[{{Fynbo} {et~al.}(2005){Fynbo}, {Diget}, {Bergmann}, {Borge},
  {Cederk{\" a}ll}, {Dendooven}, {Fraile}, {Franchoo}, {Fedosseev}, {Fulton},
  {Huang}, {Huikari}, {Jeppesen}, {Jokinen}, {Jones}, {Jonson}, {K{\" o}ster},
  {Langanke}, {Meister}, {Nilsson}, {Nyman}, {Prezado}, {Riisager},
  {Rinta-Antila}, {Tengblad}, {Turrion}, {Wang}, {Weissman}, {Wilhelmsen}, {{\"
  A}yst{\" o}}, \& {The ISOLDE Collaboration}}]{fynbo:05}
{Fynbo}, H.~O.~U., {et~al.} 2005, Nature, 433, 136

\bibitem[{{Goriely}(1999)}]{goriely:99}
{Goriely}, S. 1999, \aap, 342, 881

\bibitem[{{Goriely} {et~al.}(2002){Goriely}, {Jos{\'e}}, {Hernanz}, {Rayet}, \&
  {Arnould}}]{goriely:02}
{Goriely}, S., {Jos{\'e}}, J., {Hernanz}, M., {Rayet}, M., \& {Arnould}, M.
  2002, \aap, 383, L27

\bibitem[{{Grevesse} \& {Noels}(1993)}]{grevesse:93}
{Grevesse}, N., \& {Noels}, A. 1993, in Origin and Evolution of the Elements,
  ed. {N.~Prantzos, E.~Vangioni-Flam, \& M.~Casse}, 15--25

\bibitem[{{Heil} {et~al.}(2008){Heil}, {Winckler}, {Dababneh}, {K{\"a}ppeler},
  {Wisshak}, {Bisterzo}, {Gallino}, {Davis}, \& {Rauscher}}]{heil:08}
{Heil}, M., {et~al.} 2008, ApJ, 673, 434

\bibitem[{{Herwig}(2000)}]{herwig:00}
{Herwig}, F. 2000, \aap, 360, 952

\bibitem[{{Herwig}(2001)}]{herwig:00a}
---. 2001, APSS, 275, 15

\bibitem[{{Herwig}(2005)}]{herwig:05}
---. 2005, \araa, 43, 435

\bibitem[{Herwig {et~al.}(1999)Herwig, Blocker, Langer, \&
  Driebe}]{Herwig:1999uf}
Herwig, F., Blocker, T., Langer, N., \& Driebe, T. 1999, A{\&}A, 349, L5

\bibitem[{{Herwig} {et~al.}(1999){Herwig}, {Bl{\"o}cker}, {Langer}, \&
  {Driebe}}]{herwig:99}
{Herwig}, F., {Bl{\"o}cker}, T., {Langer}, N., \& {Driebe}, T. 1999, \aap, 349,
  L5

\bibitem[{Herwig {et~al.}(2011)Herwig, Pignatari, Woodward, Porter,
  Rockefeller, Fryer, Bennett, \& Hirschi}]{Herwig:2011dj}
Herwig, F., Pignatari, M., Woodward, P.~R., Porter, D.~H., Rockefeller, G.,
  Fryer, C.~L., Bennett, M., \& Hirschi, R. 2011, ApJ, 727, 89

\bibitem[{{Herwig} {et~al.}(2011){Herwig}, {Pignatari}, {Woodward}, {Porter},
  {Rockefeller}, {Fryer}, {Bennett}, \& {Hirschi}}]{herwig:11}
{Herwig}, F., {Pignatari}, M., {Woodward}, P.~R., {Porter}, D.~H.,
  {Rockefeller}, G., {Fryer}, C.~L., {Bennett}, M., \& {Hirschi}, R. 2011,
  \apj, 727, 89

\bibitem[{Herwig {et~al.}(2014)Herwig, Woodward, Lin, Knox, \&
  Fryer}]{Herwig:2014cx}
Herwig, F., Woodward, P.~R., Lin, P.-H., Knox, M., \& Fryer, C. 2014, ApJ, 792,
  L3

\bibitem[{{Herwig} {et~al.}(2014){Herwig}, {Woodward}, {Lin}, {Knox}, \&
  {Fryer}}]{herwig:14}
{Herwig}, F., {Woodward}, P.~R., {Lin}, P.-H., {Knox}, M., \& {Fryer}, C. 2014,
  \apjl, 792, L3

\bibitem[{{Hillebrandt} {et~al.}(2013){Hillebrandt}, {Kromer}, {R{\"o}pke}, \&
  {Ruiter}}]{hillebrandt:13}
{Hillebrandt}, W., {Kromer}, M., {R{\"o}pke}, F.~K., \& {Ruiter}, A.~J. 2013,
  Frontiers of Physics, 8, 116

\bibitem[{{Howard} {et~al.}(1991){Howard}, {Meyer}, \& {Woosley}}]{howard:91}
{Howard}, W.~M., {Meyer}, B.~S., \& {Woosley}, S.~E. 1991, \apjl, 373, L5

\bibitem[{{Iben}(1981)}]{iben:81}
{Iben}, Jr., I. 1981, \apj, 243, 987

\bibitem[{{Iben} \& {Tutukov}(1984)}]{iben:84c}
{Iben}, I., J., \& {Tutukov}, A.~V. 1984, \apjs, 54, 335

\bibitem[{{Iglesias} \& {Rogers}(1996)}]{Iglesias:96}
{Iglesias}, C.~A., \& {Rogers}, F.~J. 1996, \apj, 464, 943

\bibitem[{{Iliadis} {et~al.}(2001){Iliadis}, {D'Auria}, {Starrfield},
  {Thompson}, \& {Wiescher}}]{iliadis:01}
{Iliadis}, C., {D'Auria}, J.~M., {Starrfield}, S., {Thompson}, W.~J., \&
  {Wiescher}, M. 2001, ApJS, 134, 151

\bibitem[{{Imbriani} {et~al.}(2004){Imbriani}, {Costantini}, {Formicola},
  {Bemmerer}, {Bonetti}, {Broggini}, {Corvisiero}, {Cruz}, {F{\" u}l{\" o}p},
  {Gervino}, {Guglielmetti}, {Gustavino}, {Gy{\" u}rky}, {Jesus}, {Junker},
  {Lemut}, {Menegazzo}, {Prati}, {Roca}, {Rolfs}, {Romano}, {Rossi Alvarez},
  {Sch{\" u}mann}, {Somorjai}, {Straniero}, {Strieder}, {Terrasi},
  {Trautvetter}, {Vomiero}, \& {Zavatarelli}}]{imbriani:04}
{Imbriani}, G., {et~al.} 2004, A\&A, 420, 625

\bibitem[{{Imbriani} {et~al.}(2005){Imbriani}, {Costantini}, {Formicola},
  {Vomiero}, {Angulo}, {Bemmerer}, {Bonetti}, {Broggini}, {Confortola},
  {Corvisiero}, {Cruz}, {Descouvemont}, {F{\"u}l{\"o}p}, {Gervino},
  {Guglielmetti}, {Gustavino}, {Gy{\"u}rky}, {Jesus}, {Junker}, {Klug},
  {Lemut}, {Menegazzo}, {Prati}, {Roca}, {Rolfs}, {Romano}, {Rossi-Alvarez},
  {Sch{\"u}mann}, {Sch{\"u}rmann}, {Somorjai}, {Straniero}, {Strieder},
  {Terrasi}, \& {Trautvetter}}]{imbriani:05}
---. 2005, European Physical Journal A, 25, 455

\bibitem[{{Iwamoto} {et~al.}(1999){Iwamoto}, {Brachwitz}, {Nomoto},
  {Kishimoto}, {Umeda}, {Hix}, \& {Thielemann}}]{iwamoto:99}
{Iwamoto}, K., {Brachwitz}, F., {Nomoto}, K., {Kishimoto}, N., {Umeda}, H.,
  {Hix}, W.~R., \& {Thielemann}, F.-K. 1999, \apjs, 125, 439

\bibitem[{{Jaeger} {et~al.}(2001){Jaeger}, {Kunz}, {Mayer}, {Hammer}, {Staudt},
  {Kratz}, \& {Pfeiffer}}]{jaeger:01}
{Jaeger}, M., {Kunz}, R., {Mayer}, A., {Hammer}, J.~W., {Staudt}, G., {Kratz},
  K.~L., \& {Pfeiffer}, B. 2001, Physical Review Letters, 87, 202501

\bibitem[{{Johansson} {et~al.}(2016){Johansson}, {Woods}, {Gilfanov}, {Sarzi},
  {Chen}, \& {Oh}}]{johansson:16}
{Johansson}, J., {Woods}, T.~E., {Gilfanov}, M., {Sarzi}, M., {Chen}, Y.-M., \&
  {Oh}, K. 2016, \mnras, 461, 4505

\bibitem[{{Jose} {et~al.}(1993){Jose}, {Hernanz}, \& {Isern}}]{jose:93}
{Jose}, J., {Hernanz}, M., \& {Isern}, J. 1993, \aap, 269, 291

\bibitem[{{K{\"a}ppeler} {et~al.}(2011){K{\"a}ppeler}, {Gallino}, {Bisterzo},
  \& {Aoki}}]{kaeppeler:11}
{K{\"a}ppeler}, F., {Gallino}, R., {Bisterzo}, S., \& {Aoki}, W. 2011, Reviews
  of Modern Physics, 83, 157

\bibitem[{{Karakas} \& {Lattanzio}(2014)}]{karakas:14}
{Karakas}, A.~I., \& {Lattanzio}, J.~C. 2014, \pasa, 31, 30

\bibitem[{{Kato} \& {Hachisu}(1999)}]{kato:99}
{Kato}, M., \& {Hachisu}, I. 1999, \apjl, 513, L41

\bibitem[{Kippenhahn \& Weigert(1990)}]{kippenhahn:90}
Kippenhahn, R., \& Weigert, A. 1990, Stellar structure and evolution (Berlin:
  Springer)

\bibitem[{{Kobayashi} {et~al.}(2019){Kobayashi}, {Leung}, \&
  {Nomoto}}]{kobayashi:20}
{Kobayashi}, C., {Leung}, S.-C., \& {Nomoto}, K. 2019, arXiv e-prints,
  arXiv:1906.09980

\bibitem[{{Kunz} {et~al.}(2002){Kunz}, {Fey}, {Jaeger}, {Mayer}, {Hammer},
  {Staudt}, {Harissopulos}, \& {Paradellis}}]{kunz:02}
{Kunz}, R., {Fey}, M., {Jaeger}, M., {Mayer}, A., {Hammer}, J.~W., {Staudt},
  G., {Harissopulos}, S., \& {Paradellis}, T. 2002, \apj, 567, 643

\bibitem[{{Kusakabe} {et~al.}(2011){Kusakabe}, {Iwamoto}, \&
  {Nomoto}}]{kusakabe:11}
{Kusakabe}, M., {Iwamoto}, N., \& {Nomoto}, K. 2011, \apj, 726, 25

\bibitem[{{Langanke} \& {Mart{\'{\i}}nez-Pinedo}(2000)}]{langanke:00}
{Langanke}, K., \& {Mart{\'{\i}}nez-Pinedo}, G. 2000, Nuclear Physics A, 673,
  481

\bibitem[{{Langer} {et~al.}(2000){Langer}, {Deutschmann}, {Wellstein}, \&
  {H{\"o}flich}}]{langer:00}
{Langer}, N., {Deutschmann}, A., {Wellstein}, S., \& {H{\"o}flich}, P. 2000,
  \aap, 362, 1046

\bibitem[{{Lugaro} {et~al.}(2014){Lugaro}, {Tagliente}, {Karakas}, {Milazzo},
  {K{\"a}ppeler}, {Davis}, \& {Savina}}]{lugaro:14}
{Lugaro}, M., {Tagliente}, G., {Karakas}, A.~I., {Milazzo}, P.~M.,
  {K{\"a}ppeler}, F., {Davis}, A.~M., \& {Savina}, M.~R. 2014, \apj, 780, 95

\bibitem[{{Ma} {et~al.}(2013){Ma}, {Chen}, {Chen}, {Denissenkov}, \&
  {Han}}]{ma:13}
{Ma}, X., {Chen}, X., {Chen}, H.-l., {Denissenkov}, P.~A., \& {Han}, Z. 2013,
  \apjl, 778, L32

\bibitem[{{MacDonald}(1983)}]{macdonald:83}
{MacDonald}, J. 1983, \apj, 273, 289

\bibitem[{{Maoz} {et~al.}(2014){Maoz}, {Mannucci}, \& {Nelemans}}]{maoz:14}
{Maoz}, D., {Mannucci}, F., \& {Nelemans}, G. 2014, \araa, 52, 107

\bibitem[{{Mazzali} {et~al.}(2007){Mazzali}, {R{\"o}pke}, \&
  {Hillebrandt}}]{mazzali:07}
{Mazzali}, P., {R{\"o}pke}, F.K..~{Benetti}, S., \& {Hillebrandt}. 2007, 315,
  825

\bibitem[{{Mestel}(1952)}]{mestel:52}
{Mestel}, L. 1952, \mnras, 112, 598

\bibitem[{{Miller Bertolami} \& {Althaus}(2007)}]{miller-bertolami:07}
{Miller Bertolami}, M.~M., \& {Althaus}, L.~G. 2007, MNRAS, 380, 763

\bibitem[{Miller~Bertolami {et~al.}(2006)Miller~Bertolami, Althaus, Serenelli,
  \& Panei}]{MillerBertolami:2006dr}
Miller~Bertolami, M.~M., Althaus, L.~G., Serenelli, A.~M., \& Panei, J.~A.
  2006, A{\&}A, 449, 313

\bibitem[{{Mohr} {et~al.}(2007){Mohr}, {K{\"a}ppeler}, \& {Gallino}}]{mohr:07}
{Mohr}, P., {K{\"a}ppeler}, F., \& {Gallino}, R. 2007, Phys.\ Rev.\ C., 75,
  012802

\bibitem[{{Nemeth} {et~al.}(1994){Nemeth}, {Kaeppeler}, {Theis}, {Belgya}, \&
  {Yates}}]{nemeth:94}
{Nemeth}, Z., {Kaeppeler}, F., {Theis}, C., {Belgya}, T., \& {Yates}, S.~W.
  1994, \apj, 426, 357

\bibitem[{{Nishimura} {et~al.}(2019){Nishimura}, {Rauscher}, {Hirschi},
  {Cescutti}, {Murphy}, \& {Fr{\"o}hlich}}]{nishimura:19}
{Nishimura}, N., {Rauscher}, T., {Hirschi}, R., {Cescutti}, G., {Murphy}, A.
  S.~J., \& {Fr{\"o}hlich}, C. 2019, \mnras, 489, 1379

\bibitem[{{Nishimura} {et~al.}(2018){Nishimura}, {Rauscher}, {Hirschi},
  {Murphy}, {Cescutti}, \& {Travaglio}}]{nishimura:18}
{Nishimura}, N., {Rauscher}, T., {Hirschi}, R., {Murphy}, A. S.~J., {Cescutti},
  G., \& {Travaglio}, C. 2018, \mnras, 474, 3133

\bibitem[{{Nomoto}(1982)}]{nomoto:82}
{Nomoto}, K. 1982, \apj, 253, 798

\bibitem[{{Nomoto}(1984)}]{nomoto:84}
---. 1984, ApJ, 277, 791

\bibitem[{{Nomoto} {et~al.}(2007){Nomoto}, {Saio}, {Kato}, \&
  {Hachisu}}]{nomoto:07}
{Nomoto}, K., {Saio}, H., {Kato}, M., \& {Hachisu}, I. 2007, \apj, 663, 1269

\bibitem[{{Nomoto} {et~al.}(1984){Nomoto}, {Thielemann}, \&
  {Yokoi}}]{nomoto:84b}
{Nomoto}, K., {Thielemann}, F.~K., \& {Yokoi}, K. 1984, \apj, 286, 644

\bibitem[{{Oda} {et~al.}(1994){Oda}, {Hino}, {Muto}, {Takahara}, \&
  {Sato}}]{oda:94}
{Oda}, T., {Hino}, M., {Muto}, K., {Takahara}, M., \& {Sato}, K. 1994, Atomic
  Data and Nuclear Data Tables, 56, 231

\bibitem[{{Paczynski}(1974)}]{paczynski:74}
{Paczynski}, B. 1974, \apj, 192, 483

\bibitem[{{Paczynski}(1983)}]{paczynski:83}
---. 1983, \apj, 264, 282

\bibitem[{{Paczynski} \& {Proszynski}(1986)}]{paczynski:86}
{Paczynski}, B., \& {Proszynski}, M. 1986, \apj, 302, 519

\bibitem[{{Pakmor} {et~al.}(2012){Pakmor}, {Kromer}, {Taubenberger}, {Sim},
  {R{\"o}pke}, \& {Hillebrandt}}]{palkmor:12}
{Pakmor}, R., {Kromer}, M., {Taubenberger}, S., {Sim}, S.~A., {R{\"o}pke},
  F.~K., \& {Hillebrandt}, W. 2012, \apjl, 747, L10

\bibitem[{Paxton {et~al.}(2010)Paxton, Bildsten, Timmes, Nelson, Lesaffre,
  Herwig, Dotter, VandenBerg, Sigur?son, Hirschi, \& Tomshine}]{mesa}
Paxton, P., {et~al.} 2010, {MESA:} modules for experiments in stellar
  astrophysics, http://mesa.sourceforge.net

\bibitem[{{Piersanti} {et~al.}(2014){Piersanti}, {Tornamb{\'e}}, \&
  {Yungelson}}]{piersanti:14}
{Piersanti}, L., {Tornamb{\'e}}, A., \& {Yungelson}, L.~R. 2014, \mnras, 445,
  3239

\bibitem[{{Pignatari} {et~al.}(2010){Pignatari}, {Gallino}, {Heil}, {Wiescher},
  {K{\"a}ppeler}, {Herwig}, \& {Bisterzo}}]{pignatari:10}
{Pignatari}, M., {Gallino}, R., {Heil}, M., {Wiescher}, M., {K{\"a}ppeler}, F.,
  {Herwig}, F., \& {Bisterzo}, S. 2010, \apj, 710, 1557

\bibitem[{{Pignatari} {et~al.}(2016){Pignatari}, {Herwig}, {Hirschi},
  {Bennett}, {Rockefeller}, {Fryer}, {Timmes}, {Ritter}, {Heger}, {Jones},
  {Battino}, {Dotter}, {Trappitsch}, {Diehl}, {Frischknecht}, {Hungerford},
  {Magkotsios}, {Travaglio}, \& {Young}}]{pignatari:16}
{Pignatari}, M., {et~al.} 2016, \apjs, 225, 24

\bibitem[{{Piro} \& {Bildsten}(2008)}]{piro:08}
{Piro}, A.~L., \& {Bildsten}, L. 2008, \apj, 673, 1009

\bibitem[{{Rauscher} {et~al.}(2002){Rauscher}, {Heger}, {Hoffman}, \&
  {Woosley}}]{rauscher:02}
{Rauscher}, T., {Heger}, A., {Hoffman}, R.~D., \& {Woosley}, S.~E. 2002, \apj,
  576, 323

\bibitem[{Rauscher {et~al.}(2020)Rauscher, Nishimura, Cescutti, Hirschi,
  Murphy, \& Fröhlich}]{rauscher:20}
Rauscher, T., Nishimura, N., Cescutti, G., Hirschi, R., Murphy, A. S.~J., \&
  Fröhlich, C. 2020, Impact of Uncertainties in Astrophysical Reaction Rates
  on Nucleosynthesis in the <italic>νp</italic> Process

\bibitem[{{Rauscher} {et~al.}(1997){Rauscher}, {Thielemann}, \&
  {Kratz}}]{rauscher:97}
{Rauscher}, T., {Thielemann}, F.-K., \& {Kratz}, K.-L. 1997, \prc, 56, 1613

\bibitem[{{Ritter} {et~al.}(2018){Ritter}, {Andrassy}, {C{\^o}t{\'e}},
  {Herwig}, {Woodward}, {Pignatari}, \& {Jones}}]{ritter:18}
{Ritter}, C., {Andrassy}, R., {C{\^o}t{\'e}}, B., {Herwig}, F., {Woodward},
  P.~R., {Pignatari}, M., \& {Jones}, S. 2018, \mnras, 474, L1

\bibitem[{{R{\"o}pke} {et~al.}(2012){R{\"o}pke}, {Kromer}, {Seitenzahl},
  {Pakmor}, {Sim}, {Taubenberger}, {Ciaraldi-Schoolmann}, {Hillebrandt},
  {Aldering}, {Antilogus}, {Baltay}, {Benitez-Herrera}, {Bongard}, {Buton},
  {Canto}, {Cellier-Holzem}, {Childress}, {Chotard}, {Copin}, {Fakhouri},
  {Fink}, {Fouchez}, {Gangler}, {Guy}, {Hachinger}, {Hsiao}, {Chen},
  {Kerschhaggl}, {Kowalski}, {Nugent}, {Paech}, {Pain}, {Pecontal}, {Pereira},
  {Perlmutter}, {Rabinowitz}, {Rigault}, {Runge}, {Saunders}, {Smadja},
  {Suzuki}, {Tao}, {Thomas}, {Tilquin}, \& {Wu}}]{ropke:12}
{R{\"o}pke}, F.~K., {et~al.} 2012, \apjl, 750, L19

\bibitem[{{Rybicki} \& {Lightman}(1979)}]{rybicki:79}
{Rybicki}, G.~B., \& {Lightman}, A.~P. 1979, {Radiative processes in
  astrophysics}

\bibitem[{{Seitenzahl} {et~al.}(2019){Seitenzahl}, {Ghavamian}, {Laming}, \&
  {Vogt}}]{seitenzahl:19}
{Seitenzahl}, I.~R., {Ghavamian}, P., {Laming}, J.~M., \& {Vogt}, F.~P.~A.
  2019, \prl, 123, 041101

\bibitem[{{Shen} \& {Bildsten}(2007)}]{shen:07}
{Shen}, K.~J., \& {Bildsten}, L. 2007, \apj, 660, 1444

\bibitem[{{Sion} {et~al.}(1979){Sion}, {Acierno}, \& {Tomczyk}}]{sion:79}
{Sion}, E.~M., {Acierno}, M.~J., \& {Tomczyk}, S. 1979, \apj, 230, 832

\bibitem[{{Starrfield} {et~al.}(2012){Starrfield}, {Iliadis}, {Timmes}, {Hix},
  {Arnett}, {Meakin}, \& {Sparks}}]{starrfield:12}
{Starrfield}, S., {Iliadis}, C., {Timmes}, F.~X., {Hix}, W.~R., {Arnett},
  W.~D., {Meakin}, C., \& {Sparks}, W.~M. 2012, Bulletin of the Astronomical
  Society of India, 40, 419

\bibitem[{{Tagliente} {et~al.}(2008{\natexlab{a}}){Tagliente}, {Fujii},
  {Milazzo}, {Moreau}, {Aerts}, {Abbondanno}, {{\'A}lvarez}, {Alvarez-Velarde},
  {Andriamonje}, {Andrzejewski}, {Assimakopoulos}, {Audouin}, {Badurek},
  {Baumann}, {Be{\v{c}}v{\'a}{\v{r}}}, {Berthoumieux}, {Bisterzo},
  {Calvi{\~n}o}, {Calviani}, {Cano-Ott}, {Capote}, {Carrapi{\c{c}}o},
  {Cennini}, {Chepel}, {Chiaveri}, {Colonna}, {Cortes}, {Couture}, {Cox},
  {Dahlfors}, {David}, {Dillman}, {Domingo-Pardo}, {Dridi}, {Duran},
  {Eleftheriadis}, {Embid-Segura}, {Ferrant}, {Ferrari}, {Ferreira-Marques},
  {Furman}, {Gallino}, {Goncalves}, {Gonzalez-Romero}, {Gramegna}, {Guerrero},
  {Gunsing}, {Haas}, {Haight}, {Heil}, {Herrera-Martinez}, {Igashira},
  {Jericha}, {K{\"a}ppeler}, {Kadi}, {Karadimos}, {Karamanis}, {Kerveno},
  {Koehler}, {Kossionides}, {Krti{\v{c}}ka}, {Lamboudis}, {Leeb}, {Lindote},
  {Lopes}, {Lozano}, {Lukic}, {Marganiec}, {Marrone}, {Mart{\'\i}nez},
  {Massimi}, {Mastinu}, {Mengoni}, {Mosconi}, {Neves}, {Oberhummer}, {O'Brien},
  {Pancin}, {Papachristodoulou}, {Papadopoulos}, {Paradela}, {Patronis},
  {Pavlik}, {Pavlopoulos}, {Perrot}, {Pigni}, {Plag}, {Plompen}, {Plukis},
  {Poch}, {Praena}, {Pretel}, {Quesada}, {Rauscher}, {Reifarth}, {Rubbia},
  {Rudolf}, {Rullhusen}, {Salgado}, {Santos}, {Sarchiapone}, {Savvidis},
  {Stephan}, {Tain}, {Tassan-Got}, {Tavora}, {Terlizzi}, {Vannini}, {Vaz},
  {Ventura}, {Villamarin}, {Vincente}, {Vlachoudis}, {Vlastou}, {Voss},
  {Walter}, {Wendler}, {Wiescher}, \& {Wisshak}}]{tagliente:08a}
{Tagliente}, G., {et~al.} 2008{\natexlab{a}}, \prc, 77, 035802

\bibitem[{{Tagliente} {et~al.}(2010){Tagliente}, {Milazzo}, {Fujii},
  {Abbondanno}, {Aerts}, {{\'A}lvarez}, {Alvarez-Velarde}, {Andriamonje},
  {Andrzejewski}, {Audouin}, {Badurek}, {Baumann}, {Be{\v{c}}v{\'a}{\v{r}}},
  {Belloni}, {Berthoumieux}, {Bisterzo}, {Calvi{\~n}o}, {Calviani}, {Cano-Ott},
  {Capote}, {Carrapi{\c{c}}o}, {Cennini}, {Chepel}, {Chiaveri}, {Colonna},
  {Cortes}, {Couture}, {Cox}, {Dahlfors}, {David}, {Dillmann}, {Domingo-Pardo},
  {Dridi}, {Duran}, {Eleftheriadis}, {Embid-Segura}, {Ferrari},
  {Ferreira-Marques}, {Furman}, {Gallino}, {Goncalves}, {Gonzalez-Romero},
  {Gramegna}, {Guerrero}, {Gunsing}, {Haas}, {Haight}, {Heil},
  {Herrera-Martinez}, {Igashira}, {Jericha}, {K{\"a}ppeler}, {Kadi},
  {Karadimos}, {Karamanis}, {Kerveno}, {Kossionides}, {Krti{\v{c}}ka},
  {Lamboudis}, {Leeb}, {Lindote}, {Lopes}, {Lozano}, {Lukic}, {Marganiec},
  {Marrone}, {Mart{\'\i}nez}, {Massimi}, {Mastinu}, {Mengoni}, {Moreau},
  {Mosconi}, {Neves}, {Oberhummer}, {O'Brien}, {Pancin}, {Papachristodoulou},
  {Papadopoulos}, {Paradela}, {Patronis}, {Pavlik}, {Pavlopoulos}, {Perrot},
  {Pigni}, {Plag}, {Plompen}, {Plukis}, {Poch}, {Praena}, {Pretel}, {Quesada},
  {Rauscher}, {Reifarth}, {Rosetti}, {Rubbia}, {Rudolf}, {Rullhusen},
  {Salgado}, {Santos}, {Sarchiapone}, {Savvidis}, {Stephan}, {Tain},
  {Tassan-Got}, {Tavora}, {Terlizzi}, {Vannini}, {Vaz}, {Ventura},
  {Villamarin}, {Vincente}, {Vlachoudis}, {Vlastou}, {Voss}, {Walter},
  {Wendler}, {Wiescher}, \& {Wisshak}}]{tagliente:10}
---. 2010, \prc, 81, 055801

\bibitem[{{Tagliente} {et~al.}(2011{\natexlab{a}}){Tagliente}, {Milazzo},
  {Fujii}, {Abbondanno}, {Aerts}, {{\'A}lvarez}, {Alvarez-Velarde},
  {Andriamonje}, {Andrzejewski}, {Audouin}, {Badurek}, {Baumann},
  {Be{\v{c}}v{\'a}{\v{r}}}, {Belloni}, {Berthoumieux}, {Bisterzo},
  {Calvi{\~n}o}, {Calviani}, {Cano-Ott}, {Capote}, {Carrapi{\c{c}}o},
  {Cennini}, {Chepel}, {Chiaveri}, {Colonna}, {Cortes}, {Couture}, {Cox},
  {Dahlfors}, {David}, {Dillmann}, {Domingo-Pardo}, {Dridi}, {Duran},
  {Eleftheriadis}, {Embid-Segura}, {Ferrari}, {Ferreira-Marques}, {Furman},
  {Gallino}, {Goncalves}, {Gonzalez-Romero}, {Gramegna}, {Guerrero}, {Gunsing},
  {Haas}, {Haight}, {Heil}, {Herrera-Martinez}, {Jericha}, {K{\"a}ppeler},
  {Kadi}, {Karadimos}, {Karamanis}, {Kerveno}, {Kossionides}, {Krti{\v{c}}ka},
  {Lamboudis}, {Leeb}, {Lindote}, {Lopes}, {Lozano}, {Lukic}, {Marganiec},
  {Marrone}, {Mart{\'\i}nez}, {Massimi}, {Mastinu}, {Mengoni}, {Moreau},
  {Mosconi}, {Neves}, {Oberhummer}, {O'Brien}, {Pancin}, {Papachristodoulou},
  {Papadopoulos}, {Paradela}, {Patronis}, {Pavlik}, {Pavlopoulos}, {Perrot},
  {Pigni}, {Plag}, {Plompen}, {Plukis}, {Poch}, {Praena}, {Pretel}, {Quesada},
  {Rauscher}, {Reifarth}, {Rosetti}, {Rubbia}, {Rudolf}, {Rullhusen},
  {Salgado}, {Santos}, {Sarchiapone}, {Savvidis}, {Stephan}, {Tain},
  {Tassan-Got}, {Tavora}, {Terlizzi}, {Vannini}, {Vaz}, {Ventura},
  {Villamarin}, {Vincente}, {Vlachoudis}, {Vlastou}, {Voss}, {Walter},
  {Wiescher}, \& {Wisshak}}]{tagliente:11a}
---. 2011{\natexlab{a}}, \prc, 84, 015801

\bibitem[{{Tagliente} {et~al.}(2011{\natexlab{b}}){Tagliente}, {Milazzo},
  {Fujii}, {Abbondanno}, {Aerts}, {{\'A}lvarez}, {Alvarez-Velarde},
  {Andriamonje}, {Andrzejewski}, {Audouin}, {Badurek}, {Baumann},
  {Be{\v{c}}v{\'a}{\v{r}}}, {Belloni}, {Berthoumieux}, {Calvi{\~n}o},
  {Calviani}, {Cano-Ott}, {Capote}, {Carrapi{\c{c}}o}, {Cennini}, {Chepel},
  {Chiaveri}, {Colonna}, {Cortes}, {Couture}, {Dahlfors}, {David}, {Dillmann},
  {Domingo-Pardo}, {Dridi}, {Duran}, {Eleftheriadis}, {Embid-Segura},
  {Ferrari}, {Ferreira-Marques}, {Furman}, {Goncalves}, {Gonzalez-Romero},
  {Gramegna}, {Guerrero}, {Gunsing}, {Haas}, {Haight}, {Heil},
  {Herrera-Martinez}, {Jericha}, {K{\"a}ppeler}, {Kadi}, {Karadimos},
  {Karamanis}, {Kerveno}, {Kossionides}, {Krti{\v{c}}ka}, {Lamboudis}, {Leeb},
  {Lindote}, {Lopes}, {Lukic}, {Marganiec}, {Marrone}, {Mart{\'\i}nez},
  {Massimi}, {Mastinu}, {Mengoni}, {Moreau}, {Mosconi}, {Neves}, {Oberhummer},
  {O'Brien}, {Pancin}, {Papachristodoulou}, {Papadopoulos}, {Paradela},
  {Patronis}, {Pavlik}, {Pavlopoulos}, {Perrot}, {Pigni}, {Plag}, {Plompen},
  {Plukis}, {Poch}, {Praena}, {Pretel}, {Quesada}, {Reifarth}, {Rosetti},
  {Rubbia}, {Rudolf}, {Rullhusen}, {Salgado}, {Santos}, {Sarchiapone},
  {Savvidis}, {Stephan}, {Tain}, {Tassan-Got}, {Tavora}, {Terlizzi}, {Vannini},
  {Vaz}, {Ventura}, {Villamarin}, {Vincente}, {Vlachoudis}, {Vlastou}, {Voss},
  {Walter}, {Wiescher}, \& {Wisshak}}]{tagliente:11b}
---. 2011{\natexlab{b}}, \prc, 84, 055802

\bibitem[{{Tagliente} {et~al.}(2013){Tagliente}, {Milazzo}, {Fujii},
  {Abbondanno}, {Aerts}, {{\'A}lvarez}, {Alvarez-Velarde}, {Andriamonje},
  {Andrzejewski}, {Audouin}, {Badurek}, {Baumann}, {Be{\v{c}}v{\'a}{\v{r}}},
  {Belloni}, {Berthoumieux}, {Calvi{\~n}o}, {Calviani}, {Cano-Ott}, {Capote},
  {Carrapi{\c{c}}o}, {Cennini}, {Chepel}, {Chiaveri}, {Colonna}, {Cortes},
  {Couture}, {Dahlfors}, {David}, {Dillmann}, {Domingo-Pardo}, {Dridi},
  {Duran}, {Eleftheriadis}, {Embid-Segura}, {Ferrari}, {Ferreira-Marques},
  {Furman}, {Goncalves}, {Gonzalez-Romero}, {Gramegna}, {Guerrero}, {Gunsing},
  {Haas}, {Haight}, {Heil}, {Herrera-Martinez}, {Jericha}, {K{\"a}ppeler},
  {Kadi}, {Karadimos}, {Karamanis}, {Kerveno}, {Kossionides}, {Krti{\v{c}}ka},
  {Lamboudis}, {Leeb}, {Lindote}, {Lopes}, {Lukic}, {Marganiec}, {Marrone},
  {Mart{\'\i}nez}, {Massimi}, {Mastinu}, {Mengoni}, {Moreau}, {Mosconi},
  {Neves}, {Oberhummer}, {O'Brien}, {Papachristodoulou}, {Papadopoulos},
  {Paradela}, {Patronis}, {Pavlik}, {Pavlopoulos}, {Perrot}, {Pigni}, {Plag},
  {Plompen}, {Plukis}, {Poch}, {Praena}, {Pretel}, {Quesada}, {Reifarth},
  {Rosetti}, {Rubbia}, {Rudolf}, {Rullhusen}, {Salgado}, {Santos},
  {Sarchiapone}, {Savvidis}, {Stephan}, {Tain}, {Tassan-Got}, {Tavora},
  {Terlizzi}, {Vannini}, {Vaz}, {Ventura}, {Villamarin}, {Vincente},
  {Vlachoudis}, {Vlastou}, {Voss}, {Walter}, {Wiescher}, \&
  {Wisshak}}]{tagliente:13}
---. 2013, \prc, 87, 014622

\bibitem[{{Tagliente} {et~al.}(2008{\natexlab{b}}){Tagliente}, {Milazzo},
  {Fujii}, {Aerts}, {Abbondanno}, {{\'A}lvarez}, {Alvarez-Velarde},
  {Andriamonje}, {Andrzejewski}, {Assimakopoulos}, {Audouin}, {Badurek},
  {Baumann}, {Be{\v{c}}v{\'a}{\v{r}}}, {Belloni}, {Berthoumieux},
  {Calvi{\~n}o}, {Calviani}, {Cano-Ott}, {Capote}, {Carrapi{\c{c}}o},
  {Cennini}, {Chepel}, {Colonna}, {Cortes}, {Couture}, {Cox}, {Dahlfors},
  {David}, {Dillmann}, {Domingo-Pardo}, {Dridi}, {Duran}, {Eleftheriadis},
  {Embid-Segura}, {Ferrant}, {Ferrari}, {Ferreira-Marques}, {Furman},
  {Goncalves}, {Gonzalez-Romero}, {Gramegna}, {Guerrero}, {Gunsing}, {Haas},
  {Haight}, {Heil}, {Herrera-Martinez}, {Igashira}, {Jericha}, {K{\"a}ppeler},
  {Kadi}, {Karadimos}, {Karamanis}, {Kerveno}, {Koehler}, {Kossionides},
  {Krti{\v{c}}ka}, {Lamboudis}, {Leeb}, {Lindote}, {Lopes}, {Lozano}, {Lukic},
  {Marganiec}, {Marrone}, {Mart{\'\i}nez}, {Massimi}, {Mastinu}, {Mengoni},
  {Moreau}, {Mosconi}, {Neves}, {Oberhummer}, {O'Brien}, {Pancin},
  {Papachristodoulou}, {Papadopoulos}, {Paradela}, {Patronis}, {Pavlik},
  {Pavlopoulos}, {Perrot}, {Pigni}, {Plag}, {Plompen}, {Plukis}, {Poch},
  {Praena-Rodriguez}, {Pretel}, {Quesada}, {Rauscher}, {Reifarth}, {Rubbia},
  {Rudolf}, {Rullhusen}, {Salgado}, {Santos}, {Sarchiapone}, {Savvidis},
  {Stephan}, {Tain}, {Tassan-Got}, {Tavora}, {Terlizzi}, {Vannini}, {Vaz},
  {Ventura}, {Villamarin}, {Vincente}, {Vlachoudis}, {Vlastou}, {Voss},
  {Walter}, {Wiescher}, \& {Wisshak}}]{tagliente:08b}
---. 2008{\natexlab{b}}, \prc, 78, 045804

\bibitem[{Takahashi \& Yokoi(1987)}]{takahashi:87}
Takahashi, K., \& Yokoi, K. 1987, ADNDT, 36, 375

\bibitem[{{The} {et~al.}(2007){The}, {El Eid}, \& {Meyer}}]{the:07}
{The}, L., {El Eid}, M.~F., \& {Meyer}, B.~S. 2007, ApJ, 655, 1058

\bibitem[{{Thielemann} {et~al.}(1986){Thielemann}, {Nomoto}, \&
  {Yokoi}}]{thielemann:86}
{Thielemann}, F.-K., {Nomoto}, K., \& {Yokoi}, K. 1986, \aap, 158, 17

\bibitem[{{Tornamb{\'e}} {et~al.}(2000){Tornamb{\'e}}, {Iben}, {Piersanti}, \&
  {Cassisi}}]{tornambe:00}
{Tornamb{\'e}}, A., {Iben}, Jr., I., {Piersanti}, L., \& {Cassisi}, S. 2000,
  \memsai, 71, 471

\bibitem[{{Travaglio} {et~al.}(2015){Travaglio}, {Gallino}, {Rauscher},
  {R{\"o}pke}, \& {Hillebrandt}}]{travaglio:15}
{Travaglio}, C., {Gallino}, R., {Rauscher}, T., {R{\"o}pke}, F.~K., \&
  {Hillebrandt}, W. 2015, \apj, 799, 54

\bibitem[{{Travaglio} {et~al.}(2018){Travaglio}, {Rauscher}, {Heger},
  {Pignatari}, \& {West}}]{travaglio:18}
{Travaglio}, C., {Rauscher}, T., {Heger}, A., {Pignatari}, M., \& {West}, C.
  2018, ArXiv e-prints

\bibitem[{{Travaglio} {et~al.}(2011){Travaglio}, {R{\"o}pke}, {Gallino}, \&
  {Hillebrandt}}]{travaglio:11}
{Travaglio}, C., {R{\"o}pke}, F.~K., {Gallino}, R., \& {Hillebrandt}, W. 2011,
  \apj, 739, 93

\bibitem[{{van den Heuvel} {et~al.}(1992){van den Heuvel}, {Bhattacharya},
  {Nomoto}, \& {Rappaport}}]{vandenheuvel:92}
{van den Heuvel}, E.~P.~J., {Bhattacharya}, D., {Nomoto}, K., \& {Rappaport},
  S.~A. 1992, \aap, 262, 97

\bibitem[{{Webbink}(1984)}]{webbink:84}
{Webbink}, R.~F. 1984, \apj, 277, 355

\bibitem[{{Whelan} \& {Iben}(1973)}]{whelan:73}
{Whelan}, J., \& {Iben}, Jr., I. 1973, \apj, 186, 1007

\bibitem[{{Wolf} {et~al.}(2013){Wolf}, {Bildsten}, {Brooks}, \&
  {Paxton}}]{wolf:13}
{Wolf}, W.~M., {Bildsten}, L., {Brooks}, J., \& {Paxton}, B. 2013, \apj, 777,
  136

\bibitem[{{Woods} \& {Gilfanov}(2013)}]{woods:13}
{Woods}, T.~E., \& {Gilfanov}, M. 2013, \mnras, 432, 1640

\bibitem[{{Woosley} {et~al.}(1990){Woosley}, {Hartmann}, {Hoffman}, \&
  {Haxton}}]{woosley:90}
{Woosley}, S.~E., {Hartmann}, D.~H., {Hoffman}, R.~D., \& {Haxton}, W.~C. 1990,
  \apj, 356, 272

\bibitem[{{Woosley} \& {Howard}(1978)}]{woosley:78}
{Woosley}, S.~E., \& {Howard}, W.~M. 1978, \apjs, 36, 285

\end{thebibliography}

\clearpage

\begin{table}
\begin{center}
\caption{List of accreting WD models with critical stable H-burning conditions: initial mass, initial metallicity and CBM parameterization are given. This first set of stellar models focuses specifically on the H-burning phase (i.e. they experience no He-flash instabilities).
The CBM parameterization is given by a single-exponential decreasing profile. The CBM parameter $f$ is given below the H-burning shell. 
The nuclear network adopted is denoted by $cno$ where $cno-extras.net$ was used or $nova$ where $nova.net$ was used.
}
\begin{tabular}{lcccc}
\hline
name & mass [M$_{\odot}$] & metallicity & f & network\\
\hline
\hline
M0p6.Z1m2.Hburn.bare & 0.639  & 0.01 & - & cno \\ 		
M0p8.Z1m2.Hburn.bare & 0.856  & 0.01 & - & cno \\ 	
M1p0.Z1m2.Hburn.bare & 1.025  & 0.01 & - & cno \\ 
M1p3.Z1m2.Hburn.bare & 1.316  & 0.01 & - & cno \\
\hline
M0p6.Z1m2.Hburn.cbm1   & 0.639  & 0.01 & 0.004 & nova \\ 		
M0p8.Z1m2.Hburn.cbm1   & 0.856  & 0.01 & 0.004 & nova \\ 	
M1p0.Z1m2.Hburn.cbm1   & 1.025  & 0.01 & 0.004 & nova \\ 
M1p3.Z1m2.Hburn.cbm1   & 1.316  & 0.01 & 0.004 & nova \\
\hline
M0p6.Z1m2.Hburn.cbm2   & 0.639  & 0.01 & 0.014 & nova \\ 		
M0p8.Z1m2.Hburn.cbm2   & 0.856  & 0.01 & 0.014 & nova \\ 	
M1p0.Z1m2.Hburn.cbm2   & 1.025  & 0.01 & 0.014 & nova \\ 
M1p3.Z1m2.Hburn.cbm2   & 1.316  & 0.01 & 0.014 & nova \\

\noalign{\smallskip}
\hline
\end{tabular}
\label{tab:WDmodel_name_burn_regime}
\end{center}
\end{table}

\begin{table}
\begin{center}
\tiny\begin{longtable}{lcccccc}
\caption{List of accreting WD models and their relevant parameters: initial mass, initial metallicity, CBM parameterization, number of TP computed with \MESA\ and \mppnp\ and the accretion rate adopted.
The CBM parameterization is given by the exponential decreasing profile of the diffusion coefficient, as discussed by \citet{herwig:00}.)
}\\
\hline
name & mass [M$_{\odot}$] &  metallicity & f$_{1}$ & TP \MESA\ & TP MPPNP & Accretion rate (10$^{-7}$ \msun\ ) \\
\hline
\hline
M0p856.Z1m2 & 0.856  & 0.01	 & 0.011 & 4 & 4 & 2.08 \\ 	
M1p025.Z1m2 & 1.025  & 0.01	 & 0.011 &137& 7 & 3.16 \\ 
M1p259.Z1m2 & 1.259  & 0.01	 & 0.011 & 4 & 4 & 3.70 \\
M1p316.Z1m2 & 1.316  & 0.01	 & 0.011 & 8 & 5 & 6.00 \\
M1p376.Z1m2 & 1.376  & 0.01      & 0.011 & 2 & 2 & 6.00 \\
\hline
\label{tab:WDmodel_name}
\end{longtable}
\end{center}
\end{table}

\begin{table}
\begin{center}
\tiny\begin{longtable}{ccccccc}
\caption{Evolution properties of stellar models presented in this work.
}\\
\hline
 TP &  $t_{TP}$ & $T_{FBOT}$ &  $m_{FBOT}$ & $m_{HTP}$ & $M_{\ast}$ & $M_{\ast \ast}$\\
 & [$yrs$] & [$K$] & [$M_{\odot}$] & [$M_{\odot}$] & [$M_{\odot}$] & [$M_{\odot}$] \\
\hline
\multicolumn{7}{c}{M0p856.Z1m2} \\
\hline
1 & 0.00E+00  &  8.58 &  0.8380 &  0.8608 &  0.861 &  0.855       \\
2 & 2.60E+04  &  8.58 &  0.8368 &  0.8606 &  0.861 &  0.856       \\
3 & 5.21E+04  &  8.59 &  0.8354 &  0.8609 &  0.861 &  0.855       \\
4 & 8.02E+04  &  8.59 &  0.8351 &  0.8609 &  0.861 &  0.855       \\

\hline
\multicolumn{7}{c}{M1p025.Z1m2} \\
\hline
1 & 0.00E+00  &  8.65 &  1.0144 &  1.0265 &  1.026 &  1.024       \\
2 & 5.53E+03  &  8.64 &  1.0149 &  1.0261 &  1.026 &  1.025       \\
3 & 1.11E+04  &  8.64 &  1.0148 &  1.0262 &  1.026 &  1.025       \\
4 & 1.68E+04  &  8.64 &  1.0145 &  1.0263 &  1.026 &  1.025       \\
5 & 2.24E+04  &  8.64 &  1.0145 &  1.0264 &  1.026 &  1.025       \\
6 & 2.81E+04  &  8.64 &  1.0144 &  1.0264 &  1.026 &  1.025       \\
7 & 3.38E+04  &  8.64 &  1.0141 &  1.0265 &  1.026 &  1.025       \\
  &           &       &         &         &        &              \\
136 & 7.88E+05  &  8.66 &  1.0155 &  1.0308 &  1.031 &  1.029       \\
137 & 7.94E+05  &  8.66 &  1.0155 &  1.0309 &  1.031 &  1.029       \\

\hline
\multicolumn{7}{c}{M1p259.Z1m2} \\
\hline
1 & 0.00E+00  &  8.72 &  1.25914 &  1.25950 &  1.25951 & 1.25926       \\
2 & 3.80E+02  &  8.69 &  1.25918 &  1.25941 &  1.25943 & 1.25928       \\
3 & 7.40E+02  &  8.68 &  1.25921 &  1.25943 &  1.25944 & 1.25931       \\
4 & 1.09E+03  &  8.68 &  1.25924 &  1.25945 &  1.25947 & 1.25933       \\

\hline
\multicolumn{7}{c}{M1p316.Z1m2} \\
\hline
1 & 0.00E+00  &  8.64 &  1.316007 &  1.316101 &  1.316103 & 1.316089 \\
2 & 8.20E+01  &  8.61 &  1.316019 &  1.316130 &  1.316131 & 1.316117 \\
3 & 1.58E+02  &  8.61 &  1.316037 &  1.316154 &  1.316156 & 1.316142 \\
4 & 2.33E+02  &  8.61 &  1.316038 &  1.316179 &  1.316181 & 1.316167 \\
5 & 3.10E+02  &  8.61 &  1.316039 &  1.316204 &  1.316206 & 1.316191 \\
6 & 3.86E+02  &  8.61 &  1.316042 &  1.316229 &  1.316230 & 1.316216 \\
7 & 4.62E+02  &  8.61 &  1.316100 &  1.316254 &  1.316255 & 1.316240 \\
8 & 5.45E+02  &  8.61 &  1.316127 &  1.316279 &  1.316280 & 1.316265 \\

\hline
\multicolumn{7}{c}{M1p376.Z1m2} \\
\hline
1 & 0.00E+00  &  8.77 &  1.37616 &  1.37619 &  1.37619 & 1.37618   \\
2 & 2.50E+01  &  8.73 &  1.37618 &  1.37620 &  1.37620 & -         \\

\hline
\multicolumn{7}{l}{TP: TP number.} \\
\multicolumn{7}{l}{ $t_{TP}$: Time since first TP.} \\
\multicolumn{7}{l}{$T_{FBOT}$: Largest temperature (in logarithm) at the bottom of the flash-convective zone.} \\
\multicolumn{7}{l}{$m_{FBOT}$: Mass coordinate at the bottom of the He-flash convective zone.} \\
\multicolumn{7}{l}{$m_{HTP}$: Mass coordinate at the top of the He-flash convective zone.} \\
\multicolumn{7}{l}{ $M_{\ast}$: Stellar mass at the TP.} \\ 
\multicolumn{7}{l}{ $M_{\ast \ast}$: Stellar mass after the TP.} \\ 
\label{tab:WDm3z2m2_TPprop}
\end{longtable}
\end{center}
\end{table}

\begin{table}
\begin{center}
\tiny\begin{longtable}{cccccc}
\caption{Isotopic production factors of selected isotopes resulting from the listed models. The full table is available online or upon request 
} \\
\hline
 Specie & M0.856Z0.0100 & M1.025Z0.0100 & M1.259Z0.0100 & M1.316Z0.0100 & M1.376Z0.0100 \\
\hline
 C  12  & 8.454E+01     & 1.099E+02     & 6.118E+01     & 1.070E+02     & 5.285E+01     \\ 
 C  13  & 3.014E-09     & 4.838E-11     & 1.774E-09     & 4.897E-09     & 5.501E-09     \\ 
 N  14  & 1.252E-05     & 3.665E-10     & 2.646E-13     & 8.528E-13     & 7.233E-14     \\ 
 O  16  & 2.544E+00     & 3.342E+00     & 8.236E+00     & 2.035E+01     & 1.446E+00     \\ 
 NE 20  & 8.191E-01     & 2.778E+00     & 2.037E+01     & 4.726E+01     & 5.008E+00     \\ 
 NE 22  & 2.908E+01     & 2.150E+00     & 1.762E-07     & 2.872E-07     & 2.713E-08     \\ 
 MG 24  & 1.807E-01     & 2.801E+00     & 1.013E+03     & 5.412E+02     & 5.649E+02     \\ 
 MG 25  & 3.439E+01     & 6.383E+01     & 2.229E+02     & 1.955E+02     & 2.365E+02     \\ 
 SI 28  & 8.315E-01     & 1.202E+00     & 3.518E+01     & 3.867E+00     & 5.637E+02     \\ 
 S  32  & 2.639E-01     & 2.076E-01     & 4.903E-01     & 3.657E-01     & 6.946E+00     \\ 
 CA 40  & 2.020E-01    & 1.382E-01     & 4.283E-01     & 2.728E-01     & 1.672E-01    \\ 
 ZN 70  & 2.397E+01     & 7.556E+01     & 2.990E+01     & 5.392E+01     & 1.536E+02     \\ 
 GE 72  & 2.298E+01     & 5.016E+01     & 1.084E+01     & 4.271E+01     & 4.420E+01     \\ 
 GE 73  & 2.431E+01     & 5.463E+01     & 1.268E+01     & 4.837E+01     & 5.316E+01     \\ 
 GE 74  & 2.557E+01     & 5.954E+01     & 1.400E+01     & 5.517E+01     & 4.316E+01     \\ 
 GE 76  & 1.428E+01     & 5.715E+01     & 2.902E+01     & 5.466E+01     & 9.378E+01     \\ 
 SE 82  & 9.012E+00     & 5.441E+01     & 5.978E+01     & 6.250E+01     & 1.836E+02     \\ 
 KR 80  & 3.176E+00     & 1.367E+01     & 2.872E+01     & 7.853E+01     & 2.432E+01     \\ 
 KR 82  &  2.797E+01     & 8.648E+01     & 4.002E+01     & 1.100E+02     & 7.674E+01    \\ 
 KR 86  & 3.360E+01     & 1.717E+02     & 1.592E+02     & 3.667E+02     & 3.640E+02     \\ 
 RB 85  & 2.044E+01     & 7.732E+01     & 2.142E+01     & 9.561E+01     & 7.747E+01     \\ 
 RB 87  & 6.639E+01     & 3.126E+02     & 2.768E+02     & 7.389E+02     & 4.903E+02     \\ 
 SR 88  & 5.568E+00     & 2.623E+01     & 3.594E+01     & 9.610E+01     & 7.792E+01     \\ 
 Y  89  & 4.926E+00     & 2.563E+01     & 4.926E+01     & 1.093E+02     & 1.256E+02     \\ 
 ZR 96  & 1.912E+01     & 1.410E+02     & 8.821E+02     & 1.782E+03     & 3.734E+03     \\ 
 MO100  & 1.004E+00     & 6.555E+00     & 8.355E+01     & 1.045E+02     & 1.976E+02     \\ 
 SB123  & 8.262E-01     & 2.020E+00     & 2.900E+01     & 9.487E+01     & 2.654E+02     \\ 
 BA136  & 9.003E-01     & 1.224E+00     & 2.996E+01     & 1.333E+02     & 4.668E+02     \\
 BA138  & 8.352E-01     & 7.401E-01     & 4.807E+00     & 2.327E+02     & 6.383E+02     \\
 HG204  & 3.557E+00     & 3.709E+00     & 2.170E+00     & 1.216E+03     & 2.809E+03   \\
 PB208  & 1.041E+00     & 1.349E+00     & 7.924E-01     & 1.081E+04     & 1.107E+04   \\
 BI209  & 7.343E-01     & 9.915E-01     & 7.004E-01     & 1.141E+04     & 1.157E+04   \\ 

\hline
\label{tab:pf}
\end{longtable}
\end{center}
\end{table}

\begin{table}
\begin{center}
\tiny\begin{longtable}{ccc}
\caption{Pre- (taken from the outermost layer of the WD) and post-explosive ejected abundances normalized to solar of all the 30 $p$-only isotopes
} \\
\hline
 Specie & X/X$_{\odot}$ Pre-explosive & X/X$_{\odot}$ Post-explosive \\
\hline
 SE 74  & 1.240E-02 & 9.418E+01  \\ 
 KR 78  & 8.860E-03 & 6.279E+01  \\ 
 SR 84  & 9.325E-03 & 1.256E+02  \\ 
 MO 92  & 2.560E-01 & 8.790E+01  \\ 
 MO 94  & 9.970E-01 & 9.417E+01  \\ 
 RU 96  & 5.250E-02 & 1.005E+02  \\ 
 RU 98  & 4.967E-01 & 2.511E+02  \\ 
 PD102  & 9.394E-03 & 1.256E+03  \\ 
 CD106  & 1.792E-02 & 1.570E+03  \\ 
 CD108  & 6.273E+00 & 7.533E+02  \\ 
 SN112  & 3.995E-02 & 1.569E+03  \\ 
 SN114  & 1.555E+00 & 1.256E+03  \\ 
 TE120  & 2.696E-03 & 1.255E+03  \\ 
 XE124  & 1.216E-03 & 5.023E+03  \\ 
 XE126  & 2.353E-02 & 5.651E+03  \\ 
 BA130  & 7.751E-04 & 2.198E+03  \\ 
 BA132  & 1.554E-02 & 1.884E+03  \\ 
 LA138  & 8.211E-03 & 3.704E+00  \\ 
 CE136  & 1.612E-02 & 1.450E+03  \\ 
 CE138  & 1.212E-01 & 1.689E+03  \\ 
 SM144  & 1.980E-01 & 3.704E+03  \\ 
 DY156  & 1.976E-04 & 2.574E+03  \\ 
 DY158  & 1.431E-01 & 1.249E+03  \\ 
 ER162  & 1.810E-04 & 2.511E+03  \\ 
 YB168  & 2.509E-04 & 6.907E+03  \\ 
 HF174  & 3.189E-04 & 6.279E+03  \\ 
 W 180  & 2.598E+00 & 1.256E+04  \\ 
 OS184  & 1.310E-03 & 5.023E+03  \\ 
 PT190  & 2.682E-03 & 2.198E+03  \\ 
 HG196  & 5.216E-02 & 7.848E+03  \\ 
\hline
\label{tab:prepost}
\end{longtable}
\end{center}
\end{table}


\end{document}